\documentclass[10pt]{article}
\pdfoutput=1
\usepackage[utf8]{inputenc}
\usepackage{cite}
\usepackage{amsmath,amssymb,amsbsy,amstext,amsthm,simplewick,amsfonts}
\usepackage{graphicx}
\usepackage{wrapfig}
\usepackage{upgreek}
\usepackage{bm} 
\usepackage{framed}
\usepackage{bbm}
\usepackage{textcomp}
\usepackage{tikz}
\usepackage{pifont}
\usetikzlibrary{matrix,shapes,fit,tikzmark,calc}
\usepackage{adjustbox}
\usepackage{makecell}
\usepackage{tcolorbox}
\usepackage{physics}
\usepackage{empheq}
\usepackage[normalem]{ulem}
\usepackage{enumitem}
\usepackage{braket} 
\usepackage{array}
\usepackage{dsfont}
\usepackage{ulem}
\usepackage{titlesec}
\titleformat{\section}{\normalfont\fontsize{12}{16}\bfseries}{\thesection}{1em}{}

\numberwithin{equation}{section}

\def\be{\begin{equation}}
\def\ee{\end{equation}}

\def\e{\epsilon}

\def\ba{\begin{eqnarray}}
\def\ea{\end{eqnarray}}
\def\k{\kappa}

\def\O{\Omega}
\def\bfx{\textbf{x}}

\def\vpi{\varphi}
\def\bfk{\textbf{k}}
\def\bfs{\textbf{s}}
\def\bft{\textbf{t}}
\def\bfu{\textbf{u}}
\def\bfx{\textbf{x}}

\def\bfs{\textbf{s}}

\usepackage{caption}
\usepackage{subcaption}
\usepackage[framemethod=default]{mdframed}
\newmdenv[skipabove=7pt,
skipbelow=7pt,
rightline=false,
leftline=false,
topline=false,
bottomline=false,
backgroundcolor=gray!10,
linecolor=gray,
innerleftmargin=5pt,
innerrightmargin=5pt,
innertopmargin=5pt,
innerbottommargin=5pt,
leftmargin=0cm,
rightmargin=0cm,
linewidth=4pt]{eBox}
\newmdenv[skipabove=7pt,
skipbelow=7pt,
rightline=false,
leftline=false,
topline=false,
bottomline=false,
backgroundcolor=gray!10,
linecolor=gray,
innerleftmargin=5pt,
innerrightmargin=5pt,
innertopmargin=-5pt,
innerbottommargin=5pt,
leftmargin=0cm,
rightmargin=0cm,
linewidth=4pt]{eBox2}

\definecolor{blue3}{RGB}{31,119,180}
\definecolor{red3}{RGB}{214,39,40}
\definecolor{orange3}{RGB}{255,127,14}
\definecolor{green3}{RGB}{44,160,44}

\newcommand{\then}{\quad\Rightarrow\quad}

\usepackage{colortbl}
\definecolor{lightgreen}{cmyk}{0.2, 0, 0.2, 0.2}
\definecolor{lightgray}{cmyk}{0.1,0.2,0,0.1}
\definecolor{lightgray2}{cmyk}{0.1,0.1,0,0.1}

\usepackage{hyperref}
\hypersetup{colorlinks=true,linkcolor=teal,citecolor=orange3,urlcolor=green3,pdfencoding=auto}

\makeatletter
\newlength{\apb@width}
\newcommand{\autoparbox}[2][c]{\settowidth{\apb@width}{#2}\parbox[#1]{\apb@width}{#2}}

\makeatother


\def\k{{\bf k}}

\def\O{{\cal O}}

\def\bfp{\textbf{p}}

\def\beq{\begin{equation}}
\def\eeq{\end{equation}}
\newcommand{\ex}[1]{\langle #1 \rangle}

\newcommand{\vk}{\mathbf{k}}
\newcommand{\BPE}{B^{\text{PE}}}
\newcommand{\BPO}{B^{\text{PO}}}

\allowdisplaybreaks[1]
\begin{document}


\begin{titlepage}
\setcounter{page}{1} \baselineskip=15.5pt

\thispagestyle{empty}

\renewcommand*{\thefootnote}{\fnsymbol{footnote}}

\begin{center}

{\fontsize
{20}{20} \bf Parity violation in the scalar trispectrum: \\ \vspace{0.1cm}
no-go theorems and yes-go examples \;} \\
\end{center}

\vskip 18pt
\begin{center}
\noindent
{\fontsize{12}{18}\selectfont Giovanni Cabass\footnote{\tt gcabass@ias.edu}$^{,a}$, Sadra Jazayeri\footnote{\tt jazayeri@iap.fr}$^{,b}$, Enrico Pajer\footnote{\tt enrico.pajer@gmail.com}$^{,c}$ and David Stefanyszyn\footnote{\tt david.stefanyszyn@nottingham.ac.uk}$^{,d,e}$}
\end{center}

\begin{center}
\vskip 8pt
$a$\textit{ School of Natural Sciences, Institute for Advanced Study, Princeton, NJ 08540, United States} \\ 
$b$  \textit{ Institut d'Astrophysique de Paris, GReCO, UMR 7095 du CNRS et de Sorbonne Universit\'{e},\\ 98bis
boulevard Arago, 75014 Paris, France} \\ 
$c$\textit{ Department of Applied Mathematics and Theoretical Physics, University of Cambridge, Wilberforce Road, Cambridge, CB3 0WA, UK} \\
$d$ \textit{Nottingham Centre of Gravity, University of Nottingham, University Park, Nottingham, NG7 2RD, UK} \\
$e$ \textit{School of Mathematical Sciences \& School of Physics and Astronomy,
University of Nottingham, University Park, Nottingham, NG7 2RD, UK} \\
\end{center}


\vspace{1.4cm}

\noindent We derive a set of no-go theorems and yes-go examples for the parity-odd primordial trispectrum of curvature perturbations. We work at tree-level in the decoupling limit of the Effective Field Theory of Inflation and assume scale invariance and a Bunch-Davies vacuum. We show that the parity-odd scalar trispectrum vanishes in the presence of any number of scalar fields with arbitrary mass and any parity-odd scalar correlator vanishes in the presence of any number of spinning fields with massless de Sitter mode functions, in agreement with the findings of Liu, Tong, Wang and Xianyu \cite{Liu:2019fag}. The same is true for correlators with an odd number of conformally-coupled external fields. We derive these results using both the (boostless) cosmological bootstrap, in particular the Cosmological Optical Theorem, and explicit perturbative calculations. We then discuss a series of yes-go examples by relaxing the above assumptions one at the time. In particular, we provide explicit results for the parity-odd trispectrum for (i) violations of scale invariance in single-clock inflation, (ii) the modified dispersion relation of the ghost condensate (non-Bunch-Davies vacuum), and (iii) interactions with massive spinning fields. Our results establish the parity-odd trispectrum as an exceptionally sensitive probe of new physics beyond vanilla inflation.


\end{titlepage}


\setcounter{tocdepth}{2}
{
\hypersetup{linkcolor=black}
\tableofcontents
}

\renewcommand*{\thefootnote}{\arabic{footnote}}
\setcounter{footnote}{0}

\section{Introduction}

Usually we try to constrain the laws of physics at very high energies by attentively staring into the night sky. In this work, we instead stare into a mirror that stares into the night sky. If the relevant laws of physics during the primordial universe violate parity (point inversion), the image in the mirror will appear to belong to a universe that is not ours. \\

\noindent When inquiring about parity violation in the primordial universe, it is natural to ask what observables are most sensitive to this effect. The answer splits into two relevant cases: either we only have access to the scalar sector of primordial correlators, or we also observe correlators of spinning fields such as the tensor sector involving primordial gravitational waves. When considering spinning fields, parity violation can already be detected at the level of the two point function, where for example, the two helicities of the graviton can have a different power \cite{Creminelli:2014wna}. All higher order correlators involving spinning fields can also show signs of parity violation, although they are often more constrained compared to their parity-even counterparts as recently shown for the graviton bispectrum \cite{Cabass:2021fnw} (see also \cite{Orlando:2022rih}). Conversely, if we are only allowed to consider correlators of scalar fields,\footnote{By scalar field we mean a field invariant under rotations and under parity. In particular, in our nomenclature a pseudo-scalar is not a scalar field.} we have to dig much deeper. In this case, for the two- and three-point scalar correlators, namely the power spectrum and the bispectrum, invariance under rotations and translations automatically implies invariance under parity. The reason is simple: working in Fourier space, momentum conservation requires all momenta to lie on one plane and then a rotation perpendicular to this plane is identical to a parity transformation.\footnote{More generally, parity violation must vanish for any $n$-point scalar correlation function in $d$ spatial dimensions where the $n-1$ independent momenta span a subspace of dimension less than $d$. As a corollary, parity violation can appear only for $n>d$. It would be interesting to see if this has any implications for theories with extra dimensions.} The first time that parity violation can manifest itself in scalar correlators is therefore in the four-point function, a.k.a.~the trispectrum, which is the focus of this work. We will derive a set of no-go theorems and yes-go examples for parity violation in this scalar trispectrum. Other discussions of the parity-odd scalar trispectrum in the literature can be found in \cite{Shiraishi:2016mok,Liu:2019fag}, while for the parity-even trispectrum see \cite{Chen:2006dfn,Seery:2006vu,Seery:2008ax,Arroja:2008ga}. \\

\noindent Our motivation for this work is twofold. As we have just discussed, in the scalar sector the trispectrum is the leading possible signal of parity violation so understanding its properties is important. Given that parity violation in the graviton bispectrum is highly restricted \cite{Cabass:2021fnw}, one might expect similar restrictions for the scalar trispectrum and indeed we will show that this is the case. Furthermore, two recent papers \cite{Philcox:2022hkh,Hou:2022wfj} searched for signs of parity violation in the BOSS galaxy survey, and found hints of parity violation. Although further investigation is clearly necessary to exclude non-primordial sources of parity violation, we take these preliminary and tantalising findings as further motivation to ask about the microscopic underpinning of parity violation in the primordial scalar trispectrum. \\

\noindent It will be convenient to decompose a general parity violating scalar correlator into a parity odd (PO) and a parity even contribution (PE) as
\begin{align}
    \bigg\langle{\prod_{a=1}^n \phi(\vk_a)\bigg\rangle} &= (2\pi)^3 \delta^{(3)}_{\rm D}\bigg(\sum_{a=1}^n \vk_a \bigg) B_n( \{ \vk \} )\,\,, \\
    \BPE_n(\vk_1,\dots,\vk_n) &\equiv \frac{1}{2}\left[ B_n(\{ \vk \}) +B_n(-\{ \vk \})\right]\,\,, \\
  \BPO_n(\vk_1,\dots,\vk_n) &\equiv \frac{1}{2}\left[ B_n(\{ \vk \}) -B_n(-\{ \vk \})\right]\,\,,
\end{align}
where in $B_{n}$ we factor out the ever-present momentum conserving delta function that appears due to spatial homogeneity. Recall that the expectation value of Hermitian operators must be real. Real fields are Hermitian operators in position space, $\phi(\bfx)^\dagger = \phi(\bfx)$, but their Fourier transform is not. Indeed, the reality of $\phi(\bfx)$ requires $\phi(\bfk)^\dagger = \phi(-\bfk)$ and so $\phi(\bfk)$ is a parity transformation away from being Hermitian. As a consequence, a general parity-violating correlator in Fourier space is a complex number and its parity-even and parity-odd components correspond to its real and ($i$ times) imaginary part respectively i.e.
\begin{align}
\BPE_n &= \Re B_n \in \mathbb{R} \,\,, \\  \label{POi}
\BPO_n &=i \Im B_n \in i \times \mathbb{R}\,\,.
\end{align}
In this paper we study $\BPO_4$ for curvature perturbations generated during inflation around a quasi de Sitter spacetime. We work in perturbation theory at tree-level and we make ample use of many recent results derived within the cosmological bootstrap program. In particular, we use the Cosmological Optical Theorem (COT) \cite{COT}, which is a consequence of unitary time evolution and the choice of the Bunch-Davies vacuum, the Manifestly Local Test (MLT) \cite{MLT}, which is necessary condition for manifestly local interactions, and recent results that provide exact expressions for correlators involving massive fields \cite{CosmoBootstrap1,CosmoBootstrap2}. We derive a series of no-go theorems and discuss a number of yes-go examples that invalidate different assumptions of the no-go theorems. We work in the decoupling limit of the Effective Field Theory of Inflation (EFTI) \cite{Cheung:2007st}, where the small effects of dynamical gravity are ignored, so our main object of interest is the trispectrum of the EFTI Goldstone mode $\pi(\eta, \bfx)$ that non-linearly realises the broken time translations. Our no-go theorems crucially rely on scale invariance, which in turn fixes the overall scaling of cosmological correlators with the various momenta, and we will show how breaking scale invariance allows for many possible parity odd trispectra, and we give concrete examples. 
\paragraph{Summary of the results} Our main results are summarized as follows. We prove a series of no-go theorems for generating a parity-odd scalar trispectrum $\BPO_4$ and higher-point correlation functions $\BPO_n$ of curvature perturbations. In particular, we prove that $\BPO_n=0$ for any $n\geq 2$ at tree-level assuming scale invariance and a Bunch-Davies state in single-clock inflation. This had already been derived by Liu, Tong, Wang and Xianyu in \cite{Liu:2019fag} by manipulation of the explicit perturbative contributions. Our re-derivation stresses the importance of the assumption of unitarity and the choice of vacuum. We extend these results in a few different directions, always under the assumption of a Bunch-Davies vacuum and scale invariance at tree level.
\begin{itemize}
    \item $\BPO_4=0$ in the presence of any number of scalars of any mass. 
    \item $\BPO_n =0$ for any number of fields of any spin as long as they all have massless or conformally-coupled de Sitter mode functions (see \eqref{dSmode}). Crucially, this relies on the interaction being IR-finite, which corresponds to the restriction $n_i+2n_\eta\geq4$ with $n_{i,\eta}$ the number of spatial and time derivatives respectively. Interactions with only three spatial derivatives, such as the one in \eqref{ni3}, give rise to $\log[\eta_0]$ late-time divergences in the wavefunction and then unitarity demands an associated non-vanishing contribution to the $n$-point correlator (see \eqref{IRdiv}), $\BPO_n\neq 0$, as anticipated in \cite{Cabass:2021fnw}. 
    \item $\BPO_4=0$ in the presence of full de Sitter isometries, or equivalently full conformal invariance at the boundary (Sec. \ref{ssec:conf}). This is true even before imposing soft theorems, which would require this contribution to vanish anyways in single-clock inflation \cite{Green:2020ebl}. 
    \item Small corrections to the linear dispersion implied by the Bunch-Davies vacuum do not alter any of the above conclusions in perturbation theory. This applies also to quadratic mixing operators in the action. 
\end{itemize}
Then, we derive a series of yes-go examples in which one of the assumptions of the no-go results is relaxed.
\begin{itemize}
    \item For non-scale-invariant interactions we find $\BPO\neq 0$, as expected for example from the discussion in \cite{Soda:2011am}. This can happen in two ways: via IR-divergences that break scale invariance at the scale we cut-off the time integrals $\eta_{0}$, as discussed above, or because of time-dependent couplings that can generally arise in the EFTI. A non-vanishing contribution arises already at the level of a single contact interaction in single-clock models. The size of the parity-odd non-Gaussianity depends on how strongly scale invariance is broken.
    \item As an example of a non-Bunch-Davies vacuum we consider the non-linear dispersion relation of the Ghost condensate \cite{GhostCondensate,GhostInflation}, $\omega^2 \propto k^4$. We show that both the leading and the subleading (in the EFT expansion) parity-odd self-interactions can give rise to a non-zero parity-odd signal. Our final results are, respectively 
    \begin{align}
    B^\zeta_{4} &= {\frac{128i\pi^3 \Lambda^5(H\tilde{\Lambda})^{1/2}}{M_{\rm PO} \tilde{\Lambda}^5 \Gamma(\frac{3}{4})^2}}(\Delta^2_\zeta)^3\frac{(\k_2\cdot\k_3\times \k_4)(\k_1\cdot \k_3)(\k_1\cdot \k_2)(\k_3\cdot \k_4)}{k_1^{\frac{3}{2}}k_2^{\frac{3}{2}}k_3^{\frac{3}{2}}k_4^{\frac{3}{2}}}\,{\rm Im}\,{\cal T}(k_1,k_2,k_3,k_4) \nonumber \\
    &\;\;\;\; + \text{$23$ perms.}\,\,, 
    \\
    B^\zeta_{4} &= {\frac{512i\pi^3 \Lambda^5(H\tilde{\Lambda})^{3/2}}{\Lambda_{\rm PO}^2 \tilde{\Lambda}^6 \Gamma(\frac{3}{4})^2}}(\Delta^2_\zeta)^3(\k_2\cdot\k_3\times \k_4)(\k_2\cdot \k_3) k_1^{\frac{1}{2}}k_2^{-\frac{3}{2}}k_3^{\frac{1}{2}}k_4^{\frac{1}{2}}\,{\cal T}(k_1,k_2,k_3,k_4) + \text{$23$ perms.} 
    \end{align}
    In these two equations $\mathcal{T}$ is given respectively by 
    \be
    {\cal T} = \int_{0}^{+\infty}{\rm d}\lambda\, \lambda^{11}\,H^{(1)}_{{\frac{3}{4}}}(2ik^2_1\lambda^2)H^{(1)}_{{\frac{3}{4}}}(2ik^2_2\lambda^2)H^{(1)}_{{\frac{3}{4}}}(2ik^2_3\lambda^2)H^{(1)}_{{\frac{3}{4}}}(2ik^2_4\lambda^2) 
    \ee
    and 
    \be
    {\cal T} = \int_{0}^{+\infty}{\rm d}\lambda\, \lambda^{13}\,H^{(1)}_{{-\frac{1}{4}}}(2ik^2_1\lambda^2)H^{(1)}_{{\frac{3}{4}}}(2ik^2_2\lambda^2)H^{(1)}_{{\frac{3}{4}}}(2ik^2_3\lambda^2)H^{(1)}_{{\frac{3}{4}}}(2ik^2_4\lambda^2)\,\,,
    \ee
    $\Lambda$ and $\tilde\Lambda$ are the energy scales entering the non-linear dispersion relation ($\smash{\omega=\tilde{\Lambda}k^2/\Lambda^2}$ in flat space, see also Eq.~\eqref{LLt} for the quadratic action in de Sitter spacetime), $M_{\rm PO}$ and $\Lambda_{\rm{PO}}$ set the scale of the parity-odd interaction for the leading and subleading interactions (Eqs.~\eqref{quartic_action_GI_new} and \eqref{quartic_action_GI}, respectively), and finally $\Delta_\zeta^2 $ is the amplitude of the curvature power spectrum, $\smash{P_\zeta(k) = \Delta^2_\zeta/k^3}$. 
    \item The tree-level exchange of massive spinning fields leads to a non-vanishing $\BPO_4$, whose overall size depends on the mass. For the explicit example of a spin-$1$ vector field we find the final result in \eqref{FactorisedTrispectrumCompact}, which we report here 
    \begin{equation}
\begin{split}
 B_{4}^{\zeta} &= {-\left(\prod_{a=1}^4P_\zeta(k_a) \right)} \frac{c_{s}^4\lambda_{1}\lambda_{3}}{H^3} (s^2-k_{1}^2-k_{2}^2)(s^2-k_{3}^2-k_{4}^2)(k_{1}-k_{2})(k_{3}-k_{4}) (\k_3\cdot\k_4\times \k_2) \\ 
 &\;\;\;\; \times [k_{12}I_{3}(c_{s}k_{12},s)+ i c_{s}k_{1}k_{2}I_{4}(c_s k_{12},s)][k_{34}I_{4}(c_{s}k_{34},s)+ i c_{s}k_{3}k_{4}I_{5}(c_{s}k_{34},s)] \\ 
 &\;\;\;\;\times \sin \left(\frac{\pi}{2}(\nu+1/2)  \right) \cos \left(\frac{\pi}{2}(\nu+1/2)  \right) +  [ (1,2) \leftrightarrow (3,4)] \\
 &\;\;\;\; + t + u\,\,.
\end{split}
\end{equation}
Here $I_n$ are the integrals defined in Eq.~\eqref{eq:nima_integral}:
\begin{align}
I_{n}(a,b) = (-1)^{n+1} \frac{H}{\sqrt{2 b}} \left(\frac{i}{2b} \right)^{n} \frac{\Gamma \left(\alpha \right)\Gamma \left(\beta \right)}{\Gamma(1+n)} \times {}_2 F_1 \Big(\alpha, \beta; 1+n; \frac{1}{2}-\frac{a}{2b}\Big)\,\,,
\end{align}
where $\alpha=\frac{1}{2}+n-\nu$, $\beta =\frac{1}{2}+n + \nu$ and the mass of the exchanged field enters via $\smash{\nu=\sqrt{9/4-m^2/H^2}}$. The corresponding overall amplitude is shown in Fig.~\ref{fig:mass_dependence}. In agreement with the above no-go results we find that $\BPO_4=0$ when the exchanged vector field has mode functions corresponding to $m=0$ or $m^2=2H^2$. 
\end{itemize}
Before concluding, we stress that in this work we discuss exclusively tree-level processes. Remarkably, loop contributions can be the leading source of $\BPO_4$, but we defer a thorough discussion of this exciting possibility to an upcoming paper. \\

\noindent The rest of this paper is structured as follows. In Section \ref{sec:2} we set the stage by very briefly reviewing the Effective Field Theory of Inflation (EFTI), showing how the trispectrum is related to the wavefunction of the Universe, and fixing a normalisation for the trispectrum. In Section \ref{sec:3} we discuss a series of no-go theorems to produce a non-vanishing $\BPO_n$ using either scalars of any mass at tree-level or fields of any spin with massless and conformally-coupled de Sitter mode functions. We then show how relaxing different assumptions leads to different predictions for a non-vanishing $\BPO_4$. In Section \ref{sec:4} we relax the assumption of exact scale invariance and show how this can lead to non-vanishing $\BPO_4$ thanks to either IR-divergences or time-dependent couplings. In Section \ref{sec:5}, we relax the condition of a Bunch-Davies vacuum and as a concrete example we study the Ghost Condensate, which features a non-linear dispersion relation. Then, in Section \ref{sec:6} we allow for massive spinning fields and compute $\BPO_4$ due to the exchange of a massive spin-$1$ field. We conclude in Section \ref{sec:7}. \\

\noindent While this paper was being finalised, another paper appeared that also considers the effects of massive particle exchange on parity-odd inflationary correlators \cite{Qin:2022fbv}.

\paragraph{Notations and conventions} The exchanged momenta and energy in a four-point exchange diagram are defined by
\begin{align}\label{stu}
\bfs &= \bfk_{1}+\bfk_{2}\,\,, & \bft &= \bfk_{1}+\bfk_{3}\,\,, & \bfu &= \bfk_{2}+\bfk_{3} \,\,, \\
s &= |\bfk_{1}+\bfk_{2}|\,\,,&t& = |\bfk_{1}+\bfk_{3}|\,\,, & u& = |\bfk_{2}+\bfk_{3}|\,\,.
\end{align}
These energies satisfy the non-linear relation
\begin{align} 
\label{stuRelation}
\sum_{a=1}^4 k_{a}^2 = s^2 + t^2 + u^2\,\,.     
\end{align}


\section{Generalities}
\label{sec:2}

In this section we summarize some results on the wavefunction approach to quantum field theory in curved spacetime and discuss the action of the Effective Field Theory of Inflation (EFTI) in the decoupling limit where we neglect the effects of dynamical gravity. Finally, we setup the normalisation of the trispectrum that we will use throughout this paper.


\subsection{The Effective Field Theory of Inflation and the decoupling limit}

In the EFTI \cite{Cheung:2007st} (see e.g.~\cite{snowmass} for a review), operators are constructed from building blocks that are invariant under spatial diffeomorphisms. These can be constructed from $g^{00}$ and the normal $n_\mu$ to the hypersurfaces of constant time. For example, one can consider the extrinsic curvature $K_{\mu\nu}$ of these hypersurfaces. In unitary gauge the action takes the form \cite{Cheung:2007st}
\be
\label{efti_ug_action}
\begin{split}
S = \int{\rm d}^4x\,\sqrt{-g}\,\bigg[&\frac{M^2_{\rm P}}{2}R + M^2_{\rm P}\dot{H}g^{00} - M^2_{\rm P}(3H^2 + \dot{H}) + \frac{M^4_2}{2}(g^{00}+1)^2 + \frac{M^4_3}{6}(g^{00}+1)^3 + \cdots \\
&- \frac{\bar{M}^3_1}{2}(g^{00}+1)\delta\!K - \frac{\bar{M}^2_2}{2}\delta\!K^2 + \cdots +\text{higher-order operators}\bigg] 
\end{split}
\,\,, 
\ee 
where $\delta\!K_{\mu\nu} = K_{\mu\nu} - Hh_{\mu\nu}$, and $h_{\mu\nu} = g_{\mu\nu} + n_\mu n_\nu$. The first three terms correspond to the minimal action of slow-roll inflation, with Einstein gravity plus the terms required to make the background metric a consistent solution. All remaining terms in the action start at quadratic order or higher in perturbations, so tadpole cancellation is guaranteed to all orders. Here we are primarily interested in the decoupling limit of inflationary theories, which is well-suited to studying large non-Gaussianities. The ever-present fluctuations of the clock, in this limit, are better described by the scalar degree of freedom $\pi(t, \bfx)$. In the decoupling limit this scalar mode decouples from metric fluctuations, and we can view it as the Goldstone boson of the spontaneously broken time translations. Moreover, $\pi$ inherits a shift symmetry which ensures an approximately scale invariant power spectrum of scalar fluctuations, as dictated by observations. On superhorizon scales, the relationship between $\pi$ and the comoving curvature perturbation is a simple rescaling: $\zeta = {-H\pi}$. Throughout we will take the background metric to be exact de Sitter, 
\begin{align}
{\rm d}s^2 = a^2(\eta)({- {\rm d}\eta^2} + {\rm d}{\bf x}^2)\,\,, \qquad a(\eta) = -\frac{1}{\eta H}\,\,,
\end{align}
where $H$ is the approximately constant Hubble parameter during inflation. This allows us to capture the leading contributions to inflationary correlators \cite{Maldacena:2002vr}. One can obtain the action for $\pi$ via the St\"uckelberg trick, which in the decoupling limit and neglecting terms suppressed by ${-\dot{H}/H^2}$ reads \cite{Cusin:2017mzw}\footnote{We are working in the limit of exact scale invariance which means that any time dependence in the action comes from the background metric rather than time-dependent couplings. We will, however, allow for time-dependent couplings when we come to discuss the breaking of scale invariance in Section \ref{sec:4}.} 
\begin{align}
g^{00}+1 &= {-2\dot{\pi}} - \dot{\pi}^2 + \frac{({\bm \nabla}\pi)^2}{a^2}\,\,, \label{building_blocks-1} \\
n_\mu &= \frac{\delta^0_\mu + \partial_\mu\pi}{\sqrt{1+2\dot{\pi} + \dot{\pi}^2 - \frac{({\bm \nabla}\pi)^2}{a^2}}}\,\,, \label{building_blocks-2} \\
\delta\!K^{i}_{\hphantom{i}j} &= {-(1-\dot{\pi})}\frac{\delta^{ik}\partial_k\partial_j\pi}{a^2} + \frac{H}{2}\frac{({\bm \nabla}\pi)^2}{a^2}\delta^i_j - H\frac{\delta^{ik}\partial_k\pi\partial_j\pi}{a^2} + \frac{\delta^{ik}\partial_{k}\dot{\pi}\partial_{j}\pi}{a^2} + \frac{\delta^{ik}\partial_{j}\dot{\pi}\partial_{k}\pi}{a^2} + {\cal O}(\pi^3)\,\,, \label{building_blocks-3}
\end{align} 
where $\delta\!K^i_{\hphantom{i}0}$, $\delta\!K^0_{\hphantom{0}j}$ and $\delta\!K^0_{\hphantom{0}0}$ can be obtained using the relations $\delta\!K^\mu_{\hphantom{\mu}\nu}n^\nu = 0$, $\delta\!K^\mu_{\hphantom{\mu}\nu}n_\mu = 0$ and $\delta\!K^\mu_{\hphantom{\mu}\nu}n_\mu n^\nu = 0$. For example, we have $\delta \!K^0_{\hphantom{0}j}(1+\dot{\pi}) = {-\delta \!K^i_{\hphantom{i}j}\partial_i\pi}$. While the action for $\pi$ coming from Eqs.~\eqref{efti_ug_action}, \eqref{building_blocks-1}, \eqref{building_blocks-2}, and \eqref{building_blocks-3} is naturally written in cosmic time $t$, it is in conformal time $\eta$ that scale invariance is manifest and is the time coordinate most suited to computing late-time cosmological correlators. We will switch between $t$ and $\eta$ in the rest of the work, while always denoting the Goldstone mode by $\pi$. \\

\noindent Two different regimes of the free theory of $\pi$ will be of interest in this paper \cite{Cheung:2007st}:  
\begin{itemize}
    \item First, there is the case where the free theory is dictated by the tadpole $\smash{M^2_{\rm P}}\dot{H}g^{00}$ and the operator $\smash{M^4_2}$. As is well known, $\smash{M^4_2}$ leads to a linear dispersion relation $\omega = c_sk$ for $\pi$, with speed of sound $\smash{1/c^2_s = 1-M^4_2/(\dot{H}M^2_{\rm P})}$. In the limit of small $c_s$ one has large non-Gaussianities of the equilateral and orthogonal form of order $1/c^2_s$ \cite{Cheung:2007st,Senatore:2009gt,Alvarez:2014vva} (see \cite{Cabass:2022wjy,DAmico:2022gki} for recent constraints on these non-Gaussianities from BOSS data). Another case that falls under the same umbrella is the ``de Sitter limit'' $\smash{\dot{H}\to 0}$, $\smash{c_s\to 0}$ such that $\smash{{-\dot{H}}M_{\rm P}^2(1-c^2_s)/c^2_s} = 2M^4_2$ stays fixed, and the free theory is dominated by the operators $M^4_2$ and $\bar{M}^3_1$. This leads again to a linear dispersion relation with a speed of sound $c_s$ of order $H/M\ll 1$ for $M_2\sim M_1\sim M\gg H$, and non-Gaussianities of order $(M/H)^2$. As far as the results of this work are concerned, this case can be grouped with the one with tadpole and $\smash{M^4_2}$ dominating the quadratic action since for both $\omega\propto k$. 
    \item Then, we will consider the de Sitter limit in which the free theory is determined by $M^4_2$, and terms quadratic in the extrinsic curvature. This leads to a quadratic dispersion relation $\omega\propto k^2$ and non-Gaussianities that scale as $\smash{1/P_\zeta^{{2}/{5}}}$. This case corresponds to the Ghost Inflation \cite{GhostInflation} limit of the EFTI. 
\end{itemize}
Let us again emphasize that the crucial difference between these regimes of the EFTI is the dispersion relation of $\pi$. As we will see in Sections~\ref{sec:3} and \ref{sec:5}, this will have very important implications for the existence of parity violation in the scalar trispectrum. \\

\noindent With regards to interactions, in this paper quartic self-interactions of the Goldstone will be important, as will cubic interactions that involve two Goldstone modes and one other field that will contribute to exchange diagrams. Such interactions can arise as the leading ones from some covariant operator in the EFTI, or as sub-leading terms required by the non-linearly realised symmetries. For example, quartic self-interactions can arise from operators with four building blocks, each which start at linear order in $\pi$, or from operators with two or three building blocks which can start at quadratic or cubic order in perturbations. Interestingly, in the decoupling limit this former choice cannot yield quadratic or cubic terms if the covariant operator is parity-odd as there are no such operators for scalars. This means that the quartic interactions are in some sense the leading ones, in this case. This does not mean that they are degenerate with the four building block operators, rather we expect that they can be distinguished by how they realise the leading order parts of the non-linear symmetries. Quartic vertices coming from four building block operators will be ``invariants'' in the context of non-linear realisations, whereas those coming lower building block operators will be ``Wess-Zumino'' terms. As far as we are aware no classification of such Wess-Zumino terms has been performed and it is certainly an interesting avenue for future work, but given that in this work we aim to provide some examples of parity-violation in the trispectrum we will concentrate on the former possibility.


\subsection{The wavefunction, the density matrix and cosmological correlators}
\label{subsec_WFU}

\noindent To compute cosmological correlators we will use both the wavefunction and the in-in formalism. Here we review the former and discuss the latter in Section \ref{sec:3.6}. The wavefunction\footnote{As common in the phenomenological literature, here we are assuming the system is in a pure state. Given that we work with effective field theories with a limited validity in energy scale and that we study an expanding universe, one expects some (small) corrections from entanglement with high-energy modes, see e.g. \cite{Burgess:2014eoa}. It would be nice to have a systematic study of these corrections.} can be defined formally in terms of a bulk path integral with specified boundary conditions
\begin{align}\label{PI}
    \Psi[\phi,\eta_0]&=\bra{\phi} \ket{\Psi_{\eta_{0}} }= \int_{ \Omega  \; \text{at} \; \eta \to - \infty }^{\phi  (\eta_0) = \phi  } \mathcal{D} \Phi  \; e^{i S [\phi]}\,\,,
\end{align}
where $\smash{\ket{\Psi_{\eta_0}}}$ is the quantum state of the system at some late conformal time $
\eta_0\to 0$, $\Omega$ represents the initial condition at $\eta\to -\infty$,  $\bra{\phi}$ a basis of (non-normalizable) field eigenstates with eigenvalue $\phi(\bfk)$ and $S$ is some action functional of the fields that determines the theory under consideration. We will parameterize $\Psi$ in terms of wavefunction coefficients $\psi_{n}$, which contain all the dynamical information about bulk evolution and which can be written in the following way\footnote{The non-perturbative wavefunction can contain terms for which $\log \Psi$ is not an analytical functional of $\phi$ at $\phi=0$. We omit this possibility in our parameterization because no such terms appear in perturbation theory.}
\begin{align} 
\label{WFU}
\Psi[\eta_{0}, \phi] = \exp\left[{-\sum_{n=2}^{+\infty}}\frac{1}{n!} \int_{\bfk_{1} \ldots \bfk_{n}} \psi_{n}(\{ k \}, \{ \bfk \}) (2 \pi)^3 \delta^{(3)}_{\rm D} \bigg(\sum_{a=1}^n \vk_a \bigg)\phi(\bfk_{1}) \cdots \phi(\bfk_{n}) \right] \,\,.
\end{align}
Here $\{ k \}$ collectively denotes the energies\footnote{We refer to the magnitude of a spatial momentum vector as ``energy'' despite the absence of time translation symmetry in cosmology.} $k_a=|\bfk_a|$ of the $n$ external fields, $\{ \bfk \}$ collectively denotes their spatial momenta, and $\phi(\bfk)$ collectively represents all fields in the theory with indices suppressed. Notice that this parameterization does not require any saddle-point approximation of the bulk path integral that defines $\Psi$. In fact, the wavefunction coefficient can be found non-perturbatively from
\begin{equation}
\begin{split}
    \psi_n(\{ k \}, \{ \bfk \}) (2 \pi)^3 \delta^{(3)}_{\rm D} \bigg(\sum_{a=1}^n \vk_a \bigg)&=- \frac{\delta^n \log\Psi [\eta_0,\phi ]}{ \delta \phi_{\bfk_1}\cdots \delta \phi_{\bfk_n}  } \bigg|_{\phi  = 0} \,\,. 
\end{split}
\end{equation} 
Upon renormalization, $\psi_n$ can be computed to any desired order in perturbation theory including any number of loops. In this work we focus on the natural observables of the Poincar\'e patch of de Sitter and of inflationary cosmology, namely correlation functions of the equal-time product of fields at the future conformal boundary $\eta_0 \to 0$. Notice that the correlators of both massive fields and of derivatives of massless fields decay with some positive power of $\eta$. In the inflationary context, this corresponds to a suppression of these correlators by some positive powers of $e^{-N}$ with $N\simeq \O(50)$ the number of e-foldings of inflation. Hence we focus on computing only correlators of the product of massless fields, namely
\begin{align}
\bigg\langle{\prod_a^n\phi(\bfk_a)}\bigg\rangle_c=(2 \pi)^3 \delta^{(3)}_{\rm D} \bigg(\sum_{a=1}^n \vk_a \bigg) B_n(\{\bfk\})\,\,, \\
B_n(\{\bfk\})=\frac{\int D\phi \, \Psi[\phi]\Psi[\phi]^\ast \,\prod_a^n\phi(\bfk_a) }{\int D\phi \, \Psi[\phi]\Psi[\phi]^\ast }\,\,,
\end{align}
where ``$c$'' stands for connected, i.e. with a single overall momentum-conserving delta function, and $D\phi$ denotes a Euclidean three-dimensional path integral (in contrast with the $3+1$ Lorentzian path integral in \eqref{PI}). All the dynamical information that we need is now contained in $|\Psi[\phi]|^2$, which we recognize to be the diagonal of the density matrix $\rho$
\begin{align}
    \hat \rho&=\ket{\Psi}\bra{\Psi} \then \rho_{\phi\tilde\phi}=\bra{\phi} \hat \rho \ket{\tilde\phi}\,\,, \\
    |\Psi[\phi]|^2&=\Psi[\phi]\Psi[\phi]^\ast=\braket{\phi|\Psi}\braket{\Psi| \phi}= \rho_{\phi\phi}\,\,.
\end{align}
Analogously to what we did for the wavefunction, we can parameterize $\rho_{\phi\phi}$ as
\begin{align}
    \rho_{\phi\phi}=|\Psi|^2=  \exp\left[{-\sum_{n=2}^{+\infty}}\frac{1}{n!} \int_{\bfk_{1} \ldots \bfk_{n}} \rho_{n}(\{ k \}, \{ \bfk \}) (2 \pi)^3 \delta^{(3)}_{\rm D} \bigg(\sum_{a=1}^n \vk_a \bigg)\phi(\bfk_{1}) \cdots \phi(\bfk_{n}) \right] \,\,,
\end{align}
where the density matrix coefficients are related to the wavefunction coefficients by
\begin{align}
    \rho_n (\{ k \}, \{ \bfk \}) &= \psi_n (\{ k \}, \{ \bfk \}) +\psi_n^\ast (\{ k \}, \{ -\bfk \})\,\,.
\end{align}
In perturbation theory, correlators can be computed in terms of the $\rho_n$'s. For example, at tree level we have
\begin{align} \label{WFUtoCorrelators}
    B_2\equiv P&=\frac{1}{\rho_2}\,\,, &    B_n^{\text{contact}}&=-\frac{\rho_n}{\prod_{a=1}^n \rho_2(\bfk_a)} \,\,, &
    B_4&=-\frac{1}{\prod_{a=1}^4 \rho_2(\bfk_a)} \left[\rho_4 - \frac{\rho_3 \rho_3}{\rho_2} \right]\,\,,
\end{align}
where $P$ is the power spectrum. For the four-point function due to exchange processes, we refer to the contribution $\rho_3 \rho_3$ as the ``factorised contribution". While these expressions assume a single scalar field, it is straightforward to generalize them to any number of fields of any spin. In Section \ref{sec:6} we will present an explicit formula for $B_{4}$ due the exchange of spinning fields.


\subsection{Normalization of the parity-odd trispectrum} \label{sec:2.3}

Before proceeding, it is sensible to fix a normalisation for the trispectrum $\BPO_{4}$. Given that this object must be parity odd and invariant under any permutation of the four external momenta ${\bf k}_a$, it can always be written as 
\begin{align}
\label{F_definition-1}
\BPO_{4}(\{\bfk\})=\left( \bfk_1\times \bfk_2 \cdot \bfk_3\right) \, F_{123} + \text{$3$ perms.} = \left( \bfk_1\times \bfk_2 \cdot \bfk_3\right)  \, \left[ F_{123}-F_{234}+F_{134}-F_{124} \right]\,\,,
\end{align}
where $F_{abc}$ is an alternating function of three momenta
\begin{align}
\label{F_definition-2}
F_{abc}=F(\bfk_a,\bfk_b,\bfk_c)=\text{sign}(\sigma) F(\bfk_{\sigma(a)},\bfk_{\sigma(b)},\bfk_{\sigma(c)})\,\,,
\end{align}
for any permutation $\sigma$ with permutation-parity $\text{sign}(\sigma) = \pm 1$, and we have used momentum conservation to write the parity-odd part of the correlator as $\left( \bfk_1\times \bfk_2\right) \cdot \bfk_3$, without loss of generality. \\

\begin{figure}[h!]
\centering
\includegraphics[width=0.4\textwidth]{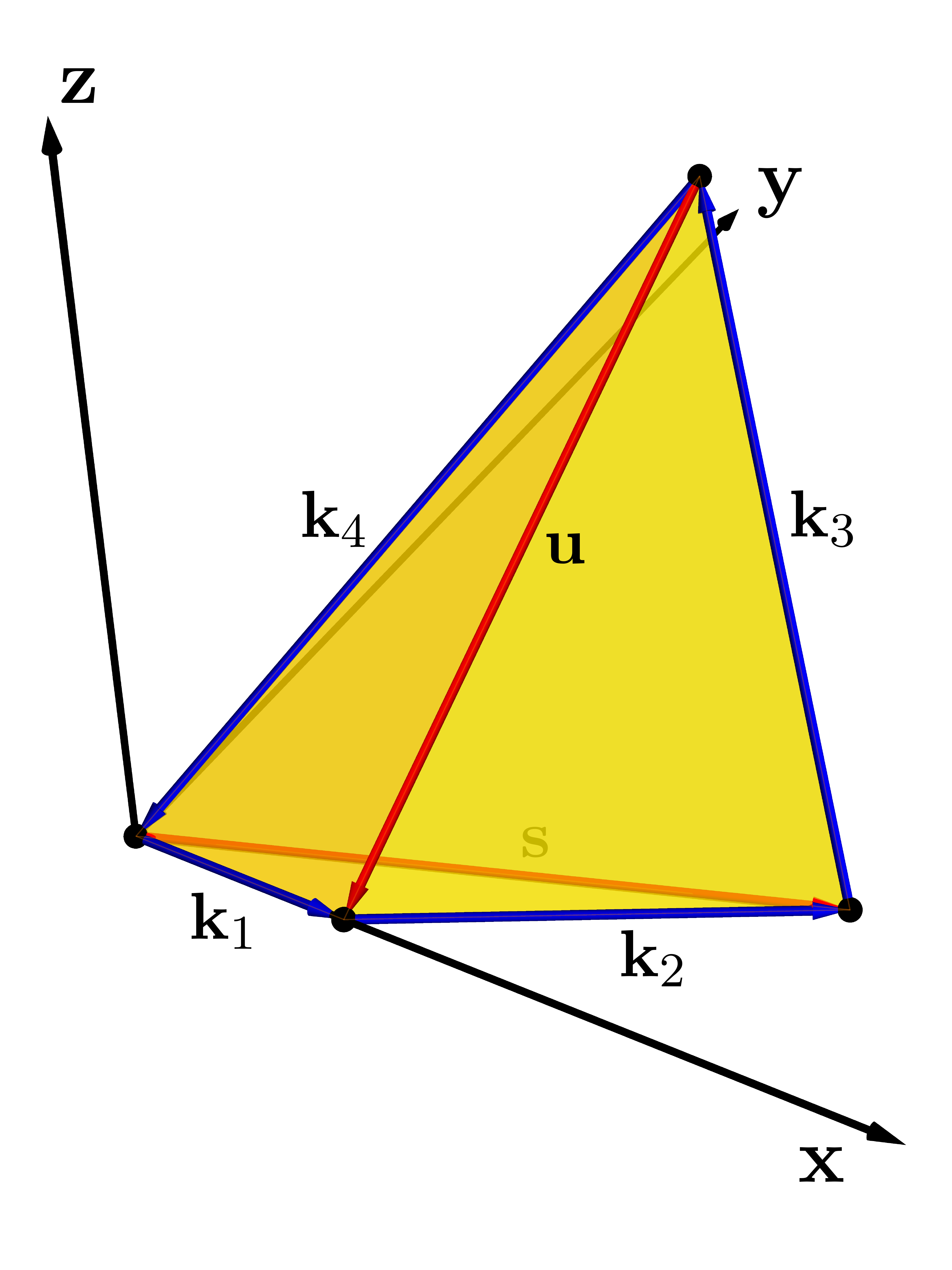}
\caption{Tetrahedron configuration \eqref{eq:tet_config} for the normalization condition and definition of $\smash{\tau_{\rm NL}^{\rm PO}}$ \eqref{normalization}.}
\label{fig:tet}
\end{figure}

\noindent To discuss the phenomenology of a given non-Gaussianity it is always useful to have a reference point in kinematic space where to specify the overall size of the signal. A highly symmetric choice of such a reference kinematic point is convenient for explicit calculations. For the bispectrum, the reference point is often taken to be the equilateral configuration where $k_1 = k_2 = k_3$. For the trispectrum we could also look for a very symmetric configuration. Let us first discuss how to visualize a given trispectrum configuration. We can draw a tetrahedron with four of the edges taken to be the ${\bf k}_a$. Momentum conservation, $\smash{\sum_{a=1}^4{\bfk}_a={\bf 0}}$, is then manifest if the $\bfk_a$ are connected to each other cyclically. The two remaining two edges of the irregular tetrahedron are then two of the Maldelstam-like variables, depending on the order in which $\bfk_a$ are chosen. In Fig.~\ref{fig:tet} we choose the ordering so that these edges are $\bf s$ and $\bf u$. We could then choose all $k_a$ to be equal, and specify the values of the diagonals $s$ and $u$ (Eq.~\eqref{stuRelation} relating the value of $t$ to them). One such option is a regular tetrahedron, corresponding to the values  
\begin{align}
k_1=k_2=k_3=k_4=s=u=\frac{t}{\sqrt{2}} \quad \text{(tetrahedron)}\,\,.
\end{align}
However, \emph{parity-odd} trispectra vanish in this configuration, so we instead look for a less symmetric one. We could not find a particularly convenient or simple choice of irregular tetrahedron that displays chirality \cite{tetrachirality}, so we settled for (see \eqref{stu} for the definitions of $s$, $t$ and $u$)
\begin{align}
k_1 = \frac{k_2}{2} = \frac{k_3}{3} = \frac{k_4}{\sqrt{14-\sqrt{3}}}= \frac{s}{\sqrt{5+2\sqrt{3}}} = \frac{t}{\sqrt{7}} = \frac{u}{\sqrt{16-3\sqrt{3}}}\,\,,
\end{align} 
which guarantees that even a parity-odd trispectrum that depends on the $k_a$ and only one of $s$, $t$ or $u$ does not vanish. We can choose the wavenumbers in this configuration to take the values
\begin{align}
\label{eq:tet_config}
\bar{\bfk}_1 &= \bar{k} (1,0,0)\,\,, &\bar{\bfk}_2 &= \bar{k} (\sqrt{3},1,0)\,\,, \\
\bar{\bfk}_3 &= \bar{k} \bigg({-\frac32},\frac32,\frac{3}{\sqrt{2}}\bigg)\,\,, &\bar{\bfk}_4 &= {-\bfk_1}-\bfk_2-\bfk_3\,\,. 
\end{align}
Here $\bar{k}$ is an arbitrary reference momentum, whose value is largely irrelevant if we assume scale invariance. As discussed above, we visualize this configuration by taking a tetrahedron whose edges are the four $\bfk_a$. Choosing one end of $\bfk_1$ as the origin we find Fig.~\ref{fig:tet}. Our definition for the overall normalization of the parity-odd trispectrum is then the following: given a trispectrum model, we isolate its imaginary part, and define 
\be
\label{normalization}
\tau_{\rm NL}^{\rm PO}(\bar{k})\equiv\frac{{\rm Im}B^\zeta_4(\bar{\bfk}_1,\bar{\bfk}_2,\bar{\bfk}_3,\bar{\bfk}_4)}{[P_\zeta(\bar{k}_1)P_\zeta(\bar{k}_2)P_\zeta(\bar{k}_3)P_\zeta(\bar{k}_4)]^{\frac{3}{4}}}\,\,. 
\ee

\section{No-go theorems} 
\label{sec:3}

\noindent In this section, we will present various properties of wavefunction coefficients that together enable us to derive no-go theorems that forbid parity violation in the scalar trispectrum (and other $n$-point functions). We will first show that having exact de Sitter symmetries forbids parity violation in the four-point function. Then we assume the Bunch-Davies vacuum and unitarity in the form of the Cosmological Optical Theorem (COT) \cite{COT}, and exact scale invariance to show that parity violation is also absent in the scalar trispectrum in a theory where all fields have massless/conformally-coupled mode functions and with boost-breaking interactions. We will work with the wavefunction in Section \ref{sec:3.2}, and directly with in-in correlators in Section \ref{sec:3.6}. \\

\noindent Many of our no-go results had already been derived by Liu, Tong, Wang and Xianyu in \cite{Liu:2019fag}. New results that we present here, and that were not explicitly discussed there, include: the connection to unitarity via the cosmological optical theorem; the no-go result from full conformal invariance at the boundary (Section~\ref{ssec:conf}); a discussion of the non-vanishing contribution to parity-odd correlators from IR-divergent interactions (see \eqref{IRdiv}); and the extension to fields of any spin and conformally-coupled mode functions.

\begin{figure}
    \centering
    \includegraphics[width=0.8\textwidth]{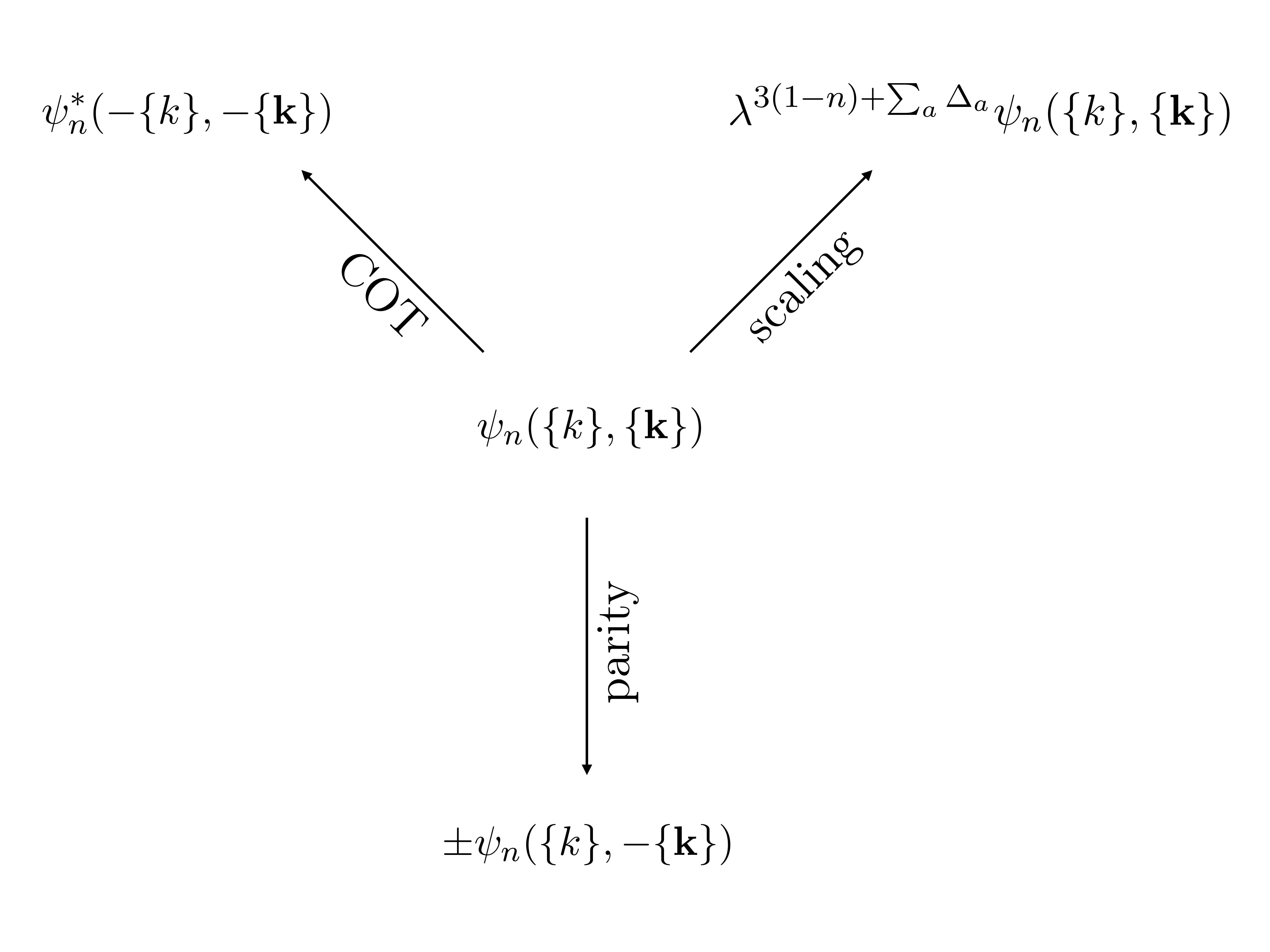}
    \caption{The flux capacitor in the figure summarises the three main relations that we use to prove a series of no-go theorems for parity odd correlators.}
    \label{fig:flux}
\end{figure}


\subsection{Conformally invariant parity-odd correlators}\label{ssec:conf}

The isometry group $SO(4,2)$ of de Sitter spacetime coincides with the Euclidean conformal group in $3$ spatial dimensions, identified with the spacelike future conformal boundary. These isometries must be broken in all cosmological models because de Sitter is maximally symmetric and as such cannot have any non-trivial history. However, it can happen that in some models the breaking is appropriately small and can be neglected in the first approximation. This is for example the case for slow-roll inflation with a canonical kinetic term or for more general inflationary models where some sector of the theory does not couple significantly to the boost-breaking inflaton background (e.g.~the graviton sector in $P(X,\phi)$ models). Then one can use (approximate) conformal symmetry to derive general results that do not rely necessarily on perturbation theory. For example, in \cite{Maldacena:2011nz} it was shown that there are only three de Sitter invariant cubic wavefunction coefficients for massless gravitons, and only two of these give a non-vanishing graviton bispectrum \cite{Soda:2011am}. This was generalized to mixed bispectra in \cite{Mata:2012bx} and to scalar bispectra in \cite{Pajer:2016ieg}. These results resonate with the expectation that conformal symmetry fixes three point functions \cite{Bzowski:2013sza}. Conversely, we expect an infinite class of conformally invariant four point functions, loosely corresponding to arbitrary functions of the conformal cross ratios. Hence, to make progress one needs to rely on perturbation theory to organize the infinitely many solutions of conformal Ward identities, as pioneered in \cite{CosmoBootstrap1} and later studied in \cite{CosmoBootstrap2,CosmoBootstrap3}. Here, we recount non-perturbative arguments that parity-odd conformal-invariant scalar four-point functions must vanish. For the first argument\footnote{This argument was explained to us by M. Mirbabayi.}, recall that conformal transformations
can bring any four points $x_a$ to the same plane (conventionally one sets three points on a line and the last point determines the plane, $x_1=0$, $x_3=1$, $x_4=\infty$ and $x_2$ arbitrary). A non-vanishing parity-odd scalar correlator requires contracting the three indices of the Levi-Civita symbol $\e_{ijk}$ with three linearly independent vectors. However, since all points are on one plane only two of them can correspond to independent vectors and hence the parity-odd contribution must vanish. A second argument relies on embedding space\footnote{This argument was explained to us by S. Jain.}. For a three-dimensional correlator we go to five dimensional embedding space. There the Levi-Civita symbol has five indices but we have only four independent points to contract them with, and once again the parity-odd contribution must vanish. \\

\noindent A non-vanishing parity-odd trispectrum of scalars must therefore arise from the breaking of de Sitter symmetries, and this motivates us to study more general situations in which de Sitter boosts are broken and time derivatives are unrelated to space derivatives. This is our focus for the rest of this paper. To develop a more general set of no-go theorems, in the following three subsections we derive some useful properties of wavefunction coefficients that hold under some mild assumptions.


\subsection{Derivation using the boostless cosmological bootstrap}\label{sec:3.2}

In this section we derive a series of no-go results for the parity-odd scalar trispectrum using techniques from the boostless cosmological bootstrap. Later, in Section \ref{sec:3.6}, we will generalize these results using explicit expressions of the in-in formalism. The main idea of the derivation, summarized in Figure \ref{fig:flux}, is to use a parity transformation and the cosmological optical theorem to show that the combination of wavefunction coefficients  appearing in the correlator (namely the diagonal density matrix) vanishes under certain assumptions. In the following we first discuss the three ingredients of the derivation in turn and then combine them together.


\paragraph{Parity transformations} Here we derive some general results involving parity, a.k.a.~point inversion, namely the simultaneous flipping of the sign of all spatial coordinates and Fourier momenta. In a generic parity-violating theory, quantities do not transform in a simple way under parity, but they can always be decomposed into a parity-even (PE) and a parity-odd (PO) component as we touched on in the introduction:
\begin{align}
    \psi_n^{\rm{PE}}(\{k\},\{\bfk\})&\equiv \frac12 \left[ \psi_n (\{k\},\{\bfk\}) +\psi_n (\{k\},\{-\bfk\})\right]\,\,, \\ \psi_n^{\rm{PO}}(\{k\},\{\bfk\})&\equiv \frac12 \left[ \psi_n (\{k\},\{\bfk\}) -\psi_n (\{k\},\{-\bfk\})\right]\,\,,
\end{align}
and similarly for correlators $B_n$. As we explained in Section \ref{sec:2}, correlators are related to wavefunction coefficients via the density matrix coefficients $\rho_n$ for which we have
\begin{align}
    \rho_n^{\rm{PO}} (\{k\},\{\bfk\}) &= \frac12 \left[ \rho_n (\{k\},\{\bfk\}) -\rho_n (\{k\},\{-\bfk\})\right] \\
    &= \frac12 \left[ \psi_n (\{k\},\{\bfk\}) +\psi_n^\ast (\{k\},\{-\bfk\})- \psi_n (\{k\},\{-\bfk\}) -\psi_n^\ast (\{k\},\{\bfk\}) \right]\,\,,
\end{align}
and so $\rho_n^{\rm{PO}}$ must purely imaginary if we are to find a non-vanishing parity-odd correlator. Similarly, one can easily see that $\rho_n^{\rm{PE}}$ must be real:
\begin{align}
    \rho_n^{\rm{PO}} &\in i \times \mathbb{R} & \rho_n^{\rm{PE}} &\in  \mathbb{R}\,\,.
\end{align}
Our no-go theorems will be based on asking when $\psi_{n}$, and therefore $\rho_n^{\rm{PO}}$, can be imaginary. \\

\noindent Furthermore, as observed in the introduction, the power spectrum and bispectrum of scalars cannot violate parity because we don't have three independent spatial momenta to contract the three indices of $\e_{ijk}$. This argument is non-perturbative. A similar argument also tells us that ``off-shell'' cubic vertices cannot break parity, irrespective of what the vertex is connected to in a perturbative diagram. This already tells us that the tree-level contribution to a four-point function from the exchange of a scalar of any mass cannot break parity.


\paragraph{Hermitian analyticity: a heritage from the Bunch-Davies vacuum} Here we derive a second important result that will enable us to relate wavefunction coefficients to their complex conjugate, which will prove useful when we come to analyse the properties of $\rho_{n}$. As emphasised in Ref.~\cite{COT}, the bulk-to-boundary propagators $K_k(\eta)$ and bulk-to-bulk propagators $G_p(\eta_1,\eta_2)$ appearing in the perturbative calculation of wavefunction coefficients enjoy the simple property\footnote{The prescription is that all real values of energies such as $k$ and $-k$ should be approached from the lower-half complex plane, so for complex energies this property becomes $K_{-k^\ast}(\eta)^\ast=K_{k}(\eta)$.}
\begin{align}\label{HA}
    K_{-k}(\eta)^\ast &=K_{k}(\eta)\,\,, & G_{-p}(\eta_1,\eta_2)^\ast & = G_p(\eta_1,\eta_2) \,\,.
\end{align}
Here we are using $k$ to denote an external energy, and $p$ to denote an internal one. This \textit{Hermitian analyticity} is easy to see by eye in the standard de Sitter massless mode functions, but actually holds very generally, namely for fields of any mass, any spin and in any FLRW spacetime, as long as one chooses the Bunch-Davies vacuum, which physically corresponds to the Minkowski vacuum at short distances \cite{COT,Goodhew:2021oqg}. Furthermore, interaction vertices with real coupling constants are also Hermitian analytic. In particular, $\partial_\bfx$ Fourier transforms to $i \bfk $, which satisfies the analog of \eqref{HA}. Time derivatives preserve the Hermitian analyticity relations in \eqref{HA} because they are linear and real operations on the propagators. In a less precise but more evocative way, we can say that \textit{perturbative unitary evolution preserves the analytic structure of the initial state}.\footnote{Here, we are equating unitary time evolution to the reality of coupling constants in the Hamiltonian. In principle, one can have imaginary couplings in a Hermitian Hamiltonian in the presence of non-Hermitian operators. One trivial example are interactions like $H_\text{int} \supset i \lambda \phi^n (\Pi\phi-\phi\Pi)$. However these can always be be re-written in terms of Hermitian operators and real couplings. It would be interesting to investigate whether there are relevant cases where imaginary couplings cannot be removed.}  \\

\noindent The Hermitian analytic properties above were used in a series of papers to derive several infinite sets of identities that go under the collective name of the Cosmological Optical Theorem \cite{COT} (see also \cite{Cespedes:2020xqq,Baumann:2021fxj}), e.g. cosmological cutting rules at loop level \cite{sCOTt} and single-cut rules at tree-level \cite{Goodhew:2021oqg}. For contact diagrams at tree-level one finds
\begin{align} 
\label{ContactCOT}
\psi_{n}(\{ k \}, \{ \bfk \}) + \psi_{n}^{\ast}( - \{ k \}, -  \{ \bfk \}) = 0 \,\,.
\end{align}
An $s$-channel four-point exchange diagram due to two cubic vertices satisfies
\begin{equation} 
\label{ExchangeCOT}
\begin{split}
&\psi_{4}(\{ k \},s,\{ \bfk \}) + \psi_{4}^{\ast}( - \{ k \},s, -  \{ \bfk \}) = \\
P_{s}&[\psi_{3,L}(k_{1},k_{2},s,\{ \bfk \}) + \psi_{3,L}^{\ast}(-k_{1},-k_{2},s,-\{ \bfk \})][\psi_{3,R}(k_{3},k_{4},s,\{ \bfk \}) + \psi_{3,R}^{\ast}(-k_{3},-k_{4},s,-\{ \bfk \})] \,\,,
\end{split}
\end{equation}
where here we have denoted the internal energy as $s$, and collectively denoted external energies and spatial momenta as $\{ k \}$ and $\{ \bfk \}$, respectively. $P_{s}$ is the power spectrum of the exchanged field. There are generalisations of this COT for exchange diagrams that apply to higher-point wavefunction coefficients, and we refer the reader to \cite{COT,Goodhew:2021oqg} for further details. The $t$ and $u$ channel expressions are simple generalisations of \eqref{ExchangeCOT}. We note that in all of these expressions, any tensor structures that depend on spatial momenta and polarisations factorise since spatial momenta always come with a factor of $i$ while polarisation vectors, and higher-spin generalisations, satisfy $e^{h}_{i}(\bfk) = [e^{h}_{i}(-\bfk)]^{\ast}$. It follows that these COT expressions constrain only the part of the wavefunction coefficients that arise due to time evolution, as expected.  \\

\noindent In summary, Hermitian analyticity, in the form of the COT, provides us with a way to relate $\psi_n$ to $\psi^{\ast}_n$, or equivalently $\Psi$ and $\Psi^\ast$. More intuitively, this gives us a relationship between the bra and the ket, which are both needed to compute correlators. The price to pay for removing the complex conjugate of $\psi_{n}$ from $\rho_{n}$ is that we have to analytically continue energies to unphysical negative values. As we will now discuss, we can go back to physical positive energies using momentum scaling as long as scale invariance is an exact symmetry of the boundary/late-time correlators.


\paragraph{Scaling symmetry} For our no-go results we will assume exact scale invariance. This is an interesting limit because scale invariance is an approximate symmetry of the observed primordial power spectrum of curvature perturbations with violations at the percent level. Deviations from scale invariance are discussed in Section \ref{sec:4}. \\

\noindent Since we want to model primordial perturbations, typically denoted by the (perturbatively) gauge invariant variable $\mathcal{R}$ or $\zeta$, we are interested in massless scalars. For IR-finite wavefunction coefficients this implies that wavefunction coefficients obey the following scaling relation
\begin{align} \label{MasslessScaling}
    \psi_n(\{\lambda k\},\{ \lambda  \bfk\})&=\lambda^3 \psi_n (\{ k\},\{\bfk\})\,\,,
\end{align}
where the $\lambda^3$ factor cancels the scaling of the Dirac delta function to ensure that the wavefunction is invariant. Since each $\psi_{n}$ comes with a single delta function in the wavefunction, this scaling holds for all $n$. This scaling relation holds when all external fields have massless de Sitter mode functions, and implies that $n$-point correlators scale as $B_{n}(\{ \lambda k \},\{ \lambda  \bfk\}) = \lambda^{3-3n} B_{n} (\{ k\},\{\bfk\})$ where the factor of $\lambda^{-3n}$ comes from the $n$ power spectra. All fields can be different and each can have any spin.\footnote{When de Sitter isometries are fully intact, the scaling dimension and the associated mode functions are fixed by the mass and spin. However, this is not the case once boosts are broken spontaneously and in the low-energy effective theory fields of any spin can have massless de Sitter mode functions, as nicely discussed in \cite{Bordin:2018pca}.} If some of the external mode functions are not the massless de Sitter ones, then the overall scaling changes to
\begin{align}\label{genscla}
    \psi_{n}(\{\lambda k\},\{ \lambda  \bfk\})=\lambda^{3(1-n)+\sum_{a}\Delta_{a}}\psi_{n}(\{ k\},\{\bfk\})\,\,,
\end{align}
where $\Delta=3/2+(9/4-m^2/H^2)^{1/2}$, so that $\Delta=3$ for $m=0$ and $\Delta=2$ for $m^2=2H^2$ (conformally-coupled). For a combination of fields with massless and conformally-coupled mode functions, one finds that the scaling is an integer. This overall integer scaling will allow us to eliminate negative energies and momenta in favour of positive ones. As we will explain in Section \ref{sec:4}, these scaling symmetries are not exact when there are IR-divergences, since we need to cut-off the time integrals at some scale $\eta_{0}$, or when the coupling constants in the action of perturbations have a non-trivial time-dependence. \\

\noindent We now have all the ingredients to derive our tree-level no-go theorems. We will consider contact diagrams and exchange diagrams separately

\paragraph{Contact diagrams in the wavefunction} There are two types of contact diagrams that can play a role in the scalar trispectrum: a quartic diagram that contributes to $B_{4}^{\rm{PO}}$ via $\rho_{4}$, and cubic contact diagrams that contribute to the factorised part via $\rho_{3} \rho_{3}$, c.f. Eq.~\eqref{WFUtoCorrelators}. The former must be a diagram with four external massless scalars, while the latter must involve two massless scalars and one other field which is the ``exchanged'' field. We will derive a no-go theorem for this factorised part when the exchanged field has massless or conformally-coupled mode functions. Later, in Section \ref{sec:6}, we will present explicit computations for the exchange of fields of generic mass. \\

\noindent First consider the quartic contact diagram where we combine the contact Cosmological Optical Theorem \eqref{ContactCOT} plus exact scale invariance \eqref{MasslessScaling} to conclude that $\psi_{4}$ is \textit{purely real}. It then directly follows that $\rho_{4}$ is also purely real and therefore there is no parity-odd contribution to the scalar trispectrum since this requires an imaginary $\rho_{4}$, as we explained in Section \ref{sec:3.2}. This observation results in our first no-go theorem:
\begin{center}
\textit{Scale invariant, IR-finite, parity-odd $n$-point correlators $\BPO_n$ from contact interactions of massless scalars with a Bunch-Davies vacuum vanish.}
\end{center}
This result was first derived in \cite{Liu:2019fag}. Our presentation emphasises the role of unitarity and the assumption of IR-finiteness, among other things. Notice that this result actually holds for all $n$-point contact diagrams since the contact COT and integer scaling symmetry apply generally. It also holds for gravitons, and was used in \cite{Cabass:2021fnw} to derive the set of highly constrained parity-odd graviton bispectra. \\

\noindent Now consider the cubic wavefunction coefficients that can contribute to the factorised part of an exchange trispectrum via the $\rho_{3}\rho_{3}$ contribution in \eqref{WFUtoCorrelators}.\footnote{Here we allowing for spinning fields on external lines and therefore the wavefunction coefficients will depend on polarisation vectors. In this case we do not simply pick up a $\pm$ when we flip the sign of all momenta since polarisation vectors do not have such a property. In this case, we simply need to first strip off all polarisation vectors and access the reality of what is left over. This is the correct thing to do since in $\rho_n$ the polarisation vectors factorise so it is the reality of the remainder that is important.} If we only have scalars, all parity-odd cubic interactions must vanish ``off-shell'' by momentum conservation. We are left with interactions with spinning fields, which we are assuming here have either massless or conformally-coupled mode functions. We need one of the cubic vertices to be parity-odd and the other to be parity-even which means that one of the $\rho_{3}$ needs to be imaginary while the other needs to be real. However, if we combine the contact COT \eqref{ContactCOT} with the fact that the cubic wavefunction coefficients have a fixed, integer scaling with the external momenta, it is simple to see that regardless of parity, the $\rho_{3}$ are always real or always imaginary. We therefore arrive at another no-go theorem: 
\begin{center}
\textit{The factorised contribution to the exchange trispectrum cannot be parity-odd under the assumptions that the constituent cubic wavefunction coefficients are IR-finite and involve fields with massless or conformally-coupled mode functions with Bunch-Davies vacuum conditions.}
\end{center}

\paragraph{Exchange diagrams in the wavefunction} Exchange contributions to the quartic wavefunction coefficient are slightly more complicated to analyse compared to their contact counterparts since the COT for exchange diagrams \eqref{ExchangeCOT} does not have a vanishing left-hand side, rather it relates the discontinuity of a quartic wavefunction coefficient to the product of discontinuities of cubic ones. Nevertheless, we can still derive a no-go theorem that states that these diagrams always vanish under the assumptions we have made throughout this discussion.  \\

\noindent The cubic wavefunction coefficients that appear on the right-hand side of Eq.~\eqref{ExchangeCOT} must satisfy the contact COT. Since we are assuming that they correspond to fields with massless or conformally-coupled mode functions, and that they are IR-finite, they have a fixed integer scaling with momentum which ensures that they are either purely real or purely imaginary, regardless of how they transformation under parity. It follows that the product of their discontinuities, and the left-hand side of \eqref{ExchangeCOT}, is always real. This by itself does not automatically imply that $\psi_{4}$ is real since an imaginary contribution could in principle cancel on the left-hand side of the COT, but let us now argue that this cannot be the case. We first recall that quartic wavefunction coefficients have a restricted set of singularities, see e.g. \cite{Arkani-Hamed:2017fdk}: the wavefunction can be singular as the total-energy goes to zero which is a property of almost all cosmological wavefunctions regardless of the diagram they come from,\footnote{Interestingly, on the leading total-energy pole we recover the flat-space scattering amplitude for the same process \cite{COT,Maldacena:2011nz,Raju:2012zr}.} but an $s$-channel exchange diagram can also yield ``partial-energy'' singularities where the partial-energies are a sum of energies that enter a sub-diagram. In this case the partial-energies are $E_{L} = k_{1}+k_{2}+s$ and $E_{R} = k_{3}+k_{4}+s$. If the constituent cubic wavefunctions are IR-finite, then the corresponding quartic wavefunction coefficient is rational so only poles can occur as we approach one of these singular kinematic points. Now as explained in \cite{MLT}, the exchange COT \eqref{ExchangeCOT} fixes \textit{all} residues of partial-energy poles since the second term on the left-hand side does not have partial-energy poles so there is no way they could be cancelled. Given that the left-hand side of the COT is fixed to be real, all of these partial-energy poles are then also real. The remaining structure of the quartic wavefunction coefficient, namely, total-energy poles and regular terms, can be fixed by the Manifestly Local Test (MLT). The MLT states that wavefunction coefficients of massless scalars (and gravitons) satisfy the following simple relation \cite{MLT}
 \begin{align}  
 \label{MLT}
    \frac{\partial }{\partial k_{c}} \psi_{4}\Big|_{k_{c}=0}=0\,\,,\qquad \forall\, c=1,2,3,4\,\,,
    \end{align}
where we hold all other energies fixed. The MLT holds when interactions are manifestly local, i.e. they do not contain inverse Laplacians. This is certainly the case for the self-interactions of $\pi$ in the decoupling limit of the EFTI. Since this is a real constraint on the energy dependence of the quartic wavefunction coefficient, all remaining parts of $\psi_{4}$ will also be real. A real $\psi_{4}$ leads to a real $\rho_{4}$ which cannot contribute to the parity-odd trispectrum. We therefore arrive at another no-go theorem:
\begin{center}
\textit{Exchange contributions to the quartic wavefunction coefficient of massless scalars with Bunch-Davies vacuum conditions cannot contribute to the parity-odd trispectrum if the exchanged field has massless or conformally-coupled mode functions, Bunch-Davies vacuum conditions and if the constituent cubic wavefunction coefficients are IR-finite.}
\end{center}

\noindent In the following subsection we derive these no-go theorems using the in-in formalism and provide some generalisations that follow from the same assumptions we have made here. In the rest of the paper, we will discuss how non-vanishing parity-odd trispectra can arise if one or more of these assumptions are violated.


\subsection{Derivation using the in-in formalism} \label{sec:3.6}

\noindent In this section we extend the no-go theorems for the parity-odd trispectrum to more general $n$-point correlators by directly using the in-in perturbative expressions for tree-level correlators, without any reference to the wavefunction or the COT. First, we briefly review the Feynman rules to compute in-in diagrams and then show that by performing an appropriate Wick rotation of all time integrals, tree-level parity odd correlators manifestly vanish. Similar results were first derived in \cite{Liu:2019fag} using this formalism. The parity-odd trispectrum from loop corrections will be discussed in a separate paper. 

\paragraph{Feynman rules} The Feynman rules for the correlators are nicely spelled out in \cite{Chen:2017ryl} (see also \cite{Musso:2006pt}):
\begin{itemize}
\item Draw a diagram and indicate all possible vertices with an $r$ if the corresponding interactions $ H_{\rm int}$ is to the right of the operator, as in $ \ex{\O(\bfk) H_{\rm int}}$, or with an $ l$ if the Hamiltonian is to the left of the operator, as in $ \ex{H_{\rm int}\O(\bfk)}$. 
\item Every vertex gets a vertex factor that depends on the theory. Every spatial derivative is $ \partial_{\bf x}\to (-i\bfk)$ because it would be a $ +i\bfp$ in the Fourier-space Hamiltonian, which then gets integrated over $ \delta^{(3)}(\bfk+\bfp)$. We will find it convenient to use the \textit{amplitude convention}\footnote{The alternative convention, used e.g. in \cite{COT,Goodhew:2021oqg,Melville:2021lst}, is to put an overall $i^{1-L}$, no $i$'s on the vertices and an extra $i$ on the propagators.} to get the crucial factors of $i$ right: no $ i$ overall and if the coupling in the Hamiltonian is $ H_{\rm int}\sim +\lambda$ then put a $-i \lambda $ on a right vertex a $ +i\lambda$ on a left vertex.
\item Now there are four possible bulk-to-bulk (B2B) propagators:
\begin{align}
G_{rr}(\eta_{1},\eta_{2},p)&=f_{p}(\eta_{1})f_{p}^{\ast}(\eta_{2})\theta(\eta_{1}-\eta_{2})+f_{p}^{\ast}(\eta_{1})f_{p}(\eta_{2})\theta(\eta_{2}-\eta_{1}) \,\,,\\
G_{lr}(\eta_{1},\eta_{2},p)&=f_{p}(\eta_{1})f_{p}^{\ast}(\eta_{2})=G_{rl}^{\ast}(\eta_{1},\eta_{2},p) \,\,,\\
G_{ll}(\eta_{1},\eta_{2},p) &=G_{rr}^{\ast}(\eta_{1},\eta_{2},p)\,\,,
\end{align}
where $f$ are the positive-frequency mode functions, for example
\begin{align}\label{dSmode}
f_{k}(\eta)&=\eta\frac{H}{\sqrt{2k}}e^{-ik\eta} \quad \text{(conformally coupled)}\,\,, \\ \label{masslessmode}
f_{k}(\eta)&=\frac{H}{\sqrt{2k^{3}}}(1+ik\eta)e^{-ik\eta} \quad \text{(massless, dS)}\,\,.
\end{align}
\item There are two bulk-to-boundary (B2b) propagators, $ G_{r}$ from $ H_{\rm int}$'s to the right of $ \O$ and $ G_{l}$ from $ H_{\rm int}$'s to the left of $ \O$:
\begin{align}
G_{r}(\eta,p)&=f_{p}(\eta_{0})f_{p}^{\ast}(\eta)\,\,, & G_{l}(\eta,p)&=f_{p}^{\ast}(\eta_{0})f_{p}(\eta)\,\,.
\end{align}
\end{itemize}
Finally we discuss a useful property that relates diagrams related by switching all right and left vertices, $r\leftrightarrow l$. Let $ D$ be a Feynman diagram with $ V$ vertices and $ \sigma_{a} $ with $ a=1,\dots,2^{V}$ all possible ordered lists of how to label the $ V$ vertices right or left. Let $ \bar \sigma_{a}$ represent the ordered list $ \sigma_{a}$ where all vertices have been flipped $ r \leftrightarrow l$. Then
\begin{align}
D[\sigma]=D[\bar \sigma]^{\ast} (-)^{n_{i}}\,\,,
\end{align}
where $ n_{i}$ is the total number of spatial derivatives in all the vertices. From this we deduce
\begin{align}
B_{n}&=\sum_{a} D[\sigma_{a}]=\frac{1}{2} \sum_{a} \left[  D[\sigma_{a}]+D[\bar \sigma_{a}]\right] \nonumber \\
&=\frac{1}{2} \sum_{a} \left[  D[\sigma_{a}]+D[\bar{\bar{\sigma}}_{a}]^{\ast}(-)^{n_{i}}\right] \\
&=\frac{1}{2} \sum_{a} \left[  D[\sigma_{a}]+D[\sigma_{a}]^{\ast}(-)^{n_{i}}\right]\,\,. \nonumber 
\end{align}
This is particularly useful when we discuss parity-even and parity-odd contributions:
\begin{align}
B^{\rm{PE}}&=2\Re \text{Diagram(color order $ n_{r}\geq n_{l}$)}\,\,, \\
B^{\rm{PO}}&=2i\Im \text{Diagram(color order $ n_{r}\geq n_{l}$)}\,\,.
\end{align}
Notice that, as anticipated in Eq.~\eqref{POi}, we find that parity-odd correlators in Fourier space are purely imaginary, as they should be. 

\begin{figure}
\centering
\includegraphics[width=\textwidth]{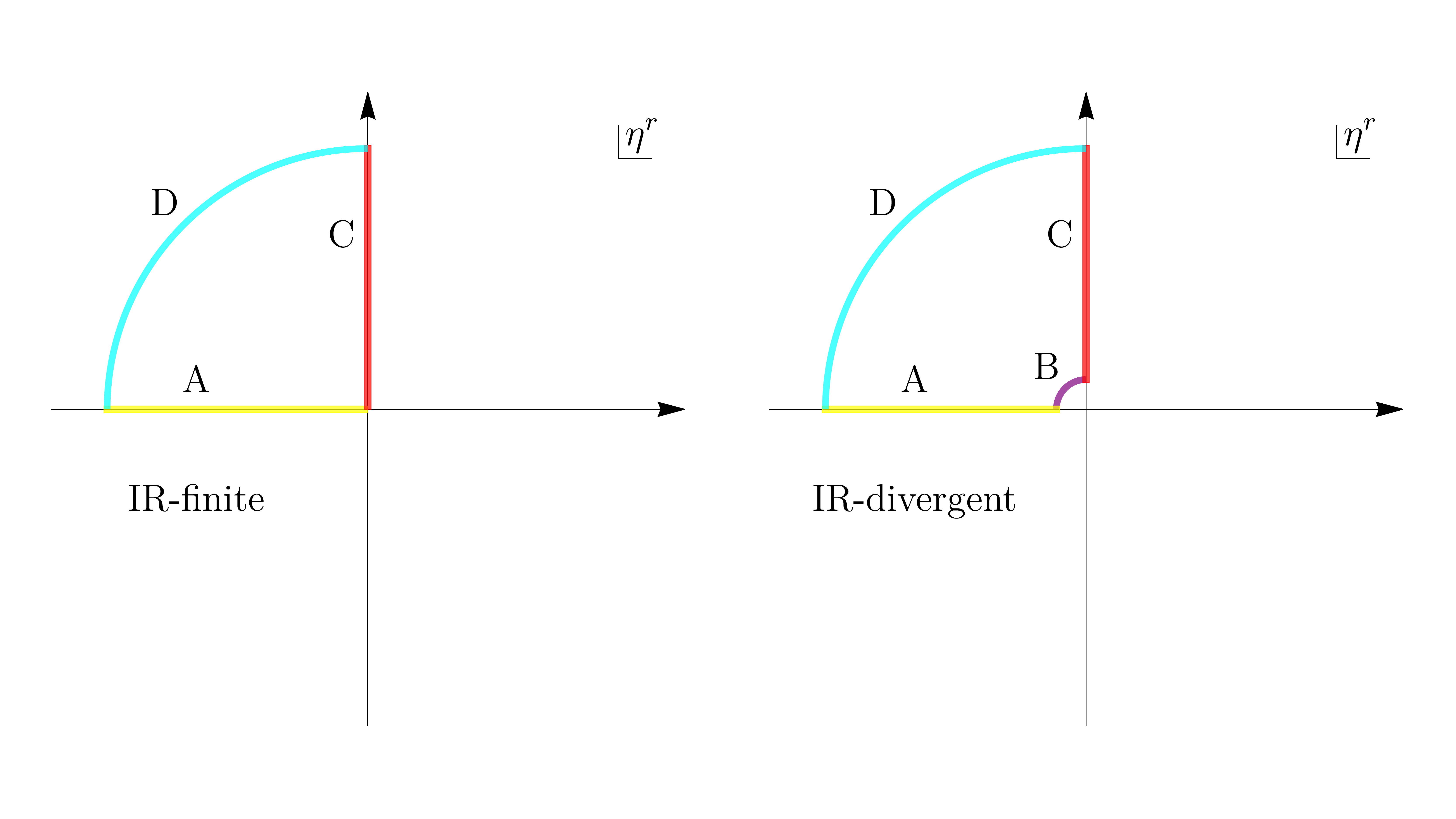}
\caption{Integration contours in complex $\eta^r$ plane. A similar contour in the lower-half complex plane applies to the left vertices.}
\label{figcont}
\end{figure}

\paragraph{Massless de Sitter mode functions} We are now in the position to prove the following no-go theorem: any tree-level parity-odd $n$-point correlation functions involving only massless de Sitter mode functions, \eqref{masslessmode}, vanishes if all time integrals are IR-convergent (in Fourier space). Here is the derivation. Consider a generic manifestly-local interaction Hamiltonian
\begin{align}\label{valency}
H_{\rm int}&=\int_{\bfp_1,\dots} \delta_{\rm D}^{(3)} \left(\sum_a \bfp_{a} \right)  \prod_{b}^{n} \left[ (-ip_b \eta)^{n_i^{(b)}} (\eta\partial_\eta)^{n_\eta^{(b)}} \phi_b(\bfp_{b}) \right]\,\,,
\end{align}
where $\smash{n_i^{(b)}}$ and $\smash{n_\eta^{(b)}}$ count the number of spatial and time derivatives in the vertex and $\phi_b$'s are fields of any spin with a \textit{massless de Sitter mode function}. The crucial point to notice is that in $H_{\rm int}$ every $i$ comes with an $\eta$ and viceversa (notice that $\eta \partial_\eta = (i\eta)\partial_{i\eta}$). In other words, $H_{\rm int}$ is a real function of the variable $i\eta$. Since the detailed values of $\smash{n_i^{(b)}}$ and $\smash{n_\eta^{(b)}}$ will be irrelevant, it is convenient to simplify our notation and rewrite this as
\begin{align}
H_{\rm int}&=\int_{\bfp_1,\dots} \delta_{\rm D}^{(3)}\left(\sum_a \bfp_{a}\right) F(i\bfk \eta) \prod_{b}^{n} \phi_b(\bfp_{b}) \,\,,
\end{align}
where now $F(i\bfk \eta)$ is a vertex factor that collects all spatial derivatives and $\phi_b$ are fields of any spin with massless de Sitter mode functions \textit{or time derivatives thereof}. The contribution to any tree-level parity-odd correlator with $V$ interactions takes the following form
\begin{align}
B_{n}&=2i \Im \left[\prod_{A=1}^V \int_{-\infty}^0 \frac{\pm i {\rm d}\eta_A}{(H\eta_A)^4} F_A(i\bfk \eta_A) \right] \left[\prod_{a=1}^n G_{X}(ik_a \eta_B) \right]\left[\prod_{m=1}^{I} G_{XX}(\eta_C,\eta_D,p_m)\right]\,\,,
\end{align}
where $X$ in each propagator can be $r$ or $l$. Each of the $I$ bulk-to-bulk propagators contains the two possible time orderings of the vertices it connects. If $I_{rr}$ and $I_{ll}$ are the numbers of $G_{rr}$ and $G_{ll}$ propagators, this gives at most $2^{I_{rr}+I_{ll}}$ terms, each corresponding to a different ordering of the right times $\eta_A^r$ and the left times $\eta_B^l$. Notice that for certain choices of right and left labelling of the vertices, there might right or left times that are not ordered with respect to each other. This is not an issue because any diagram where, say, two right vertices are not ordered with respect to each other can always be written as the sum of two diagrams that are each ordered. In this way, up to a relabelling of time integration variables, any one of the many possible left and right time orderings takes the form
\begin{equation}
\begin{split}
B_{n}&=\pm 2i \Im \int_{-\infty(1-i\e)}^0  i {\rm d}\eta_{1}^r \int_{\eta_1^r}^0 i{\rm d}\eta_2^r \dots \int_{\eta^r_{V_r-1}}^0 i{\rm d}\eta_{V_r}^r \\
& \qquad \qquad\,\,\,\, \times \int_{-\infty(1+i\e)}^0  i {\rm d}\eta_{1}^l \int_{\eta_1^l}^0 i{\rm d}\eta_2^l \dots \int_{\eta^l_{V_l-1}}^0 i{\rm d}\eta_{V_l}^l P(i \eta_A, k,\bfk)\,\,,
\end{split}
\end{equation}
where $P$ is a real-analytic function of $i \eta_A$ and the momenta (a polynomial times an exponential). To show that the argument of $\Im$ is real, we change all but two of the integration variables to $\eta_A^{r,l} = \eta_1^{r,l} \tilde \eta_A^{r,l}$ for $A=2,3,\dots,V_{r,l}$ and we rename $\eta_1^{r,l}=\eta^{r,l}$. This gives 
\begin{equation}
\begin{split}
B_{n}&=\pm 2i \Im \int_{-\infty(1-i\e)}^0  i {\rm d}\eta^r \int_{1}^0 i\eta^r {\rm d}\tilde\eta_2^r  \dots \int_{1}^0 i\eta^r {\rm d}\tilde\eta_{V_r}^r \\
&\qquad\qquad\,\,\,\, \times \int_{-\infty(1+i\e)}^0  i {\rm d}\eta^l \int_{1}^0 i\eta^l {\rm d}\tilde\eta_2^l  \dots \int_{1}^0 i\eta^l {\rm d}\tilde\eta_{V_l}^l P(i\eta^{r,l}, \tilde \eta_A^{r,l}, k,\bfk)\,\,.
\end{split}
\end{equation}
Let's discuss the converge of these integrals. No divergences can come from $\tilde \eta_A^{r,l} =1$. Convergence at $\eta^{r,l} =-\infty$ is guaranteed by the $i \epsilon$ prescription. Convergence at $ \eta^{r,l}=0$ depends on the interaction. For $n_i+2n_\eta \geq4$ there are no $\eta\to0$ divergences. We assume here that this is the case and discuss later what happens in the presence of an $\eta \to 0 $ IR divergence. Since the integrands of the ${\rm d}\eta^{r,l}$ integrals are analytic in the upper- and lower-half complex plane, respectively, we can rotate the integration contour.\footnote{This rotation is the same that relates dS calculations to Euclidean AdS calculation, as first pointed out for the wavefunction in \cite{Maldacena:2002vr} and recently elaborated in \cite{Sleight:2021plv,DiPietro:2021sjt}.} For IR-convergent interactions both integrals converge at $\eta \to 0$ by assumptions, so we only have the three contributions depicted on the left-hand side of Figure \ref{figcont}. The arc at infinity (contribution D) vanishes and so our integral on the negative real line equates the integral on the positive or negative imaginary axis for right and left times, respectively. Hence, for the right vertices we can rotate by 90$^{\circ}$ clockwise for $\eta^r$ from $ -\infty<\eta^r<0$ to $ 0<\lambda^r<+\infty$ so that $ \eta^r = i\lambda^r$. For the left contour we can rotate 90$^{\circ}$ counterclockwise to $\eta^l = i\lambda^l$ with $ 0<\lambda^l<-\infty$. Since only the combination $i\eta^{r,l}$ appeared in the integrands, the result of these rotations is manifestly real and so the parity-odd $B_n$ vanishes. \\

\noindent This proves that tree-level, parity-odd correlators involving fields of any spin with massless de Sitter mode functions vanish. A few comments are in order:
\begin{itemize}
\item This no-go result applies also to quadratic mixing of fields that are treated in perturbation theory since the above argument applies also when the valency $n$ of the interaction in \eqref{valency} is $n=2$.
\item Comparing to wavefunction calculations Minkowski space, we notice that scale invariance in de Sitter made things simpler for us. In Minkowski every additional space or time derivative brings in an extra factor of $ i$ so we have to keep track of them. Whether a parity-odd or parity-even contribution vanishes or not depends on how many derivatives we have. In contrast, in dS scale invariance forces derivatives to come in the form $ ik\eta$ or $ i\bfk \eta$ and so every $ i$ comes with an $ \eta$. We will see in Section \ref{sec:5} that even in de Sitter the situation is slightly more complicated when we have non-linear dispersion relations.
\item The above proof also generalizes to spinning fields if by ``parity odd'' we mean ``with an odd number of derivatives''. Indeed the above derivation applies unchanged if the fields $\phi_b$ have indices, as e.g. a vector $A_i$ or a graviton $\gamma_{ij}$. What changes for fields with spin is that in general they allow for different IR-divergent derivatives, with $n_i+2n_\eta <4$, as we will discuss shortly.
\end{itemize}

\paragraph{Conformally-coupled de Sitter mode functions} The above discussion can be straightforwardly generalized to include also any number of fields of any spin and conformally-coupled de Sitter mode functions, as in Eq.~\eqref{dSmode}. The only difference now is that in these mode functions one power of $\eta$ appears without a factor of $i$. Internal lines come with a bulk-to-bulk propagator that has two mode functions and so any number of conformally-coupled $G_{XX}$'s will not change our conclusion above. Conversely, we need to keep track of external conformally-coupled lines. These observations allows us to conclude that: a tree-level parity-odd correlator of an \textit{even} number of fields with conformally-coupled mode functions and any number of fields with massless mode functions vanishes, again for any spin and under the assumptions of scale invariance and a Bunch-Davies vacuum. Conversely, in the presence of an odd number of conformally-coupled external fields we expect a non-vanishing parity-odd contribution but a vanishing parity-even contribution. This extends the observation of \cite{COT} to all parity-even tree diagrams with an odd number of conformally-coupled external fields.


\section{Yes-go 1: breaking scale invariance} \label{sec:4}

In this section we will show that breaking exact scale invariance allows us to realise parity-odd trispectra. We will consider two types of symmetry breaking which both invalidate the assumption that the wavefunction coefficients have a fixed scaling with momenta: $(i)$ explicit $\eta_{0}$ dependence via IR-divergences, and $(ii)$ explicit breaking of scale invariance at the level of the action due to time-dependent couplings.

\paragraph{IR-divergences} At tree-level, the analytic properties of wavefunction coefficients with Bunch-Davies initial conditions are very constrained, irrespective of de Sitter boost invariance \cite{BBBB}. Assuming scale invariance, contact diagram contributions to the quartic wavefunction coefficient of massless scalars can yield rational terms in the external kinematics, IR-divergences in the form of poles as $\eta_{0} \rightarrow 0$, and finally IR-divergent logarithms of the form $\log(-k_{T} \eta_{0})$ where $k_{T} = k_{1}+k_{2}+k_{3}+k_{4}$ is the total-energy. The latter two possibilities break the fixed integer momentum scaling of the wavefunction coefficient. Such logs are particularly interesting in the context of parity violation since the Cosmological Optical Theorem (COT) demands that they always appear in the combination \cite{COT}
\begin{align} \label{COTandLOG}
\log(-k_{T} \eta_{0}) + \frac{i \pi}{2}\,\,,
\end{align}
so their contribution to the wavefunction is always complex. As we have discussed a number of times, parity-odd trispectra come from the imaginary part of the wavefunction so having IR-divergences in the form of logs can yield parity-odd signals. It is always the $i \pi$ that contributes to the correlator. This is familiar from in-in computations of parity violation in the gravitational sector \cite{Creminelli:2014wna,CabassBordin}, and plays a crucial role in deriving the small number of parity-odd shapes of graviton bispectra \cite{Cabass:2021fnw}. \\

\noindent Now such logs can only be multiplied by tensor structures or a polynomial in the external energies \cite{BBBB}. If we assume that what multiplies the log scales as $ \sim k^3$, combined with the fact that we require a factor of $\epsilon_{ijk}k_{1}^i k_{2}^j k_{3}^k$ to break parity, the coefficient of the log can only be a constant. We then immediately conclude from Bose symmetry that such a contribution to the wavefunction, and therefore the trispectrum, can only arise if all four scalars are different since there is no way to cancel the anti-symmetry we get from $\epsilon_{ijk}k_{1}^i k_{2}^j k_{3}^k$. We note that this case is not captured by our no-go theorem of Section \ref{sec:3} since the log invalidates the scaling in \eqref{genscla} (the wavefunction coefficient does not simply pick up an overall minus sign as we send $\{k \} \rightarrow - \{k\}$ and $\{ \bfk \} \rightarrow - \{\bfk \}$). \\ 

\noindent To be more explicit, let's consider the only parity-odd quartic interaction with $n_i+2n_\eta <4$, namely
\begin{align}\label{ni3}
    H_{\rm int}&=\lambda \phi_1 \partial_i \phi_2 \partial_j \phi_3 \partial_k \phi_4 \e_{ijk}\,\,.
\end{align}
If any two of the four fields are identical, $\phi_a=\phi_b$, then this vanishes by integration by parts. So this example requires at least four distinct scalars, in agreement with our boundary argument above. The result can be computed directly, and gives
\begin{equation}
\label{IRdiv}
\begin{split}
    B_4&=2i \lambda (\bfk_2\times\bfk_3\cdot \bfk_4) \,\Im \int_{-\infty}^{\eta_0}\frac{-i{\rm d}\eta}{(H\eta)^4} (-iH\eta)^3 \left[\prod_a^4 G_+(k_a) \right] \\
&=-i\frac{\lambda H^7 \pi}{8e_4^3} (\bfk_2\times\bfk_3\cdot \bfk_4)\,\,.
\end{split}
\end{equation}
Notice the absence of total energy poles, as discussed in \cite{Cabass:2021fnw}. This result can also be derived from the complex plane. First, note that when we rotate $\eta$ to $i\lambda$ as in Fig.~\ref{figcont}, we need to consider an additional contour corresponding to a small arch around $\eta=0$, which picks up the contribution from the pole at $ \eta_{0}=0$. This contribution is in general complex and leads to a non-vanishing $\BPO_4$. Indeed, the arc can be computed and gives
\begin{align}
\Im \int_{-\infty}^{\eta_0} \frac{{\rm d}\eta}{\eta} \,e^{i k_T \eta} \prod_a (1-ik_a \eta)=\Im \int_{\pi}^{\pi/2} \frac{|\eta_0|e^{i\theta}{\rm d}\theta}{|\eta_0|e^{i\theta}} \,e^{i k_T \eta} \left[1+\O(|\eta_0|) \right]=\frac{\pi}{2}\,\,.
\end{align}
This nicely picks up the $i\pi/2$ that accompanies the $ \log$ in the wavefunction computation.


\paragraph{Time-dependent couplings} When scale invariance is broken the coupling constants of the effective action for perturbations can depend explicitly on time. A generic time dependence leads to a non-vanishing contribution to parity-odd $n$-point correlators of fields of any spin and mass. Here we concentrate on the scalar trispectrum (see e.g. \cite{Soda:2011am,Shiraishi:2011st} for a related discussion of the graviton bispectrum). For concreteness we will consider only one of the leading parity-odd quartic interactions in the effective theory of single-clock inflation since it is enough to illustrate what is going on. We take
\begin{align}
   M(t+\pi) (g^{00}+1)\mathbf{e}^{\mu\nu\rho\sigma}n_\mu\delta\! K_{\nu\lambda}(n^\alpha\nabla_\alpha\delta\! K^\lambda_{\hphantom{\lambda}\rho})\nabla_\sigma\delta\! K & \rightarrow \frac{\lambda(\eta)}{a^{9}} \pi' \epsilon_{ijk}\partial_i\partial_l\pi\partial_l\partial_j \pi' \partial_k\partial^2{\pi}+\dots\,\,, 
\end{align}
where the $\dots$ denote terms that are higher orders in $\pi$ as required by symmetry. To present explicit expressions for $\BPO_4$, we will consider two time dependencies arising from expanding the coupling constant. First, let's assume the time evolution is well captured by a term linear in $\eta$, as in 
\begin{align}
    \lambda(\eta) &=\lambda_\ast+\lambda'_\ast (\eta-\eta_\ast)+\O\big((\eta-\eta_\ast)^2\big)\,\,.
\end{align}
The trispectrum is then
\begin{equation}
\label{etalambda}
\begin{split}
    \BPO_4&=2 i \Im \int \frac{-i\lambda(\eta) {\rm d}\eta}{(\eta H)^4}(-H\eta)^9(-i)^7 F(\bfk)\, G_r'(k_1)G_r(k_2)G_r'(k_3)G_r(k_4)+\text{$23$ perms.} \\
    &=i\lambda'_\ast 5040 F(\bfk) H^{13} \frac{k_1^2k_3^2 \left[ 90 k_2k_4+9k_T(k_2+k_4)+ k_T^2 \right]}{(k_1k_2k_3k_4)^3 k_T^{11}}+\text{$23$ perms.}\,\,,
\end{split}
\end{equation}
where the vertex $F$ is given by
\begin{align}
    F(\bfk)\equiv (\bfk_2\times \bfk_3 \cdot \bfk_4)(\bfk_2\cdot \bfk_3)k_4^2\,\,.
\end{align}
As expected, $\BPO_4$ in \eqref{etalambda} is not scale invariant since $ \BPO \sim k^{-10} $ as opposed to the expected $ k^{-9} $. \\

\noindent As a second example, we will assume that the time dependence of the coupling constant during the period in which modes cross the Hubble radius is well captured by a linear term in $t\sim \log(-\eta)$,
\begin{align}
    \lambda(\eta) &=\lambda_\ast+\lambda_{\ast,N} \log\left(\frac{\eta}{\eta_\ast}\right)+\O(\log^2)\,\,.
\end{align}
The integral in \eqref{etalambda} can now be computed with this logarithmic time dependence in terms of exponential integrals in \emph{Mathematica}, but there is a more direct analytical way to get the result. Notice that the integrand is convergent at $\eta\to 0$ even in the presence of the log. We can therefore rotate it as in the left-hand side of Fig.~\ref{figcont}. Then we have 
\begin{align}
    \log (\eta) \to \log(i\lambda)=\log\lambda+i\frac{\pi}{2}\,\,.
\end{align}
Since we only pick up the imaginary part, it is only the $i\pi/2$ term that contributes. Therefore we simply have to compute the integral in \eqref{etalambda} with $\lambda=\lambda_\ast$ constant and multiply it by $i\pi/2$. The result is
\begin{align}\label{finalBPOnoscale}
    \BPO_4&=i\lambda_{\ast,N} 630\pi F(\bfk) H^{13} \frac{k_1^2k_3^2 \left[  72 k_2k_4+8k_T(k_2+k_4)+ k_T^2\right]}{(k_1k_2k_3k_4)^3 k_T^{10}}+\text{$23$ perms.} 
\end{align}
A few comments are in order. First, notice that this $\BPO_4$ has the correct scaling dictated by scale invariance, $\BPO_4\sim k^{-9}$. This is to be expected from power counting at the level of the integral. We are therefore in a situation in which scale invariance is broken in the wavefunction by the logarithmic time dependence of a coupling constant, but the breaking is not visible to leading order in the correlator. Second, note that this $\BPO_4$ has total energy poles at $k_T\to 0$. This is in contrast to contributions arising in the scale invariant theory, such as for example \eqref{IRdiv} and the parity-odd graviton bispectra computed in \cite{CabassBordin,Cabass:2021fnw}.\\

\noindent We conclude by pointing out that these two simple examples are by no means an exhaustive list, and many other time dependencies could be considered, depending on the model under consideration. The main takeaway is that a non-vanishing $\BPO_4$ is generally produced by any deviation from scale invariance.


\section{Yes-go 2: non-Bunch-Davies vacuum} 
\label{sec:5}

\noindent In Section \ref{sec:3} we showed that contact diagram contributions to the parity-odd quartic wavefunction coefficient cannot contribute to the trispectrum when the dispersion relation is linear i.e. $\omega^2 \propto k^2$. We have also shown that our results are robust against adding small corrections to this dispersion relation. In this section we will look for a way out by considering the non-linear dispersion relation $\omega^2 \propto k^4$ which occurs in Ghost Inflation (GI) \cite{GhostInflation} which is an inflationary generalisation of the Ghost Condensate \cite{GhostCondensate} (see also \cite{Ivanov:2014yla}). Our results of the previous section do not apply to GI since the non-linear dispersion relation translates into a bulk-to-boundary propagator for the Goldstone mode that is not Hermitian analytic, as was pointed out in \cite{Goodhew:2021oqg}. In Section \ref{sec:3} we heavily relied on Hermitian analyticity of the bulk-to-boundary propagator as a way of deducing when the time integral can contain an imaginary part. In this section we will show that a large parity-odd trispectrum can indeed arise from contact diagrams in GI (in a similar way, we expect that the models studied in \cite{Ashoorioon:2018uey}, for which the dispersion relation is $\omega^2=k^{2n}$ for $n>1$, will also give a non-zero parity-violating trispectrum). \\

\noindent Before proceeding, let us emphasize that in this section we discuss GI only as an example of non-Bunch-Davies vacuum conditions, and of how deviating from a linear dispersion relation allows for parity violation in the scalar trispectrum. We do not advocate it either as an explanation of the results of Refs.~\cite{Hou:2022wfj,Philcox:2022hkh}, or as a theory of the primordial universe. Indeed, the validity of the Ghost Condensate, and of GI as an inflationary model, has been put into question in light of problems associated with black hole thermodynamics \cite{Dubovsky:2006vk} and violation of the de Sitter entropy bound \cite{Arkani-Hamed:2007ryv} (but see also \cite{Mukohyama:2009rk,Mukohyama:2009um,Jazayeri:2016jav} for a discussion of scenarios where these bounds are not violated in practice). Our result that GI can indeed yield a non-vanishing parity-odd trispectrum may motivate further model building for such non-standard dispersion relations that may overcome some of these issues. \\

\noindent Now, let us first recall that the quadratic action for the Goldstone mode in GI comes from the unitary-gauge action
\be 
S=\int{\rm d}^4x\,\sqrt{-g}\,\bigg[\frac{\Lambda^4}{4}(g^{00}+1)^2 - \frac{\Lambda^2_1}{2}\delta\!K_{\mu\nu}\delta\!K^{\mu\nu} - \frac{\Lambda^2_2}{2}\delta\!K^2\bigg]\,\,. 
\ee 
Comparing with Eq.~\eqref{efti_ug_action}, here we have defined $M^4_2 = \Lambda^4_2/2$, and we assume that these operators dominate over the minimal kinetic term $\smash{M^2_{\rm P}\dot{H}g^{00}}$ in the limit $\smash{\dot{H}\to 0}$, $\smash{c_s\to 0}$ with $\smash{{-\dot{H}}M_{\rm P}^2(1-c^2_s)/c^2_s} = \Lambda^4$ kept fixed \cite{Cheung:2007st}. 
At quadratic order, and converting to conformal time, the free theory for the Goldstone mode is therefore\footnote{Recall that $\pi$ has dimensions of length since it arises from the combination $(t + \pi)$.}
\be
\label{eq:leading_action} 
S_{\pi\pi}=\int{\rm d}^3x{\rm d}\eta\,a^4(\eta)
\bigg[\frac{\Lambda^4}{2}\frac{{\pi'}^2}{a^2(\eta)} - \frac{\tilde{\Lambda}^2}{2}\frac{(\nabla^2\pi)^2}{a^4(\eta)}\bigg]\,\,, 
\ee 
where we have defined $\tilde{\Lambda}^2\equiv\Lambda^2_1 + \Lambda^2_2$, and the number of scale factors is fixed by scale invariance. Now the bulk-to-boundary wavefunction propagator is \cite{GhostInflation}\footnote{Expanding $\eta^2 = \eta_0^2 - 2t/(a^2_0 H)$, for $t\ll H^{-1}$, we see that the bulk-to-bulk propagator goes as $\exp(i\omega(k_{\rm phys})t)$, where $\smash{\omega(k) = \tilde{\Lambda}k_{\rm phys}^2/\Lambda^2}$ and $\smash{k_{\rm phys} = k/a_0}$.}
\be
\label{eq:bulk_to_bulk_GC}
K(k,\eta) = {-\frac{e^{\frac{i\pi}{4}}\pi ({-\eta} )^{\frac{3}{2}} (\tilde{c} k)^{\frac{3}{2}} H_{\frac{3}{4}}^{(1)}({-\tilde{c}^2 k^2 \eta ^2})}{2^{\frac{3}{4}}\Gamma(\frac{3}{4})}}
\,\,, 
 \ee 
where $\tilde{c}^2=H\tilde{\Lambda}/(2\Lambda^2)$, and we drop the subscript $\pi$ on the bulk-to-boundary propagator in this section since there is no possibility of confusion. In this expression $H_{3/4}^{(1)}(z)$ is the Hankel function of the first kind and order $3/4$, and we have used 
\begin{align}
\lim_{\eta_{0} \rightarrow 0^{-}} (-\eta_{0})^{3/2}H_{\frac{3}{4}}^{(1)}({-\tilde{c}^2 k^2 \eta_{0} ^2}) = - i\frac{ 2^{3/4} \Gamma(\frac{3}{4})}{\pi (-\tilde{c}^2 k^2)^{3/4}}\,\,,
\end{align}
to eliminate all $\eta_{0}$ dependence from this propagator. One can check directly \cite{Goodhew:2021oqg} that this propagator is not Hermitian analytic i.e. $K(k, \eta) \neq K^{\ast}(-k^{\ast}, \eta)$, and therefore the COT as we wrote it in Section \ref{sec:3} does not hold. We can now compute the power spectrum and to do so we will follow the wavefunction approach since it allows us to illustrate how we deal with time integrals in GI. The quadratic wavefunction coefficient takes the form
\be\label{LLt}
\psi_2 = {-2i}\int_{-\infty(1-i\epsilon)}^0{\rm d}\eta\,a^4(\eta)\bigg[\frac{\Lambda^4}{2}\frac{K'(k,\eta)^2}{a^2(\eta)} - \frac{\tilde{\Lambda}^2}{2}\frac{k^4K(\eta,k)^2}{a^4(\eta)}\bigg]\,\,,
\ee
where the overall factor of $2$ comes from the fact that the two ``vertices'' are symmetric and as always we include an overall factor of $(-i)$. We can form a contour that goes from $-\infty(1-i\epsilon)$ to $-\infty(1-i)$, and from $-\infty(1-i)$ to $0$. Given that the contribution on the quarter-circle vanishes exponentially fast,\footnote{Recall that $H_{\nu}^{(1)}(z)\sim \sqrt{\frac{2}{\pi z}}\,e^{i \left(-\frac{\pi\nu }{2}+z-\frac{\pi}{4}\right)}$ for large $z$.} and that there are no poles inside the integration contour and we do not cross branch cuts, we have 
\be
\psi_2 = {-2i}\int_{-\infty(1-i)}^0{\rm d}\eta\,a^4(\eta)\bigg[\frac{\Lambda^4}{2}\frac{K'(k,\eta)^2}{a^2(\eta)} - \frac{\tilde{\Lambda}^2}{2}\frac{k^4K(\eta,k)^2}{a^4(\eta)}\bigg] \,\,.
\ee
The advantage of this contour will become manifest when we come to compute the quartic wavefunction coefficient and therefore the trispectrum. If we integrate by parts and use the equation of motion, we can reduce the quartic wavefunction coefficient to 
\be
\psi_2 = {-2i}\bigg[\frac{a^2(\eta)}{2}\Lambda ^4 K(k,\eta) K'(k,\eta)\bigg]_{-\infty(1-i)}^0 = \frac{e^{\frac{i\pi}{4}}k^3\pi\Lambda\tilde{\Lambda}^2}{({2H\tilde{\Lambda}})^{\frac{1}{2}}\Gamma(\frac{3}{4})^2}\,\,,
\ee
from which we can extract the power spectrum which is given by
\be
\label{eq:GC_PS}
P_\pi(k) = \frac{(H\tilde{\Lambda})^{\frac{1}{2}}\Gamma(\frac{3}{4})^2}{k^3\pi\Lambda\tilde{\Lambda}^2}\,\,. 
\ee
Before writing down quartic self-interactions for the Goldstone and computing trispectra, let us first briefly recall how to derive the scaling dimensions in GI and we refer the reader to \cite{GhostCondensate, GhostInflation, Cheung:2007st} for more details. Given that the free theory does not lead to a linear dispersion relation, counting the scaling dimension of various operators is slightly more involved. We will derive these scalings at energy scales where we can ignore the background curvature. Under a rescaling of energy (time) $E\to s E$ ($t\to s^{-1}t$), the dispersion relation $\omega^2 \propto k^4$ implies that ${\bf k}\to s^{1/2}{\bf k}$ (${\bf x}\to s^{-1/2}{\bf x}$). We then see that the quadratic action remains invariant if the scaling dimension of $\pi$ is $\pi\to s^{1/4}\pi$. We will use these scalings as guidance for finding the leading order operators in the EFT expansion. \\

\noindent As we discussed in Section \ref{sec:2}, there are two distinct ways that a quartic self-interaction can arise: $(i)$ from EFTI operators that contain four building blocks and therefore start at quartic order in perturbations and $(ii)$ from EFTI operators that contain fewer than four building blocks and can therefore start at quadratic/cubic order in perturbations. For definiteness we will work with EFTI operators that start at quartic order since these are simpler to construct and already enable us to illustrate that parity odd trispectra can indeed arise in GI. Studying other operators is certainly an interesting avenue for future work, and perhaps requires a better understanding of the non-linearly realised symmetries. 
From Eqs.~\eqref{building_blocks-1},~\eqref{building_blocks-2},~\eqref{building_blocks-3} we see that we can construct self-interactions from $\dot{\pi}$, $\partial_{i}\partial_{j} \pi$, and their derivatives. This also follows from the fact that in flat space, a superfluid non-linearly realises broken boosts as $\delta \pi = b_{i}x^{i} + \mathcal{O}(\pi)$ (see e.g. \cite{Son:2002zn,Pajer:2018egx,Nicolis:2015sra,Delacretaz:2014oxa}). The operators with the lowest scaling dimension are those with the fewest time derivatives. The leading operators with zero, one and two time derivatives are
\begin{align} 
\mathbf{e}^{\mu\nu\rho\sigma}n_\mu\delta\! K_{\alpha \beta} \delta\! K^{\alpha}{}_{\nu} \nabla^{\beta} \delta\! K_{\gamma \rho} \delta\! K^{\gamma}{}_{\sigma} & \supset a^{-9} \epsilon_{ijk}\partial_m\partial_n \pi \partial_n\partial_i {\pi}\partial_m\partial_l \partial_j {\pi} \partial_l \partial_k{\pi} \,\,, \label{ZerothPOoperator} \\ 
(g^{00}+1)\mathbf{e}^{\mu\nu\rho\sigma}n_\mu\delta\! K_{\nu\lambda}(D^2\delta\! K^\lambda_{\hphantom{\lambda}\rho})\nabla_\sigma\delta\! K & \supset a^{-9}\dot{\pi}\epsilon_{ijk}\partial_i\partial_l\pi\partial_l\partial_j\partial^2{\pi}\partial_k\partial^2{\pi} \,\,, \label{FirstPOoperator} \\ 
(g^{00}+1)\mathbf{e}^{\mu\nu\rho\sigma}n_\mu\delta\! K_{\nu\lambda}(n^\alpha\nabla_\alpha\delta\! K^\lambda_{\hphantom{\lambda}\rho})\nabla_\sigma\delta\! K & \supset a^{-7}\dot{\pi}\epsilon_{ijk}\partial_i\partial_l\pi\partial_l\partial_j\dot{\pi}\partial_k\partial^2{\pi} \label{SecondPOoperator} \,\,, 
\end{align}
where $D^2$ is the covariant Laplacian on the hypersurfaces of constant time and $\mathbf{e}_{\mu\nu\rho\sigma}$ is the volume form $\mathbf{e}_{\mu\nu\rho\sigma} = \sqrt{-g}\,\epsilon_{\mu\nu\rho\sigma}$, with $\epsilon_{0ijk} = \epsilon_{ijk}$ and $\epsilon_{0123} = 1$. We now notice that we can isolate Eq.~\eqref{FirstPOoperator} as the leading source of parity violation. Indeed, the other operators do not preserve the $t\to{-t},\pi\to{-\pi}$ symmetry of the action of the free theory Eq.~\eqref{eq:leading_action}.\footnote{One can relate this symmetry to a $t\to{-t}$ symmetry in unitary gauge. The transformation rules in unitary gauge follow from $\smash{g^{00}=g^{\mu\nu}\partial_\mu t\partial_\nu t}$, $\smash{n_\mu=-\partial_\mu t/\sqrt{-g^{00}}}$, $D_\mu = (\delta^\nu_{\hphantom{\nu}\mu} + n^\nu n_\mu)\nabla_\nu$ and $K_{\mu\nu} = \nabla_\mu n_\nu + n_\mu n^\rho\nabla_\rho n_\nu$. 
This is useful since it allows one to predict the transformation rules of a given operator at all orders in $\pi$. 
However, it is important to keep in mind that once the background of these geometric objects is covariantly subtracted to ensure tadpole cancellation, their transformation properties are spoiled. This will give rise to operators that break the $t\to{-t},\pi\to{-\pi}$ symmetry. These will generically be suppressed in the limit of large non-Gaussianities. That is, the breaking of the symmetry is small in this limit.} We will therefore initially concentrate on Eq.~\eqref{FirstPOoperator} from which we will learn a lot about when we can get a non-zero signal. We will come back to the others at the end of this section. Converting to conformal time we have
\begin{align}
\label{quartic_action_GI}
S_{\pi\pi\pi\pi} = \frac{1}{\Lambda_{3}^2} \int {\rm d}^3 x {\rm d} \eta\,a^{-6}(\eta)\,\pi' \epsilon_{ijk}\partial_i\partial_l\pi\partial_l\partial_j\partial^2{\pi}\partial_k\partial^2{\pi} \,\,,
\end{align} 
and the quartic wavefunction coefficient takes the form
\be
\begin{split}
\psi_4 &= {\frac{H^6}{\Lambda^2_{\rm PO}}} \epsilon_{ijk}k^i_2 k^j_3 k^k_4 k^l_2 k^l_3 k^2_3 k^2_4 \int_{-\infty(1-i)}^0 {\rm d}\eta \, \eta^6\,K'(k_1,\eta)K(k_2,\eta)K(k_3,\eta)K(k_4,\eta) \\
&\;\;\;\; + \text{$23$ perms.} \\
& = {\frac{\pi^4 H^6 \tilde{c}^8}{ 4 \Gamma(\frac{3}{4})^4 \Lambda^2_{\rm PO}}} \epsilon_{ijk}k^i_2 k^j_3 k^k_4 k^l_2 k^l_3 k_{1}^{\frac{7}{2}} k_{2}^{\frac{3}{2}} k^{\frac{7}{2}}_3 k^{\frac{7}{2}}_4 \\ 
&\;\;\;\; \times  \int_{-\infty(1-i)}^0 {\rm d}\eta \, \eta^{13} \,H_{-\frac{1}{4}}^{(1)}(-\tilde{c}^2 k_{1}^2 \eta^2) H_{\frac{3}{4}}^{(1)}(-\tilde{c}^2 k_{2}^2 \eta^2) H_{\frac{3}{4}}^{(1)}(-\tilde{c}^2 k_{3}^2 \eta^2) H_{\frac{3}{4}}^{(1)}(-\tilde{c}^2 k_{4}^2 \eta^2) \\
&\;\;\;\; + \text{$23$ perms.}\,\,,
\end{split}
\ee 
where the overall factor of $(-i)$ in the Feynman rules is cancelled by the factor of $(+i)$ that we get from converting the nine spatial derivatives to momentum space. Here we have used
\be
\label{eq:bulk_to_bulk_GC-bis}
K'(k,\eta) = {-\frac{2 e^{\frac{i\pi}{4}}\pi ({-\eta} )^{\frac{5}{2}} (\tilde{c} k)^{\frac{7}{2}} H_{-\frac{1}{4}}^{(1)}({-\tilde{c}^2 k^2 \eta ^2})}{2^{\frac{3}{4}}\Gamma(\frac{3}{4})}}
\,\,.
\ee 
As a consistency check we notice that the overall powers of $k$ and $\eta$ can be expressed as $(k \eta)^{14} k^3$ so we have the correct scaling required by scale invariance. We can also check that this wavefunction coefficient won't vanish when we sum over permutations: from the time integral we only need to worry about permutations in labels $(2,3,4)$ then given the overall dependence on the energies, we only need to worry about about permutations in $(3,4)$ since only these energies appear with the same power. The integrand is symmetric in these labels, the energy dependence is also symmetric. We can then write $k^{l}_{2} k_{3}^{l} = \frac{1}{2}(k^{l}_{2} k_{3}^{l}-k^{l}_{2} k_{4}^{l}-k^{l}_{2} k_{1}^{l}-k_{2}^2)$, by momentum conservation, to show that the anti-symmetric structure from the epsilon tensor is cancelled by part of this final term. We will therefore get something non-zero when we sum over permutations, as expected. Now it is useful to explicitly extract the dependence on $\tilde{c}^2$. We do this by performing a change of variables $\smash{\k_i\to\k_i/\tilde{c}}$ in the integral over $\k_1,\k_2,\k_3,\k_4$ in the wavefunction. After some algebra, and using $\smash{\pi(\k/\tilde{c}) = \tilde{c}^3\pi(\k)}$,\footnote{This follows from scale invariance of the wavefunction coefficients $\psi_n$.} 
we obtain 
\be
\psi_4= {\frac{32H^6i\pi^4}{\Lambda^2_{\rm PO}\tilde{c}^{6}\Gamma(\frac{3}{4})^4}}(\k_2\cdot\k_3\times \k_4)(\k_2\cdot \k_3) k_1^{\frac{7}{2}}k_2^{\frac{3}{2}}k_3^{\frac{7}{2}}k_4^{\frac{7}{2}}\,{\cal T}(k_1,k_2,k_3,k_4) + \text{$23$ perms.}\,\,, 
\ee
where
\be 
\label{mathcalT}
{\cal T}(k_1,k_2,k_3,k_4) = \int_{0}^{+\infty}{\rm d}\lambda\, \lambda^{13}\,H^{(1)}_{{-\frac{1}{4}}}(2ik^2_1\lambda^2)H^{(1)}_{{\frac{3}{4}}}(2ik^2_2\lambda^2)H^{(1)}_{{\frac{3}{4}}}(2ik^2_3\lambda^2)H^{(1)}_{{\frac{3}{4}}}(2ik^2_4\lambda^2)\,\,. 
\ee 
Here we notice the usefulness of the contour discussed at the beginning of this section. Not only is $\cal T$ real with this choice, thereby ensuring that the wavefunction coefficient is imaginary, the mode functions are also exponentially convergent for $\lambda\to+\infty$. The reality of the integral follows from the integral representation of the Hankel function:
\begin{align}
\label{eq:integralrepresentation}
\smash{H^{(1)}_\nu(z) = \frac{e^{\frac{-i\pi\nu}{2}}}{i\pi}}\int_{-\infty}^{+\infty}{\rm d}t\,e^{iz\cosh t-\nu t}, \quad \rm{valid ~ for} \quad 0<{\rm{arg}}\,z < \pi\,\,,
\end{align} 
and for us we have ${\rm arg}\,z = \frac{\pi}{2}$. The $\nu$-dependent factors simplify to $e^{\frac{i \pi}{8}}e^{-\frac{ 9 i \pi}{8}} =-1$ leaving us with a real $\mathcal{T}$. We remind the reader that having a non-vanishing trispectrum requires $\psi_{4}$ to contain an imaginary part which is indeed the case here thanks to the overall factor of $i$ that comes from ${\rm d} \eta \, \eta^{13} \rightarrow (i-1)^{14}\, {\rm d}\lambda \, \lambda^{13}$. Note that we are able to use this Wick rotation to assess the reality of this time integral since it is IR-finite. This ensures that we don't need to introduce an IR cut-off at $\eta_{0}$. \\

\noindent In preparation for possible future constraints on the operator in Eq.~\eqref{quartic_action_GI}, we write down the expression for the trispectrum of the comoving curvature perturbation $\zeta = {-H\pi}$. Using Eq.~\eqref{eq:GC_PS}, and the definition of $\tilde{c}^2 = H\tilde{\Lambda}/(2\Lambda^2)$, we find 
\be
\label{eq:template_GI}
B^\zeta_{4} = {\frac{512i\pi^3 \Lambda^5(H\tilde{\Lambda})^{3/2}}{\Lambda_{3}^2 \tilde{\Lambda}^6 \Gamma(\frac{3}{4})^2}}(\Delta^2_\zeta)^3(\k_2\cdot\k_3\times \k_4)(\k_2\cdot \k_3) k_1^{\frac{1}{2}}k_2^{-\frac{3}{2}}k_3^{\frac{1}{2}}k_4^{\frac{1}{2}}\,{\cal T}(k_1,k_2,k_3,k_4) + \text{$23$ perms.}\,\,, 
\ee 
where we defined 
\begin{align}
\Delta^2_\zeta\equiv k^3P_\zeta(k)\,.    
\end{align}
It is useful to check that this result conforms to our expectations. To this end, we notice that the size of the trispectrum can be estimated by comparing the quartic Lagrangian to the quadratic one at horizon crossing, and using the substitutions $\partial_\eta \sim H$, $k^2\sim H\Lambda^2/\tilde\Lambda$ and $\pi=\zeta/H$. Then we have
\be
\label{eq:overall_size}
\frac{B^\zeta_4}{P_\zeta^2}\sim\frac{{\cal L}_4}{{\cal L}_2}\bigg|_{\rm crossing}\sim\bigg(\frac{H^{\frac{3}{2}}\Lambda^5}{\Lambda^2_{\rm PO}\tilde{\Lambda}^{\frac{9}{2}}}\bigg)\zeta^2\,\,,
\ee 
which indeed agrees with Eq.~\eqref{eq:GC_PS} and \eqref{eq:template_GI}. 
We can also compute $\smash{\tau^{\rm PO}_{\rm NL}}$, which is given by $\smash{\sim B^\zeta_4/P_\zeta^3}$, in terms of the observed power spectrum of $\zeta$, $\smash{\Delta^2_\zeta\approx 4\times 10^{-8}}$. Using Eq.~\eqref{normalization} (the explicit choice of reference scale $\bar{k}$ is irrelevant due to scale invariance) we find 
\begin{align}
\tau^{\rm PO}_{\rm NL} \approx {-\frac{4\times 10^{-7}}{\Lambda^2_{\rm PO}}\bigg(\frac{\Lambda^{28}}{\tilde{\Lambda}^{18}}\bigg)^{\frac{1}{5}}}\,\,. 
\end{align}
This can be made large for small $\Lambda_{\rm PO}$, with the caveat that a too small $\Lambda_{\rm PO}$ could lead to $\smash{\tau^{\rm PO}_{\rm NL}\times\Delta^2_\zeta}$ becoming close to $1$ which would jeopardize perturbativity. Given the fact that we propose GI only as an example of how (non-perturbative) deviations from the Bunch-Davies vacuum provide a counterexample to the results of Section~\ref{sec:3}, we leave a more detailed investigation of naturalness constraints to future work. It would also be interesting to carry out the analysis of \cite{Philcox:2022hkh} using the template of Eq.~\eqref{eq:template_GI}, in order to confirm whether current observational bounds from BOSS on $\tau^{\rm PO}_{\rm NL}$ are actually competitive with simple bounds coming from the requirement of perturbativity. \\

\noindent We have learned quite a lot from this calculation. Consider the operator in Eq.~\eqref{ZerothPOoperator} which only differs from Eq.~\eqref{FirstPOoperator} by having one less time derivative. Adding time derivatives cannot alter the reality of the wavefunction coefficient because they come in the combination $\eta \partial_\eta$ by scale invariance. The final operator that we wrote, Eq.~\eqref{SecondPOoperator}, however, has a different number of spatial derivatives compared to the other two: its had seven while the others have nine. While the time derivatives do not alter the reality of the wavefunction coefficient, spatial derivatives can. Indeed, by removing two spatial derivatives we lose two powers of $\eta$ in the integrand so when we rotate we pick up a factor of $i$. This means that if an operator with nine spatial derivatives yields an imaginary wavefunction and therefore a non-zero correlator, an operator with seven spatial derivatives will yield a real wavefunction and therefore a vanishing correlator. We conclude that Eqs.~\eqref{ZerothPOoperator} and~\eqref{FirstPOoperator} yield a non-zero signal, while Eq.~\eqref{SecondPOoperator} yields a vanishing signal. \\

\noindent More generally, $\BPO_4 \neq 0$ in GI only if we have $5+4n$ spatial derivatives, regardless of the number of time derivatives.\footnote{One should be able to derive this result using the boostless bootstrap derivation in Section \ref{sec:3.2}, but using a modified COT that applies to Ghost Inflation. Such COT for Ghost Inflation was briefly discussed in \cite{Goodhew:2021oqg}.} It is easy to convince oneself that in the case with five spatial derivatives, the necessary number of time derivatives required to find a non-zero operator ensures that the scaling dimension is never less than $3$ i.e.~it is never less than the scaling dimension of the operator in Eq.~\eqref{ZerothPOoperator}. This is true for quartic interactions that come from four building block operators, and also those with fewer building blocks. With this in mind, let us also write down the trispectrum for curvature perturbations that comes from Eq.~\eqref{ZerothPOoperator}. For a quartic action in conformal time given by 
\be
\label{quartic_action_GI_new}
S_{\pi\pi\pi\pi} = \frac{1}{M_{\rm PO}} \int {\rm d}^3 x {\rm d} \eta\,a^{-5}(\eta) \epsilon_{ijk}\partial_m\partial_n \pi \partial_n\partial_i {\pi}\partial_m\partial_l \partial_j {\pi} \partial_l \partial_k{\pi}\,\,, 
\ee
where we have introduced the scale $M_{\rm PO}$ to distinguish this trispectrum from that coming from Eq.~\eqref{FirstPOoperator}, we find 
\be
\begin{split}
\psi_4 &= {-\frac{8\pi^4 H^5}{\Gamma(\frac{3}{4})^4 M_{\rm PO}\tilde{c}^6}} \epsilon_{ijk}k^i_2 k^j_3 k^k_4 k^m_1k^m_3 k^n_1k^n_2 k^l_3 k^l_4 k_{1}^{\frac{3}{2}} k_{2}^{\frac{3}{2}} k^{\frac{3}{2}}_3 k^{\frac{3}{2}}_4 {\cal T}(k_1,k_2,k_3,k_4) + \text{$23$ perms.}\,\,, 
\end{split}
\ee 
where now we have 
\be
\label{mathcalTnew}
{\cal T}(k_1,k_2,k_3,k_4) = \int_{0}^{+\infty}{\rm d}\lambda\, \lambda^{11}\,H^{(1)}_{{\frac{3}{4}}}(2ik^2_1\lambda^2)H^{(1)}_{{\frac{3}{4}}}(2ik^2_2\lambda^2)H^{(1)}_{{\frac{3}{4}}}(2ik^2_3\lambda^2)H^{(1)}_{{\frac{3}{4}}}(2ik^2_4\lambda^2)\,\,. 
\ee 
One can check, using the integral representation of Eq.~\eqref{eq:integralrepresentation}, that ${\cal T}$ is purely imaginary, leading to an imaginary $\psi_4$. The final expression for the trispectrum of the comoving curvature perturbation is 
\be
\label{new_GI_trispectrum}
\begin{split}
B^\zeta_{4} = {\frac{128i\pi^3 \Lambda^5(H\tilde{\Lambda})^{1/2}}{M_{\rm PO} \tilde{\Lambda}^5 \Gamma(\frac{3}{4})^2}}(\Delta^2_\zeta)^3\frac{(\k_2\cdot\k_3\times \k_4)(\k_1\cdot \k_3)(\k_1\cdot \k_2)(\k_3\cdot \k_4)}{k_1^{\frac{3}{2}}k_2^{\frac{3}{2}}k_3^{\frac{3}{2}}k_4^{\frac{3}{2}}}\,{\rm Im}\,{\cal T}(k_1,k_2,k_3,k_4) + \text{$23$ perms.} 
\end{split}
\ee 

\noindent We can now estimate the size of the non-Gaussianities from this operator. We take the ratio between $\smash{B^\zeta_4}$ from the $M_{\rm PO}$ operator and the one from the $\smash{\Lambda^2_{\rm PO}}$ operator in the $\smash{\tau^{\rm NL}_{\rm PO}}$ configuration of Eq.~\eqref{eq:tet_config}, finding 
\be
\frac{4\,\Lambda^2_{\rm PO}}{H M_{\rm PO}}\approx \frac{3\times10^3\,\Lambda^2_{\rm PO}}{M_{\rm PO}\Lambda^{\frac{2}{5}}\tilde{\Lambda}^{\frac{3}{5}}}\,\,, 
\ee 
where we have used Eq.~\eqref{eq:GC_PS} to express the Hubble rate in terms of $\smash{\Delta^2_\zeta\approx4\times10^{-8}}$. 
We emphasize that we do this comparison only to have a vague idea of the size of the trispectrum from the operator \eqref{ZerothPOoperator}: a proper analysis following \cite{Hou:2022wfj,Philcox:2022hkh} is needed in order to assess the importance of the difference in the shapes. Nevertheless, it is important to stress that in terms of operators with four building blocks we expect that Eq.~\eqref{new_GI_trispectrum} is the leading signal, and we will investigate other operators with the same scaling dimension in the future.


\section{Yes-go 3: exchanging massive spinning fields} 
\label{sec:6}

\noindent In this section we consider a different setup in which large parity violation can be obtained in the scalar trispectrum at tree level due to the exchange of a massive spinning field, which we denote by $\sigma$, with masses in the range $0 < m_{\sigma}/H < 3/2$. We go back to assuming a linear dispersion relation for the Goldstone mode, and consider cubic couplings within the EFTI following the set-up of \cite{Bordin:2018pca}. We will first introduce this formalism. We will then show that the exchange diagram in the wavefunction calculation is purely real so it does not contribute to $\rho_{4}$ and therefore does not contribute to the parity-odd trispectrum. We show this by two complementary methods: using weight-shifting operators, and using a Wick rotation of the nested time integrals (see Appendix \ref{sec:app}). Conversely, we show that the factorized contribution $\BPO_4 \sim \rho_3 \rho_3$ yields a non-zero signal.  

\paragraph{Spinning fields in the EFTI} To treat spinning fields in the Effective Field Theory of Inflation (EFTI) we follow the nice formalism of \cite{Bordin:2018pca} and exemplify the discussion in the case of a spin-$1$ field. Within this set-up one can avoid constraints on the mass of spinning fields in the form of the Higuchi bound\footnote{In the limit where we have full de Sitter symmetries, the Higuchi bound sets a lower bound on the mass of spinning fields by demanding that their longitudinal mode is not a ghost thereby ensuring the fields adhere to unitary representations of the de Sitter group.} \cite{Higuchi:1986py} by allowing for sizeable couplings between spinning fields and the inflaton background. This is a natural set-up for us in this work as we are not assuming exact or approximate invariance under de Sitter boosts. We will be working with spin-$1$ where the mass range we choose does not actually violate the Higuchi bound, however this is primarily for illustrative purposes and our results are easily generalised to higher-spins where the allowed range of masses here is large than those allowed by the Higuchi bound. During inflation fields are classified according to their transformation under the unbroken group of rotations, but to couple them to four-dimensional fields within the EFTI, we introduce a fictitious four-vector that when coupled to gravity and the foliation ensures that the resulting EFT has the correct linear and non-linear symmetries. For spin-$1$, the object that transforms as a vector under all diffeomorphisms is \cite{Bordin:2018pca}
\begin{align}
    \Sigma^\mu(\Sigma^i,\pi)=\left( -\frac{\Sigma^i \partial_i \pi}{1+\dot \pi},\Sigma^i \right)\,\,.
\end{align}
\noindent It is convenient to make all scale factors manifest by contracting spatial indices with $\delta_{ij}$ as opposed to $g_{ij}$. To do so we introduce $\sigma^i = a \Sigma^i$ such that the free theory for the spin-$1$ field is \cite{Bordin:2018pca}
\begin{align} 
    \label{sigmafree}
    S_2=\frac{1}{2}\int {\rm d}^3x {\rm d}\eta\, a^2(\eta)\left[(\sigma_i')^2- c_1^2 (\partial_i \sigma_j)^2 -(c_0^2-c_1^2) (\partial_i\sigma^i)^2-a(\eta)^2 m_{\sigma}^2(\sigma^i)^2 \right]\,\,,
\end{align}
where $c_{0,1}$ are the speeds of sound of the longitudinal and transverse components of $\sigma^i$ and $m_{\sigma}$ is an arbitrary mass.\footnote{Note that our $m_{\sigma}$ differs from that in \cite{Bordin:2018pca} because we wrote the action in terms of conformal time.} From now on all spatial indices are raised and lowered with $\delta_{ij}$. Crucially, in this set-up the kinetic term of the longitudinal mode is not related to the mass parameter which is why the Higuchi bound can be avoided. We note that the symmetries of EFTI dictate that this free action must also come with interactions of the form $\pi \sigma \sigma$ since $\Sigma \Sigma \supset \sigma \sigma + \pi \sigma \sigma$. Such interactions cannot contribute to the scalar trispectrum at tree level so they are not of interest for us in this work. Converting to Fourier space, we write $\smash{\sigma_{i}(\bfk, \eta) = \sum_{h} \sigma^{(h)}_{k}(\eta) \epsilon^{h}_{i}(\bfk)}$ where we normalise the polarisation vector according to $\epsilon_{i}^{h}(\bfk) \epsilon_{i}^{h'}({-\bfk}) = \delta_{h h'}$ (in this way Eq.~\eqref{sigmafree} is already the canonically-normalized quadratic action). We can easily see from the free theory that the mode functions will be equivalent to those of a massive scalar field in de Sitter. Assuming Bunch-Davies initial conditions, we have 
\begin{align} \label{MassiveModes}
\sigma_k^{(h)}=\frac{H\sqrt{\pi}}{2} e^{i\pi(\nu+1/2)/2} (-\eta)^{3/2} H_\nu^{(1)}(-c_h k \eta)\,\,, \quad \text{with} \quad \nu\equiv\sqrt{\frac{9}{4}-\frac{m_\sigma^2}{H^2}}\,\,.
\end{align}
Therefore the bulk-to-boundary propagator for this spin-$1$ field is\footnote{Recall that the bulk-to-boundary propagator is fixed by $[\sigma_{k}^{(h)}]^{\ast}$ which is why it depends on the Hankel function of the second kind.} 
\begin{align}
\label{massive_bulk_to_boundary}
K^{(h)}_{\sigma} = \left( \frac{\eta}{\eta_0} \right)^{3/2} \frac{ H_\nu^{(2)}(-c_h k \eta)}{H_\nu^{(2)}(-c_h k \eta_0)}\,\,,
\end{align}
and the power spectrum is 
\begin{equation}
\label{sigma_PS_definition}
P^{(h)}_{\sigma}(k) = \frac{\pi H^2}{4}({-\eta_0})^3 H_\nu^{(1)}(-c_h k \eta_{0}) H_\nu^{(2)}(-c_h k \eta_{0}) \,\,,
\end{equation} 
which has the correct mass dimension for a canonically-normalized field. This propagator simplifies for two special choices of the mass: for the massless case $m_{\sigma}^2 = 0$ ($\nu=3/2$) and the conformally-coupled case $m_{\sigma}^2 = 2 H^2$ ($\nu=1/2$) we have
\begin{align}
m_{\sigma}^2 = 0 &\then K_\sigma=(1-ic_h k\eta) e^{i c_hk\eta} \label{MasslessK} \,\,, \\
m_{\sigma}^2 = 2H^2  &\then K_\sigma=\frac{\eta}{\eta_0} e^{i c_hk\eta} \label{CCK}\,\,.
\end{align}
In these limiting cases our results from Section \ref{sec:3} show that no parity violation can occur due to the exchange of such a massive field so in the remainder of this section we will concentrate on more general light masses where $0<m_{\sigma}/H<3/2$. \\

\begin{figure}
    \centering
    \includegraphics[width=8cm]{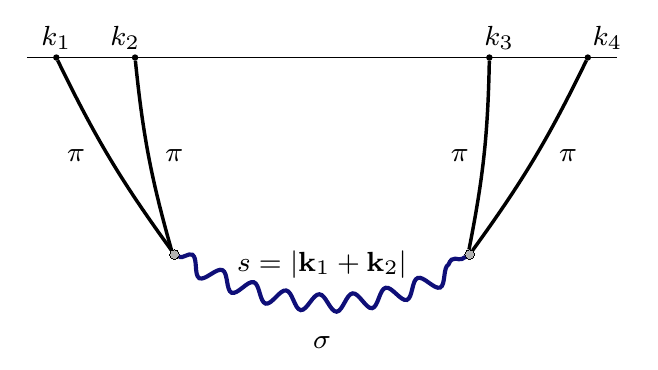}
    \caption{$s$-channel exchange diagram for the exchange of a massive spinning field.}
    \label{fig:exchange_diagram}
\end{figure}

\noindent Let's now turn our attention to interactions of the form $\pi \pi \sigma$ that can contribute to the scalar trispectrum via particle exchange. To realise a parity-odd trispectrum, we need one of the two vertices to have an odd number of spatial momenta, and the other to have an even number. To construct these actions we couple $\Sigma^{\mu}$ to the building blocks of the EFTI (see e.g.~Eqs.~\eqref{building_blocks-1},~\eqref{building_blocks-2},~\eqref{building_blocks-3} for those with the lowest number of derivatives). Denoting schematically by $\mathcal{O}$ these building blocks, there are two ways we can construct $\pi \pi \sigma$ interactions: (i) operators of the form  $\mathcal{O}\Sigma$ induce quadratic mixing terms where $\pi$ mixes with the longitudinal mode of $\sigma$, then the non-linearly realized symmetries also demands the presence of $\pi \pi \sigma$ couplings; (ii) operators of the form $\mathcal{O} \mathcal{O} \Sigma$ start at cubic order in fluctuations so the lowest order terms are just $\pi \pi \sigma$. For the purpose of this paper, where we are aiming to provide some yes-go examples for parity violation in the scalar trispectrum, we will work with the simplest set-up where there is no quadratic mixing. This ensures that there are no other Feynman diagrams to consider beyond the single-exchange diagram shown in Fig.~\ref{fig:exchange_diagram}, 
where we show only the $s$-channel for simplicity. Furthermore, we take $\sigma_{i}$ to be transverse since the longitudinal mode cannot contribute to parity-odd exchange. This means that we can take $\nabla_\mu\Sigma^\mu = 0$ and consequently there is no loss of generality in taking the Goldstone $\pi$ to appear only via $\partial_{i}\partial_{j} \pi$, $\ddot{\pi}$ and $\partial_i\dot{\pi}$ at leading order in derivatives, since we can only use $\smash{\delta\!K_{\mu\nu}}$ and $\smash{\nabla_\mu g^{00}}$. In terms of symmetries, these are operators that are invariant under the leading part of the non-linearly realised symmetries. At zeroth order in fields, $\sigma^i$ does not transform which resonates with the fact that it can be thought of as a matter field in the CCWZ construct of effective actions \cite{Bordin:2018pca}. Indeed, only Goldstones transform at this order and only $\pi$ is a Goldstone. The leading cubic action in conformal time is then given by 
\begin{align} 
\label{CubicAction}
S_{\pi \pi \sigma} = \int {\rm d}^3 x {\rm d}\eta\,\Big[ \lambda_{1} \partial_i \pi' \partial_i \partial_j \pi \sigma^j +  \lambda_{2} \pi'' \partial_i \pi' \sigma^i + \lambda_{3} a^{-1} \epsilon_{ijk} \partial_i \partial_l \pi \partial_j \partial_l \pi' \sigma^k + \lambda_{4} a^{-1} \epsilon_{ijk} \partial_i \pi'' \partial_j \pi' \sigma^k\Big]\,\,.
\end{align}
In the presence of only the first and the second operators, the action would be invariant under parity if $\sigma^i$ transformed as a vector. Conversely, the third and fourth interactions are compatible with parity only if $\sigma^i$ is a pseudo-vector, $P \sigma^i(\bfk) P = + \sigma^i(\bfk) $. Hence, any process that involves both $\lambda_1$ or $\lambda_2$ and $\lambda_3$ or $\lambda_4$ leads to parity violation. For definiteness we will concentrate on the $\mathcal{O}(\lambda_{1} \lambda_{3})$ contribution to the trispectrum but many of the following results hold more generally. This action comes from the following unitary-gauge operators: 
\begin{align}
\lambda_{1}& \to (\nabla_{\mu} \delta g^{00}) \delta\!K^\mu_{\hphantom{\mu}\nu}\Sigma^\nu\,\,, \\
\lambda_{2} &\to \Sigma^\mu(n^\nu\nabla_\nu\delta g^{00})\nabla_\mu\delta g^{00}\,\,, \\
\lambda_{3} &\to {\bf e}^{\mu\nu\rho\sigma}n_\mu\delta\!K_{\nu\lambda}(n^\alpha\nabla_\alpha\delta\!K^\lambda_{\hphantom{\lambda}\rho})\Sigma_\sigma\,\,, \\
\lambda_{4} &\to{\bf e}^{\mu\nu\rho\sigma}n_\mu\nabla_\nu(n^\alpha\nabla_\alpha\delta g^{00})(\nabla_\rho\delta g^{00})\Sigma_\sigma\,\,. 
\end{align}

\noindent Let us stress that within this set-up where we look for parity violation due to particle exchange, it is not possible to have a large parity-odd signal in combination with a small parity-even one. Indeed, parity-even signals can come from both $\mathcal{O}(\lambda_{1}^2)$ and $\mathcal{O}(\lambda_{3}^2)$ contributions to the wavefunction and if the parity-odd signal is made large by having large $\lambda_{1}$ or $\lambda_{3}$, one of these parity-even signals will also be large. We therefore expect a detection of a parity-odd trispectrum due to particle exchange to be accompanied by a detection of a parity-even trispectrum, unless there is some kinematical configuration where the parity-even shape is small rather than the overall coupling. We will also discuss the constraints on the couplings coming from the requirements of perturbativity in a moment.


\paragraph{Exchange contribution to the quartic wavefunction} Let us first consider the exchange diagram wavefunction contribution $\psi_{4}$ to the trispectrum $\BPO_4$, before moving onto the factorised contribution. Within the range of masses that we are interested in, we are restricting to cases where $\nu$ is real c.f. Eq.~\eqref{MassiveModes}, and throughout the remainder of this section we will fix the speed of the exchanged field to unity, $c_{1}=1$, while keeping the speed of the Goldstone mode, $c_{s}$, general. The $s$-channel exchange diagram at $\mathcal{O}(\lambda_{1} \lambda_{3})$ is given by
\begin{equation}
\begin{split}
\psi_{4,s} &= \sum_{h = \pm 1}(- i) \times \Big[{- i} \lambda_{1} k_1^{i}k_2^{i}k^{j}_2 \epsilon^{h}_{j}({\bf{s}})\Big] \times\Big[{- \lambda_{3}} H \epsilon_{ijk} k^{i}_3 k^{l}_3 k^{j}_4 k^{l}_4 \epsilon^{h}_{k}(- {\bf{s}})\Big] \\
&\hphantom{\;\;\sum_{h = \pm 1}(- i) \times} \times \int {\rm d} \eta {\rm d} \eta'\, \eta' K'_{\pi}( k_{1},\eta)K_{\pi}(k_{2},\eta) G_{\sigma}(\eta, \eta', s) K_{\pi}( k_{3},\eta')K'_{\pi}( k_{4},\eta') + \text{$7$ perms.} \\
&= \frac{\lambda_{1} \lambda_{3} H}{4} (s^2 -k_{1}^2 - k_{2}^2)(s^2-k_{3}^2-k_{4}^2) \epsilon_{ijk} k^{i}_3  k^{j}_4 k^{m}_2 \sum_{h = \pm 1} \epsilon^{h}_{m}({\bf{s}}) \epsilon^{h}_{k}(- {\bf{s}}) \\
&\;\;\;\; \times \int {\rm d} \eta {\rm d} \eta'\, \eta' K'_{\pi}( k_{1},\eta)K_{\pi}( k_{2},\eta) G_{\sigma}(\eta, \eta', s) K_{\pi}( k_{3},\eta')K'_{\pi}( k_{4},\eta') + \text{$7$ perms.}\,\,,
\end{split}    
\end{equation}
where we have used momentum conservation at each vertex. Once we sum over the remaining two channels, we get the correct number of $4!$ permutations. We also sum over the helicities of the exchanged field, and we restrict to the transverse components since the longitudinal mode cannot give rise to a non-zero parity-odd wavefunction. Using known polarisation sums, see e.g. \cite{CosmoBootstrap2,CosmoBootstrap3, WFCtoCorrelators2}, we can eliminate all reference to the internal field. Using $\sum_{h = \pm 1} \epsilon^{h}_{i}({\bf{s}}) \epsilon^{h}_{j}(- {\bf{s}}) = \Pi_{ij}({\bf{s}}) = \delta_{ij} - s_{i}s_{j}/s^2$,\footnote{This projection tensor is transverse as it should be: $\smash{s_{i}\Pi_{ij}({\bf{s}}) = s_{j} - s_{j} s^2/s^2 = 0}$, and has the correct normalisation since $\smash{\delta^{ij} \Pi_{ij}({\bf{s}}) = 2}$. The $2$ comes from summing over the two helicities, using $\smash{\epsilon_{i}^{h}(\bfs) \epsilon_{i}^{h'}({-\bfs}) = \delta_{h h'}}$.} we then have 
\begin{equation}
\label{psi4s}
\begin{split}
\psi_{4,s} &= \frac{\lambda_{1} \lambda_{3} H}{4} (s^2 -k_{1}^2 - k_{2}^2)(s^2-k_{3}^2-k_{4}^2) \epsilon_{ijk} k^{i}_3 k^{j}_4 k^{k}_2 \mathcal{I}_{E}(k_{1},k_{2},k_{3},k_{4},s)  + \text{$7$ perms.}\,\,,
\end{split}
\end{equation}
where we have defined 
\begin{align} \label{MassiveTimeIntegral}
\mathcal{I}_{E}(k_{1},k_{2},k_{3},k_{4},s) = \int {\rm d} \eta {\rm d} \eta'\, \eta' K'_{\pi}( k_{1},\eta)K_{\pi}(k_{2},\eta) G_{\sigma}(\eta, \eta', s) K_{\pi}( k_{3},\eta')K'_{\pi}( k_{4},\eta')\,\,.
\end{align}
We note that the second term in the polarisation sum does not contribute since $\epsilon_{ijk}k^{i}_3 k^{j}_4 s^{k} = 0$ by momentum conservation. It is useful to write the bulk-bulk propagator of the massive field explicitly in terms of the Hankel functions. We have
\begin{align}
G_{\sigma}(\eta, \eta', s) = \frac{i \pi H^2}{4}\left[ \theta(\eta -\eta') (\eta'\eta)^{3/2}H_{\nu}^{(2)}(-s \eta')[H_{\nu}^{(2)}(-s \eta)+H_{\nu}^{(1)}(-s \eta)] + (\eta \leftrightarrow \eta') \right]\,\,,
\end{align}
where we have used
\begin{align}
\frac{P_{\sigma}(s)}{[(-\eta_{0})^{3/2} H^{(2)}_{\nu}(-s \eta_{0})]^2} &= \frac{\pi H^2}{4} \frac{H^{(1)}_{\nu}(-s \eta_{0})}{H^{(2)}_{\nu}(-s \eta_{0})} \rightarrow -\frac{\pi H^2}{4}\,\,, \label{PowerSpectrumRation} \\
{\rm{Im}}\, K(s, \eta) &= -\frac{i}{2} \frac{(-\eta)^{3/2}}{(-\eta_{0})^{3/2} H_{\nu}^{(2)}(-s \eta_{0})}[H_{\nu}^{(2)}(-s \eta)+H_{\nu}^{(1)}(-s \eta)] \,\,,
\end{align}
and have assumed real $\nu$ which is the case for our masses of interest. \\

\noindent As we have explained a number of times, whether $\psi_{4,s}$ contributes to the trispectrum or not depends on the properties of $\mathcal{I}_{E}$: it must contain an imaginary part, such that $\psi_{4,s}$ is itself imaginary, for there to be a non-vanishing contribution. We will now study the properties of this nested time integral.
It turns out that $\mathcal{I}_{E}$ can be related to a simpler building block, namely the four-point function of a conformally-coupled field that interacts with a massive scalar field $\sigma$ through the trilinear vertex $\vpi^2 \sigma$. The Lagrangian of the $\vpi$-$\sigma$ theory is given by
\begin{align}
    S=\int{\rm d}^3x {\rm d}\eta \,a^4(\eta)\,\left(-\dfrac{1}{2}(\partial_\mu \vpi)^2-H^2 \vpi^2-\dfrac{1}{2}(\partial_\mu \sigma)^2-\dfrac{1}{2}m_\sigma^2 \sigma^2-g\vpi^2 \sigma\right)\,\,,
\end{align}
where $\vpi$ is a conformally-coupled field and $\sigma$ is a scalar field with the same mass as $\sigma^i$. The four-point wavefunction coefficient of $\vpi$ induced by the exchange of $\sigma$ is given by
\begin{align}
    \psi_{4,s}^{\vpi}=-\dfrac{4 i g^2}{\eta_0^4}\int \dfrac{{\rm d}\eta {\rm d}\eta'}{\eta^2 \eta'^2} e^{i(k_1+k_2)\eta} e^{i(k_3+k_4)\eta'}G_\sigma(s,\eta,\eta')\,\,.
\end{align}
This four-point coefficient is convergent in the $\eta_0\to 0$ limit \textit{iff} $0\leq \nu <\frac{1}{2}$. 
Let us define $F(u,v)=-\frac{\eta_0^4 s}{4g^2}\psi_4$ which only depends on the dimensionless quantities $u=\frac{s}{k_1+k_2}$ and $v=\frac{s}{k_3+k_4}$. A similar quantity was bootstrapped in \cite{Arkani-Hamed:2018bjr}. Here we follow the conventions laid out in Section 4.2 of \cite{Goodhew:2021oqg}, where $F(u,v)$ was found to be
    \begin{equation}
    \label{fuv}
        F(u,v)=\begin{cases}\displaystyle \sum_{m,n=0}^\infty c_{mn} u^{2m+1}\left(\frac{u}{v}\right)^n+\frac{\pi}{2\cos(\pi\nu)}g(u,v)\,\,, \quad \lvert u\rvert \leq \lvert v\rvert \\\displaystyle\sum_{m,n=0}^\infty c_{mn} v^{2m+1}\left(\frac{v}{u}\right)^n+\frac{\pi}{2\cos(\pi\nu)}g(v,u)\,\,, \quad \lvert v\rvert \leq \lvert u\rvert \end{cases}\,\,.
    \end{equation}
Here, the $c_{mn}$'s are a set of real numbers given by
\begin{align}
    c_{mn}=\dfrac{(-1)^n (n+1)(n+2)\dots (n+2m)}{[(n+\frac{1}{2})^2-\nu^2][(n+\frac{5}{2})^2-\nu^2]\dots [(n+\frac{1}{2}+2m)^2-\nu^2]}\,\,,
\end{align}
and 
\begin{align}
    g(u,v)=\kappa(\nu)P_{\nu-\frac{1}{2}}\left(\frac{1}{u}\right)-Q_{\nu-\frac{1}{2}}\left(\frac{1}{u}\right)+Q_{-\nu-\frac{1}{2}}\left(\frac{1}{u}\right)\,\,,
\end{align}
with $P_\nu$ and $Q_\nu$ the associated Legendre functions of the first and second type, respectively, and 
\begin{align}
    \kappa(\nu)=i\pi+\dfrac{\pi}{\cos (\pi\nu)}\dfrac{\sigma(s,\eta_0)}{\sigma^*(s,\eta_0)}=\pi \tan(\pi\nu)\,\,.
\end{align}
Above, $\sigma(s,\eta)$ is the positive-frequency mode function of the massive field, which is the same as \eqref{MassiveModes} with $c_h=1$. Putting everything together, we see that for physical configurations, namely for $0\leq u\leq 1$, the entire $\psi^{\vpi}_{4,s}$ is a real quantity. $\mathcal{I}_{E}$ will inherit this reality as it is related to $\psi_4$ through a Hermitian weight-shifting operator.\footnote{Weight-shifting operators are not always Hermitian, but for our interaction vertices the weight-shifting procedure does not alter the reality of the time integral.} This operator can be reconstructed from the knowledge of the type of interactions inserted at each vertex. The result is given by
\begin{equation}
\begin{split}
    \mathcal{I}_{E}(k_1,k_2,k_3,k_4,s)=&-\dfrac{k_1^2 k_4^2}{c_s^3}\left(\dfrac{\partial^3}{\partial (k_1+k_2)^3}-k_2 \dfrac{\partial^4}{\partial (k_1+k_2)^4}\right)\left(\dfrac{\partial^4}{\partial (k_3+k_4)^4}-k_2 \dfrac{\partial^4}{\partial (k_1+k_2)^5}\right) \\ 
&\times F\left(\dfrac{s}{c_s(k_1+k_2)},\dfrac{s}{c_s(k_3+k_4)}\right)\,\,.
\end{split}
\end{equation}
In this relation, $s$ is held fixed when the partial derivatives $\frac{\partial}{\partial (k_1+k_2)}$ and $\frac{\partial}{\partial (k_3+k_4)}$ operate. The indicated weight-shifting operator in front of $F(u,v)$ turns the external conformally-coupled, massive states into massless ones, hence the name (see \cite{CosmoBootstrap2} for the systematic study of such operators in de Sitter-isometric situations and \cite{Jazayeri:2022kjy,Pimentel:2022fsc}, when de Sitter boosts are broken). It follows from the reality of this weight-shifting operator and that of the seed function $F(u,v)$ that $\mathcal{I}_{E}$ has to be real. Two technical comments are in order: $(i)$ The expression for $F(u,v)$ in \eqref{fuv} on the upper (lower) line contains a series expansion that converges only for $|u|<1$ ($|v|<1$). However, for a subluminal $c_s<1$, both ratios $u=\frac{s}{c_s(k_1+k_2)}$ and $v=\frac{s}{c_s(k_3+k_4)}$ can take values beyond the unit disk where the aforementioned series expansion is not applicable.\footnote{This situation was studied in depth in \cite{Jazayeri:2022kjy}, and an explicit expression for the four-point function was found as a series expansion that converges outside the corresponding unit disk $|u|<1$ (or $|v|<1$).} Nevertheless, the reality of $\mathcal{I}_{E}$ across the region defined by $0<u<1$ and $0<v<1$ carries over to the entire $u,v>0$ region because $F(u,v)$ is analytic around $u=1$ (for arbitrary $v$) and $v=1$ (for arbitrary $u$).\footnote{This is a direct consequence of having a Bunch-Davies initial condition which implies regularity at the collinear limit (i.e. $u=1$ or $v=1$).} 
$(ii)$ Our proof so far only applies to $0<\nu<1/2$ for which the four-point $\psi_{4,s}$ is IR-convergent. Nevertheless, for lighter states, namely $\frac{1}{2}\leq \nu<\frac{3}{2}$, $\mathcal{I}_{E}$ is still convergent and analytic as a function of $\nu>0$. Consequently, the reality of $\mathcal{I}_{E}$ follows for an arbitrary positive $\nu$ from its reality across $0<\nu<1/2$. 
In Appendix \ref{sec:app} we present a complementary proof that the time integral $\mathcal{I}_{E}$ is purely real by performing Wick rotations on the two time variables.


\paragraph{Factorised contribution to the trispectrum} Having shown that there is no exchange contribution to the trispectrum for our range of masses, we now move on to the factorised contribution for which we need to compute the two cubic wavefunction coefficients. We are considering the $\mathcal{O}(\lambda_{1} \lambda_{3})$ trispectrum. The cubic wavefunction coefficient due to the $\lambda_{1}$ vertex is given by (throughout we are suppressing the integration limits which are always the same)
\begin{equation}
\begin{split}
\psi_{3,1} &= - \lambda_{1}k^{i}_{1}k^{i}_{2}k^{j}_{2} \epsilon_{j}(\bfk_{3}) \int {\rm d} \eta\, K'_{\pi}(k_{1}, \eta)K_{\pi}(k_{2}, \eta)K_{\sigma}(k_{3}, \eta) + (1 \leftrightarrow 2) \\
&= - \frac{\lambda_{1}}{2} (k_{3}^2-k_{1}^2-k_{2}^2)k^{i}_{2} \epsilon_{i}(\bfk_{3})[\mathcal{I}_{1}(k_{1},k_{2},k_{3})-\mathcal{I}_{1}(k_{2},k_{1},k_{3})] \,\,,
\end{split}
\end{equation}
where the overall factor of $(-i)$ in the Feynman rules combines with the factor of $(-i)$ we get from converting the three spatial derivatives to momentum space to give an overall factor of $(-1)$, and in the final line we have used momentum conservation and transversality of the spin-$1$ field to write $\smash{k_{1}^{i} \epsilon_{i}(\bfk_{3}) = -k_{2}^{i} \epsilon_{i}(\bfk_{3})}$. We have also defined 
\begin{align}
\mathcal{I}_{1}(k_{1},k_{2},k_{3}) = \int {\rm d}\eta\, K'_{\pi}(k_{1}, \eta)K_{\pi}(k_{2}, \eta)K_{\sigma}(k_{3}, \eta)\,\,.
\end{align}
Similarly, for the cubic wavefunction coefficient due to the $\lambda_{3}$ vertex we have
\begin{align}
\psi_{3,3} = \frac{i H\lambda_{3}}{2}(k_{3}^2-k_{1}^2-k_{2}^2) \epsilon_{ijk}k^{i}_{1}k^{j}_{2} \epsilon_{k}(\bfk_{3})[\mathcal{I}_{3}(k_{1},k_{2},k_{3})-\mathcal{I}_{3}(k_{2},k_{1},k_{3})] \,\,,
\end{align}
where we have defined
\begin{align}
\mathcal{I}_{3}(k_{1},k_{2},k_{3}) = \int {\rm d} \eta\,\eta\,K_{\pi}(k_{1}, \eta)K'_{\pi}(k_{2}, \eta)K_{\sigma}(k_{3}, \eta)\,\,.
\end{align}
Note that there is an additional factor of $\eta$ in $\mathcal{I}_{3}$ that distinguishes it from $\mathcal{I}_{1}$. Now it is the combination $\rho_{3} = \psi_{3}(\{ k \}, \{ \bfk \}) +\psi^{\ast}_{3}(\{ k \}, \{- \bfk \})$ that contributes to the trispectrum and for these two wavefunction coefficients we have
\begin{align}
\rho_{3,1} &= - i \lambda_{1} (k_{3}^2-k_{1}^2-k_{2}^2)k^{i}_{2} \epsilon_{i}(\bfk_{3}){\rm{Im}}\,[ \mathcal{I}_{1}(k_{1},k_{2},k_{3})-\mathcal{I}_{1}(k_{2},k_{1},k_{3})] \,\,, \\
\rho_{3,3} &= - H \lambda_{3} (k_{3}^2-k_{1}^2-k_{2}^2) \epsilon_{ijk} k^{i}_{1}k^{j}_{2} \epsilon_{k}(\bfk_{3}){\rm{Im}}~ [\mathcal{I}_{3}(k_{1},k_{2},k_{3})- \mathcal{I}_{3}(k_{2},k_{1},k_{3})] \,\,.
\end{align}
As expected, we see that for the interaction vertex with an odd number of spatial momenta, we have an imaginary $\rho_{3}$, while the interaction with an even number of spatial momenta yields a real $\rho_{3}$. \\

\noindent Now, the $s$-channel contribution to this factorised part of the trispectrum takes the form
\begin{align}
B_{4,s}^{\pi} = \prod_{a=1}^4P_\pi(k_a) \sum_{h= \pm 1}P_{\sigma}^{h}(s) \rho_{3,3}(\bfk_{1},\bfk_{2},\bfs)\rho_{1,3}(\bfk_{3},\bfk_{4}-\bfs) +[ (1,2) \leftrightarrow (3,4)] \,\,,
\end{align}
where we only add one permutation since we have already explicitly summed over some permutations in arriving at the above expressions for $\rho_{3,1}$ and $\rho_{3,3}$, and as always we don't include the spin zero ``exchange'' since this will always give a vanishing result. We use an equal sign for this $s$-channel exchange since as we showed above, the quartic wavefunction coefficient does not contribute to the correlator. Computing this product and summing over the helicities of the exchanged field (recalling that only the $\delta_{ij}$ in the polarisation sum contributes) yields
\begin{equation} 
\begin{split}
\label{FactorisedTrispectrumIntermediate}
B_{4,s}^{\pi} &= \left(\prod_{a=1}^4P_\pi(k_a) \right)  i H \lambda_{1} \lambda_{3} P_{\sigma}(s)(s^2-k_{1}^2-k_{2}^2)(s^2-k_{3}^2-k_{4}^2) \epsilon_{ijk}k^{i}_{3}k^{j}_{4}k^{k}_{2} \\ &\;\;\;\; \times{\rm{Im}}~[\mathcal{I}_{{3}}(k_{1},k_{2},s)- \mathcal{I}_{3}(k_{2},k_{1},s)]~ {\rm{Im}}~ [\mathcal{I}_{1}(k_{3},k_{4},s)-\mathcal{I}_{1}(k_{4},k_{3},s)]  +  [ (1,2) \leftrightarrow (3,4)]\,\,. 
\end{split}
\end{equation}
Let us first convince ourselves that this contribution is indeed non-zero. By construction it is invariant under $(1 \leftrightarrow 2)$ and $(3 \leftrightarrow 4)$. This is manifest for $(3 \leftrightarrow 4)$ due to the anti-symmetric nature of $\epsilon_{ijk}$, while it can be made manifest for $(1 \leftrightarrow 2)$ by writing $\epsilon_{ijk} k_{3}^{i}k_{4}^{k}k_{2}^{k} =\epsilon_{ijk} k_{3}^{i}k_{4}^{k}(k_{2}^{k}-k^{k}_{1})/2$. The final sum over replacing $(1,2) \leftrightarrow (3,4)$ cannot yield a vanishing result since $\mathcal{I}_{1}(k_{1},k_{2},s) \neq \mathcal{I}_{3}(k_{1},k_{2},s)$. So generically we expect this contribution to the trispectrum to be non-zero. \\

\noindent Let's now study the time integrals in more detail. First consider the time integral for the $\lambda_{1} $interaction, and we remind the reader that throughout this discussion we take $\nu$ to be real. We have 
\begin{align}
\mathcal{I}_{1}(k_{1},k_{2},s) =- \frac{1}{(-\eta_{0})^{3/2}H_{\nu}^{(2)}(-s \eta_{0})} \int {\rm d} \eta\,(-\eta)^{5/2} ~  c_{s}^2 k_{1}^2(1-i c_{s} k_{2}\eta)e^{i c_{s} k_{12} \eta} H_{\nu}^{(2)}(-s \eta) \,\,,
\end{align}
and therefore
\begin{align}
&\mathcal{I}_{1}(k_{1},k_{2},s) -\mathcal{I}_{1}(k_{2},k_{1},s)= \nonumber \\ -& \frac{c_{s}^2}{(-\eta_{0})^{3/2}H_{\nu}^{(2)}(-s \eta_{0})} \int {\rm d} \eta\,(-\eta)^{5/2} ~  [k_{1}^2(1-i c_{s} k_{2}\eta) - k_{2}^2(1-i c_{s} k_{1}\eta)]e^{i c_{s} k_{12} \eta} H_{\nu}^{(2)}(-s \eta) \,\,.
\end{align}
Here we have defined $k_{12} = k_{1}+k_{2}$. A very similar integral\footnote{It would be interesting to use the recently developed techniques in \cite{Tong:2021wai} to extract the cosmological collider oscillations implied by this signal.}, that we can use to compute this one, was computed in Appendix B of \cite{CosmoBootstrap1}. Indeed in that work it was shown that the integral\footnote{We are actually working with the complex conjugate of the integral that was computed in \cite{CosmoBootstrap1} since our integral contains $\smash{H_{\nu}^{(2)}}$ rather than $\smash{H_{\nu}^{(1)}}$.}
\begin{align}
I_{n}(a,b) = \frac{H \sqrt{\pi}}{2} e^{-\frac{i \pi}{2}(\nu + 1/2)} \int_{-\infty}^{0}  {\rm d} \eta\,(- \eta)^{n-1/2} e^{i a \eta} H_{\nu}^{(2)}(-b \eta) 
\end{align}
is given by
\begin{align}
\label{eq:nima_integral}
I_{n}(a,b) = (-1)^{n+1} \frac{H}{\sqrt{2 b}} \left(\frac{i}{2b} \right)^{n} \frac{\Gamma \left(\alpha \right)\Gamma \left(\beta \right)}{\Gamma(1+n)} \times {}_2 F_1 \Big(\alpha, \beta; 1+n; \frac{1}{2}-\frac{a}{2b}\Big)\,\,,
\end{align}
where we have defined $\alpha=\frac{1}{2}+n-\nu$ and $\beta =\frac{1}{2}+n + \nu$. We can therefore write
\begin{equation}
\begin{split}
\mathcal{I}_{1}(k_{1},k_{2},s) -\mathcal{I}_{1}(k_{2},k_{1},s) &= 
{-\frac{2 c_{s}^2}{H \sqrt{\pi}}}\frac{e^{\frac{i \pi}{2}(\nu + 1/2)}(k_{1}-k_{2})}{(-\eta_{0})^{3/2}H_{\nu}^{(2)}(-s \eta_{0})}  [k_{12}I_{3}(c_{s}k_{12},s)+ i c_{s}k_{1}k_{2} I_{4}(c_{s}k_{12},s)] \,\,,
\end{split}
\end{equation}
which neatly gives us a closed-form expression. Similarly, for the $\lambda_{3}$ interaction we have \begin{equation}
\mathcal{I}_{3}(k_{3},k_{4},s) -\mathcal{I}_{3}(k_{4},k_{3},s)= 
{-\frac{2 c_{s}^2}{H \sqrt{\pi}}}\frac{e^{\frac{i \pi}{2}(\nu + 1/2)}(k_{3}-k_{4})}{(-\eta_{0})^{3/2}H_{\nu}^{(2)}(-s \eta_{0})}  [k_{34}I_{4}(c_{s}k_{34},s)+ i c_{s}k_{3}k_{4} I_{5}(c_{s}k_{34},s)] \,\,.
\end{equation}
We can now write a more compact expression for Eq.~\eqref{FactorisedTrispectrumIntermediate}, the full $s$-channel trispectrum, which is given by 
\begin{equation}
\label{FactorisedTrispectrumCompact}
\begin{split}
 B_{4,s}^{\pi} &= {-\left(\prod_{a=1}^4P_\pi(k_a) \right)} c_{s}^4 H \lambda_{1} \lambda_{3} (s^2-k_{1}^2-k_{2}^2)(s^2-k_{3}^2-k_{4}^2)(k_{1}-k_{2})(k_{3}-k_{4}) \epsilon_{ijk}k^{i}_{3}k^{j}_{4}k^{k}_{2} \\ 
 &\;\;\;\; \times [k_{12}I_{3}(c_{s}k_{12},s)+ i c_{s}k_{1}k_{2}I_{4}(c_s k_{12},s)][k_{34}I_{4}(c_{s}k_{34},s)+ i c_{s}k_{3}k_{4}I_{5}(c_{s}k_{34},s)] \\ 
 &\;\;\;\;\times \sin \left(\frac{\pi}{2}(\nu+1/2)  \right) \cos \left(\frac{\pi}{2}(\nu+1/2)  \right) +  [ (1,2) \leftrightarrow (3,4)]\,\,,
\end{split}
\end{equation}
where, as we have seen before, all $\eta_{0}$ dependence has cancelled out to leave us with an expression that is IR-finite. There are a few quick checks we can do on this result. We notice that it is purely imaginary, as it should be, due to the $I_{3}$ and $i I_{4}$ terms, and we see that for $\nu=1/2$ and $\nu = 3/2$ this trispectrum vanishes thereby reproducing our results in Section \ref{sec:3}. We also see that the result is non-zero for the intermediate range of masses that we are interested in here. Finally, we see that the overall scaling with momenta is $k^{-9}$ as it should be for external massless scalars. Once we sum over the remaining two permutations (the $t$ and $u$ channels), and we divide by $H^4$ this gives us our final result for the parity-odd trispectrum of the comoving curvature perturbation $\zeta$ due to the exchange of a massive (but light) spin-$1$ field during inflation:  
\begin{equation}
\label{FactorisedTrispectrumCompact_zeta}
\begin{split}
 B_{4}^{\zeta} &= {-\left(\prod_{a=1}^4P_\zeta(k_a) \right)} \frac{c_{s}^4\lambda_{1}\lambda_{3}}{H^3} (s^2-k_{1}^2-k_{2}^2)(s^2-k_{3}^2-k_{4}^2)(k_{1}-k_{2})(k_{3}-k_{4}) \epsilon_{ijk}k^{i}_{3}k^{j}_{4}k^{k}_{2} \\ 
 &\;\;\;\; \times [k_{12}I_{3}(c_{s}k_{12},s)+ i c_{s}k_{1}k_{2}I_{4}(c_{s} k_{12},s)][k_{34}I_{4}(c_{s}k_{34},s)+ i c_{s}k_{3}k_{4}I_{5}(c_{s}k_{34},s)] \\ 
 &\;\;\;\;\times \sin \left(\frac{\pi}{2}(\nu+1/2)  \right) \cos \left(\frac{\pi}{2}(\nu+1/2)  \right) +  [ (1,2) \leftrightarrow (3,4)] + t + u\,\,.
\end{split}
\end{equation}

\noindent Let us now discuss the constraints on the couplings $\lambda_{1}$, $\lambda_{3}$ from the requirements of perturbativity. For $c_s$ close to $1$, i.e.~in the case that the spinning field and the Goldstone move with approximately the same speed, we can use the fact that $I_n\sim H$ (up to dimensionful functions of $k_a$ which are fixed by scale invariance and an innocuous hypergeometric function). From this we get that $B_{4}^{\zeta}\sim \lambda_{1}\lambda_{3}\Delta^8_\zeta/H$, which gives $\smash{\tau_{\rm NL}^{\rm PO}\sim B_{4}^{\zeta}/P^3_\zeta}$ of order\footnote{Notice that $\lambda_3$ is dimensionless while $\lambda_1$ has dimension of mass.}
\be
\label{eq:spinning_tauNL_estimate}
\tau_{\rm NL}^{\rm PO}\sim\frac{\Delta^2_\zeta\lambda_{1}\lambda_{3}}{H}\,\,.
\ee 
Let us then estimate the unitarity cutoff of the theory. For $c_s$ close to $1$ this is straightforward. Once we canonically normalize $\pi$ we find that the $\lambda_{1}$ interaction is a dimension $7$ operator suppressed by $\smash{\Lambda_{1}\sim H^{4/3}/(\Delta^2_\zeta\lambda_{1})^{1/3}}$, while the $\lambda_{3}$ one is a dimension $8$ operator suppressed by $\smash{\Lambda_{3}\sim H/(\Delta^2_\zeta\lambda_{3})^{1/4}}$. Requiring that at crossing we are below the cutoff, $H\ll \Lambda_{1,3}$ leads to
\begin{align}
\lambda_3 \Delta_\zeta^2 & \ll 1 \,\,, & \lambda_1 \Delta_\zeta^2 & \ll H\,\,.    
\end{align}
To relate this to the overall size $\tau^{\rm PO}_{\rm NL}$, let's focus on the regime $\Lambda_{1}\sim\Lambda_{3}$. 
Using this we can re-write $\lambda_1$ in terms of $\lambda_3$ and obtain
\be
\tau^{\rm PO}_{\rm NL}\sim\lambda_{3}(\Delta^2_\zeta\lambda_{3})^{\frac{3}{4}}\quad\text{(assuming $\Lambda_{1}\sim\Lambda_{3}$)}\,\,. 
\ee
From this we see that it is possible to choose $\lambda_{3}$ in order for the theory to be weakly coupled at horizon crossing and to have $\smash{\tau_{\rm NL}^{\rm PO}\gg 1}$. \\

\begin{figure}[h!]
\centering
\includegraphics[width=0.75\textwidth]{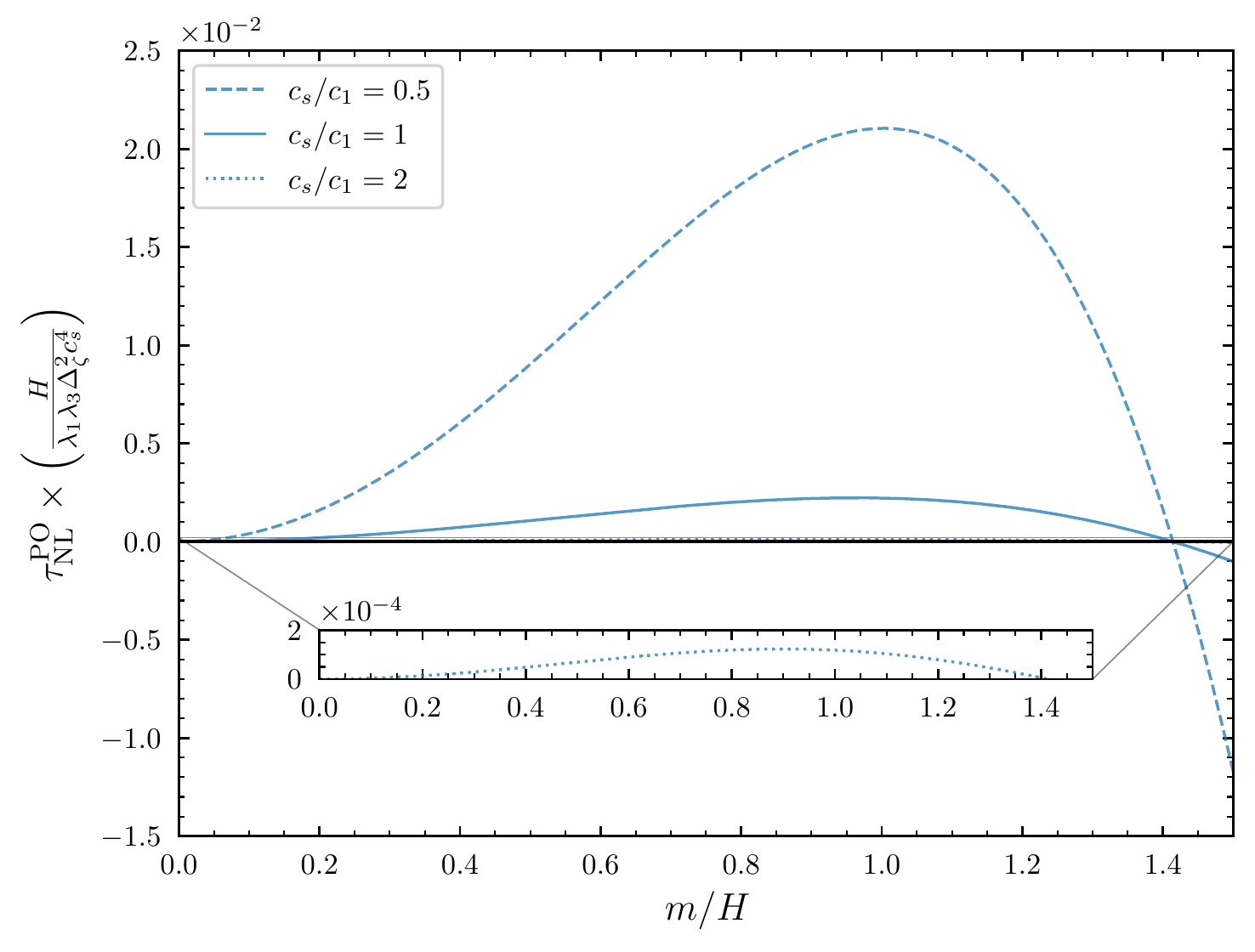}
\caption{$\smash{\tau_{\rm NL}^{\rm PO}}$ for the curvature perturbation $\zeta$ from the exchange of a spin-$1$ field with mass $m$ and interactions $\lambda_{1}$ and $\lambda_{3}$ in Eq.~\eqref{CubicAction}. Notice that the trispectrum vanishes for the conformally-coupled mass $m/H=\sqrt{2}$, in agreement with our general no-go theorems.} 
\label{fig:mass_dependence}
\end{figure}

\noindent Let us then see what happens if $c_s\gg 1$ or $c_s\ll 1$. In both cases we need to keep track of the hypergeometric function in Eq.~\eqref{FactorisedTrispectrumCompact_zeta}. Let us consider first the limit $c_s\gg 1$. We are interested in ${_2F_1}(\alpha,\beta;\gamma;z)$ at large negative $z$. In order to isolate the scaling with $c_s$ we can focus for simplicity on the case $\nu=3/2$, disregarding the fact that the overall trispectrum is zero in this case: the scaling with $c_s$ is unaffected by $\sigma$ having a non-zero mass. Hence, we have $\beta-\alpha=3$, and we can use the relation 
\be
{_2F_1}(\alpha,\beta;\gamma;z) \sim \frac{2(1-z)^{-\alpha}}{\Gamma(\beta)\Gamma(\gamma-\alpha)}\,\,, 
\ee
valid for $|z-1|>1$, $|\arg(1-z)|<\pi$ and $\beta-\alpha\in\mathbb{N}$. From this we see that the square brackets in the second line of Eq.~\eqref{FactorisedTrispectrumCompact_zeta} yield a factor of $\smash{H^2/c_s^5}$ in the limit of large $c_s$, giving
\be
\label{eq:spinning_tauNL_estimate_large_cs}
\tau_{\rm NL}^{\rm PO}\sim\frac{\Delta^2_\zeta\lambda_{1}\lambda_{3}}{Hc_s}\,\,. 
\ee 
In the case of $c_s\ll 1$, given the regularity of the hypergeometric function at $z=1/2$, we see that 
\be
\label{eq:spinning_tauNL_estimate_small_cs}
\tau_{\rm NL}^{\rm PO}\sim\frac{\Delta^2_\zeta c^4_s\lambda_{1}\lambda_{3}}{H}\,\,. 
\ee 
It is more difficult to estimate, in this case, the unitarity cutoff of the theory without doing an explicit calculation. We leave this, and a more detailed discussion on how large $\smash{\tau_{\rm NL}^{\rm PO}}$ must be in order to explain the measurement of Refs.~\cite{Hou:2022wfj,Philcox:2022hkh}, to a future work: we only notice that having a slow $\sigma^i$ suppresses non-Gaussianities at fixed $\lambda_{1}$ and $\lambda_{3}$. In Fig.~\ref{fig:mass_dependence} we plot $\smash{\tau_{\rm NL}^{\rm PO}}$ as defined by our normalization condition of Eq.~\eqref{normalization} at varying mass of the exchanged spinning particle and $c_s$.

\section{Summary and future directions}
\label{sec:7}

In this work we have studied signatures of parity violation from the inflationary primordial universe. Until tensor modes are detected, our best hope is to study the statistics of scalar fluctuations imprinted in primordial curvature perturbations. Parity violation cannot appear in the power spectrum and bispectrum, even at the full non-perturbative level, hence in this paper we have focused here on the scalar, parity-odd trispectrum $\BPO_4$. We have re-derived and extended previous no-go results \cite{Liu:2019fag} that rule out parity violation at tree-level in the presence of any number of scalars of any mass, or fields of any spin with massless de Sitter mode functions. Crucially, these no-go results assume scale invariance and a Bunch-Davies initial state. The \emph{raison d'{\^ e}tre} of these no-go theorems is to help us identify the classes of models in which we \textit{can} get a non-vanishing $\BPO_4$ and to determine the physical implications of a possible discovery of this signal. It is particularly timely to pursue this goal, in light of the recent measurements of the parity-odd four-point function of BOSS galaxies of Refs.~\cite{Hou:2022wfj,Philcox:2022hkh}. While it is important to keep in mind that it is possible that the signal detected (to different levels of significance) by these two groups could be due to systematics instead of fundamental physics, it is equivalently important to stress that this is the first example in a long time where it is data that drives the theory, as far as the study of the inflationary universe is concerned. \\

\noindent To the end of identifying models where a non-vanishing parity-odd trispectrum can be obtained, we have then relaxed these assumptions and derived explicitly the parity-odd trispectra that one can generate in more general classes of models. In Section \ref{sec:4}, we showed that general deviations from scale invariance, which show up in an explicit time dependence of the coupling constants for perturbations, leads to a non-vanishing $\BPO_4$. We stress that, for a time-dependence that is logarithmic in conformal time, the resulting $\BPO_4$ is actually scale-invariant, see e.g. Eq.~\eqref{finalBPOnoscale}. In this case, deviations from scale invariance are present in the wavefunction, but drop out when we compute the trispectrum, which for parity-odd interactions is the imaginary part of the wavefunction. In Section \ref{sec:5} we discussed a different scenario where a non-zero parity-odd trispectrum can arise: while maintaining scale invariance, we deviate from the massless de Sitter mode functions and consider those of Ghost Inflation (GI). In the infinite past, the GI mode functions do not reduce to the standard $e^{\pm i kt}$ mode functions in Minkowski, and as such they avoid the no-go theorems we derived in Section \ref{sec:3}. We concentrated on self-interactions for the Goldstone boson that are ``invariant'' under the leading part of the non-linear boost symmetry (as opposed to Wess-Zumino terms), and presented two examples that lead to a non-zero signal. The resulting $\BPO_4$ are given in \eqref{eq:template_GI} and \eqref{new_GI_trispectrum}, for two different parity-odd interactions. Despite the fact that we concentrated on these operators that arise from four building block covariant ones, we have shown that other operators cannot yield interactions with a lower scaling dimension than that of Eq.~\eqref{ZerothPOoperator}. Finally, we have considered the case in which massive fields are present. For scalars of any mass the primordial parity-odd trispectrum vanishes at tree-level, but in the presence of massive spinning fields it can be non-vanishing. As an example, we have computed $\BPO_4$ for the exchange of massive vector and the result is in Eq.~\eqref{FactorisedTrispectrumCompact_zeta} and plotted as function of mass for a specific configuration in Fig.~\ref{fig:mass_dependence}. \\

\noindent There are other ways to generate a non-vanishing $\BPO_4$, which we don't discuss here. One possibility is a process in which four external scalars exchange a spinning field whose power spectrum is chiral, i.e. the power in one helicity is different from the power in the other. Such a situation arises in models of axions coupled to a $U(1)$ gauge sector in the presence of an $f(\phi)(FF+F\tilde F)$ coupling, as discussed e.g. in \cite{Shiraishi:2016mok}, however sometimes with the added feature of the breaking of statistical isotropy. Other examples of parity-odd scalar trispectra were given in \cite{Liu:2019fag}. \\

\noindent Our work could be extended in a number of directions:
\begin{itemize}
    \item Loop corrections can contribute to $\BPO_4$ already in single-clock inflation with massless de Sitter mode functions. In the scale invariant limit this is an interesting setup where loop effects can be larger than tree-level ones. We will discuss this in detail in an upcoming paper.
    \item It would be interesting to assess what happens when we go beyond the decoupling limit i.e. where we include the effects of dynamical gravity. This can yield new exchange diagrams where curvature perturbations exchange the transverse, traceless part of the graviton, but also new contact diagrams where now the self-interactions can include inverse Laplacians which arise when we integrate out the non-dynamical parts of the metric. Such gravitational effects are expected to be too small to aid with an explanation of the signal potentially found in \cite{Philcox:2022hkh,Hou:2022wfj}, yet it is interesting to understand if such corrections can avoid our no-go theorems.
    \item In Section \ref{sec:6} where we considered the exchange of a massive spinning field, we choose EFTI interactions between the Goldstone mode and the new massive field that did not come with any quadratic mixing terms. However, it would be interesting to extend our analysis to include such mixings that will introduce new exchange diagrams as was considered in \cite{Lee:2016vti} for parity-even bispectra. Such quadratic mixing terms come with a number of spatial derivatives that is dictated by the spin of the massive field. This leads to distinctive signatures in the squeezed limit of cosmological correlators, and potentially new shapes of parity-violating trispectra.
\end{itemize}
In summary, the no-go results and yes-go examples in our work contribute to establish the parity-odd trispectrum as a particularly sensitive probe of new physics during inflation.


\paragraph{Acknowledgements} We thank Harry Goodhew, Jiamin Hou, Misha Ivanov, Sachin Jain, Mehrdad Mirbabayi, Oliver Philcox, S{\'e}bastien Renaux-Petel, Fabian Schmidt, Marko Simonovi{\' c}, Jakub Supe{\l{}} and Matias Zaldarriaga for useful discussions. G.C. acknowledges support from the Institute for Advanced Study. The work of G.C. was initiated at the Aspen Center for Physics, which is supported by National Science Foundation grant PHY-1607611. E.P. has been supported in part by the research program VIDI with Project No. 680-47-535, which is (partly) financed by the Netherlands Organisation for Scientific Research (NWO). This work has been partially supported by STFC consolidated grant ST/T000694/1.  S.J. is supported by the European Research Council under the European Union's Horizon 2020 research and innovation programme (grant agreement No. 758792, project GEODESI). D.S. is supported by a UKRI Stephen Hawking Fellowship [grant number EP/W005441/1] and a Nottingham Research Fellowship from the University of Nottingham. For the purpose of open access, the authors have applied a CC BY public copyright licence to any Author Accepted Manuscript version arising.


\appendix


\section{Wick rotation for \texorpdfstring{$\psi_4$}{\textbackslash psi \_4} due to massive exchange} 
\label{sec:app}

\noindent In this appendix we study the exchange time integral \eqref{MassiveTimeIntegral} and show that it is purely real. In turn this tells us that the associated $\psi_{4,s}$ is real and therefore does not contribute to the trispectrum. \\

\noindent We will actually work with a more general time integral of the form
\be
\label{exchange_integral}
\mathcal{I}_{E}(k_{a},s) = \int {\rm d} \eta {\rm d} \eta'\, \eta^{n_L-4}\eta'^{n_R-4} f_L(k_1,k_2,\eta) f_R(k_3,k_4,\eta')G_{\sigma}(\eta, \eta', s) \,\,, 
\ee
arising from two vertices with $n_{L}$ and $n_{R}$ derivatives. Here the function $f_L$ ($f_R$) depends on $K_\pi(k_1,\eta)$, $K_\pi(k_2,\eta)$ ($K_\pi(k_3,\eta)$, $K_\pi(k_4,\eta)$) and their time derivatives. Using the relation 
\be
\frac{{\rm d}^n}{{\rm d}\eta^n}K_\pi(k,\eta) = e^{ic_s k\eta}(i c_s k)^n(1-n-ic_s k\eta) \,\,
\ee 
we can, for example, write
\be
f_L(k_1,k_2,\eta) = (i c_s k_1)^{n_1}(i c_s k_2)^{n_2}\,{\rm poly}_{\!L}(ic_s k_1\eta,ic_s k_2\eta)e^{ic_s k_{12}\eta}\,\,, 
\ee
where $k_{12} = k_1+k_2$, ${\rm poly}_{\!L}$ is some polynomial, and $n_1+n_2\leq n_L$. A similar expression can be derived for $f_R$, with a corresponding ${\rm poly}_{\!R}$ that depends on $\eta'$. We will assume that $n_{L,R} \geq 4$ such that there are no IR divergences at $\eta \to 0$. We can confirm this by focusing on the case of a massless exchange, $\nu=3/2$. In this case, we have that the bulk-bulk propagator is IR-finite (see Eq.~\eqref{app_BB_prop} below), hence as long as we have a number of derivatives equal to or higher than $4$ on each vertex the integral is guaranteed to converge at late times. \\

\noindent Now recall that for real $\nu$, which we assume throughout this appendix, we can write the bulk-bulk propagator as 
\begin{align}
\label{app_BB_prop}
G_{\sigma}(\eta, \eta', s)=\frac{i \pi H^2}{2}\left[ \theta(\eta -\eta') (-\eta')^{3/2}(-\eta)^{3/2}H_{\nu}^{(2)}(-s \eta')J_{\nu}(-s \eta) + (\eta \leftrightarrow \eta') \right]\,\,.
\end{align}
Let's concentrate on the case where $\eta > \eta'$ meaning that we pick up the first Heaviside theta function in the bulk-bulk propagator. We integrate $\eta'$ from the far past (where we Wick rotate) up to $\eta$, then integrate $\eta$ from the far past to the late-time boundary. To aid with assessing the reality of this integral, we Wick rotate to the path $\eta(\lambda_1)=i\lambda_1$, $\eta'(\lambda_1,\lambda_2)=\eta(\lambda_1)+i\lambda_2$, with $\lambda_1,\lambda_2$ from $+\infty$ to $0$ (the path for the second contribution in the bulk-bulk propagator is readily obtained by $\eta\leftrightarrow\eta'$). With this rotation we have exponential convergence at infinity and one does not have to worry about the inner limit of integration depending on the outer integration variable. We now have
\be
\label{easy_result}
\begin{split}
\mathcal{I}_{E}(k_{a},s) &= (c_s k_1)^{n_1}(c_s k_2)^{n_2}(c_s k_3)^{n_3}(c_s k_4)^{n_4} i^{n_1+n_2+n_3+n_4+n_L+n_R} \\ 
&\;\;\;\;\times\int_0^{+\infty}\int_0^{+\infty}{\rm d}\lambda_1{\rm d}\lambda_2\,F(k_1,k_2,k_3,k_4,c_s,\lambda_1,\lambda_2)\\
&\;\;\;\;\hphantom{\times\int_0^{+\infty}\int_0^{+\infty}{\rm d}\lambda_1{\rm d}\lambda_2}\,\times H_\nu^{(2)}({-is\lambda_{12}})J_\nu(-is\lambda_1) + (\eta \rightarrow \eta')\,\,, 
\end{split}
\ee
where $\lambda_{12} = \lambda_1+\lambda_2$ and $F$ is a real function. Crucially, we then see that the reality of $\smash{\mathcal{I}_{E}(k_{a},s)}$ is determined by the combination of Hankel and Bessel functions, and the number of derivatives in each vertex. Using the following integral representations of the Hankel and Bessel functions:
\be
\label{integral_representations}
H_\nu^{(2)}(z) = {-\frac{e^{\frac{i\pi\nu}{2}}}{i\pi}}\int_{-\infty}^{+\infty}{\rm d}t\,e^{-iz\cosh t-\nu t} \quad\text{and}\quad J_\nu({-iz}) = \frac{2^{1-\nu}({-iz})^\nu}{\sqrt{\pi}\,\Gamma(\nu+\frac{1}{2})}\int_0^1{\rm d}t\,(1-t^2)^{\nu-\frac{1}{2}}\cosh(zt)\,\,,
\ee
valid for $-\pi<{\rm arg}\,z<0$ and $\smash{{\rm Re}\,\nu>{-\frac{1}{2}}}$,\footnote{Both of these conditions are satisfied for us since for $z = -i s \lambda_{12}$ we have ${\rm arg}\,z = -\frac{\pi}{2}$ and for $0 < m_{\sigma}^2 < 2H^2$ we always have $\smash{{\rm Re}\,\nu>{-\frac{1}{2}}}$.} respectively, we see that the product $H_\nu^{(2)}({-is\lambda_{12}})J_\nu(-is\lambda_1)$ is purely imaginary. Indeed, using fact that $\nu$ is real, we have $H_\nu^{(2)}(-i s \lambda_{12}) = i e^{\frac{i \nu \pi}{2}} \times (\rm {Real})$ and $J_\nu(-i s \lambda_{1}) = e^{-\frac{i \nu \pi}{2}} \times (\rm {Real})$. We therefore have
\be
\mathcal{I}_{E}(k_{a},s) = i^{n_1+n_2+n_3+n_4+n_L+n_R+1}\times (\rm{Real})\,\,. 
\ee
Now to find a parity-odd $\psi_{4}$, we need one vertex to have an even number of spatial momenta and the other to have an odd number. Let's take $n_{L}$ to contain an odd number and $n_{R}$ to contain an even number. It follows that $n_{1}+n_{2}+n_{L}$ is odd while $n_{3}+n_{4}+n_{R}$ is even. It then follows that $\mathcal{I}_{E}(k_{a},s)$, and therefore $\psi_{4}$, are purely real. This means that the parity odd quartic wavefunction coefficient due to the exchange of a massive spinning field with real $\nu$ does not contribute to the trispectrum. This proof complements the one we have in Section \ref{sec:6} using weight-shifting operators.

\bibliographystyle{JHEP}
\bibliography{refs}

\providecommand{\href}[2]{#2}\begingroup\raggedright\begin{thebibliography}{10}

\bibitem{Liu:2019fag}
T.~Liu, X.~Tong, Y.~Wang and Z.-Z.~Xianyu, \emph{{Probing P and CP Violations
  on the Cosmological Collider}},
  \href{https://doi.org/10.1007/JHEP04(2020)189}{\emph{JHEP} {\bfseries 04}
  (2020) 189} [\href{https://arxiv.org/abs/1909.01819}{{\ttfamily
  1909.01819}}].

\bibitem{Creminelli:2014wna}
P.~Creminelli, J.~Gleyzes, J.~Nore\~na and F.~Vernizzi, \emph{{Resilience of
  the standard predictions for primordial tensor modes}},
  \href{https://doi.org/10.1103/PhysRevLett.113.231301}{\emph{Phys. Rev. Lett.}
  {\bfseries 113} (2014) 231301}
  [\href{https://arxiv.org/abs/1407.8439}{{\ttfamily 1407.8439}}].

\bibitem{Cabass:2021fnw}
G.~Cabass, E.~Pajer, D.~Stefanyszyn and J.~Supe\l{}, \emph{{Bootstrapping large
  graviton non-Gaussianities}},
  \href{https://doi.org/10.1007/JHEP05(2022)077}{\emph{JHEP} {\bfseries 05}
  (2022) 077} [\href{https://arxiv.org/abs/2109.10189}{{\ttfamily
  2109.10189}}].

\bibitem{Orlando:2022rih}
G.~Orlando, \emph{{Probing parity-odd bispectra with anisotropies of GW $V$
  modes}},  \href{https://arxiv.org/abs/2206.14173}{{\ttfamily 2206.14173}}.

\bibitem{Shiraishi:2016mok}
M.~Shiraishi, \emph{{Parity violation in the CMB trispectrum from the scalar
  sector}}, \href{https://doi.org/10.1103/PhysRevD.94.083503}{\emph{Phys. Rev.
  D} {\bfseries 94} (2016) 083503}
  [\href{https://arxiv.org/abs/1608.00368}{{\ttfamily 1608.00368}}].

\bibitem{Chen:2006dfn}
X.~Chen, M.-x.~Huang and G.~Shiu, \emph{{The Inflationary Trispectrum for
  Models with Large Non-Gaussianities}},
  \href{https://doi.org/10.1103/PhysRevD.74.121301}{\emph{Phys. Rev. D}
  {\bfseries 74} (2006) 121301}
  [\href{https://arxiv.org/abs/hep-th/0610235}{{\ttfamily hep-th/0610235}}].

\bibitem{Seery:2006vu}
D.~Seery, J.E.~Lidsey and M.S.~Sloth, \emph{{The inflationary trispectrum}},
  \href{https://doi.org/10.1088/1475-7516/2007/01/027}{\emph{JCAP} {\bfseries
  01} (2007) 027} [\href{https://arxiv.org/abs/astro-ph/0610210}{{\ttfamily
  astro-ph/0610210}}].

\bibitem{Seery:2008ax}
D.~Seery, M.S.~Sloth and F.~Vernizzi, \emph{{Inflationary trispectrum from
  graviton exchange}},
  \href{https://doi.org/10.1088/1475-7516/2009/03/018}{\emph{JCAP} {\bfseries
  03} (2009) 018} [\href{https://arxiv.org/abs/0811.3934}{{\ttfamily
  0811.3934}}].

\bibitem{Arroja:2008ga}
F.~Arroja and K.~Koyama, \emph{{Non-gaussianity from the trispectrum in general
  single field inflation}},
  \href{https://doi.org/10.1103/PhysRevD.77.083517}{\emph{Phys. Rev. D}
  {\bfseries 77} (2008) 083517}
  [\href{https://arxiv.org/abs/0802.1167}{{\ttfamily 0802.1167}}].

\bibitem{Philcox:2022hkh}
O.H.E.~Philcox, \emph{{Probing parity violation with the four-point correlation
  function of BOSS galaxies}},
  \href{https://doi.org/10.1103/PhysRevD.106.063501}{\emph{Phys. Rev. D}
  {\bfseries 106} (2022) 063501}
  [\href{https://arxiv.org/abs/2206.04227}{{\ttfamily 2206.04227}}].

\bibitem{Hou:2022wfj}
J.~Hou, Z.~Slepian and R.N.~Cahn, \emph{{Measurement of Parity-Odd Modes in the
  Large-Scale 4-Point Correlation Function of SDSS BOSS DR12 CMASS and LOWZ
  Galaxies}},  \href{https://arxiv.org/abs/2206.03625}{{\ttfamily 2206.03625}}.

\bibitem{COT}
H.~Goodhew, S.~Jazayeri and E.~Pajer, \emph{{The Cosmological Optical
  Theorem}}, \href{https://doi.org/10.1088/1475-7516/2021/04/021}{\emph{JCAP}
  {\bfseries 04} (2021) 021}
  [\href{https://arxiv.org/abs/2009.02898}{{\ttfamily 2009.02898}}].

\bibitem{MLT}
S.~Jazayeri, E.~Pajer and D.~Stefanyszyn, \emph{{From locality and unitarity to
  cosmological correlators}},
  \href{https://doi.org/10.1007/JHEP10(2021)065}{\emph{JHEP} {\bfseries 10}
  (2021) 065} [\href{https://arxiv.org/abs/2103.08649}{{\ttfamily
  2103.08649}}].

\bibitem{CosmoBootstrap1}
N.~Arkani-Hamed, D.~Baumann, H.~Lee and G.L.~Pimentel, \emph{{The Cosmological
  Bootstrap: Inflationary Correlators from Symmetries and Singularities}},
  \href{https://doi.org/10.1007/JHEP04(2020)105}{\emph{JHEP} {\bfseries 04}
  (2020) 105} [\href{https://arxiv.org/abs/1811.00024}{{\ttfamily
  1811.00024}}].

\bibitem{CosmoBootstrap2}
D.~Baumann, C.~Duaso~Pueyo, A.~Joyce, H.~Lee and G.L.~Pimentel, \emph{{The
  cosmological bootstrap: weight-shifting operators and scalar seeds}},
  \href{https://doi.org/10.1007/JHEP12(2020)204}{\emph{JHEP} {\bfseries 12}
  (2020) 204} [\href{https://arxiv.org/abs/1910.14051}{{\ttfamily
  1910.14051}}].

\bibitem{Cheung:2007st}
C.~Cheung, P.~Creminelli, A.L.~Fitzpatrick, J.~Kaplan and L.~Senatore,
  \emph{{The Effective Field Theory of Inflation}},
  \href{https://doi.org/10.1088/1126-6708/2008/03/014}{\emph{JHEP} {\bfseries
  03} (2008) 014} [\href{https://arxiv.org/abs/0709.0293}{{\ttfamily
  0709.0293}}].

\bibitem{Green:2020ebl}
D.~Green and E.~Pajer, \emph{{On the Symmetries of Cosmological
  Perturbations}},  \href{https://arxiv.org/abs/2004.09587}{{\ttfamily
  2004.09587}}.

\bibitem{Soda:2011am}
J.~Soda, H.~Kodama and M.~Nozawa, \emph{{Parity Violation in Graviton
  Non-gaussianity}}, \href{https://doi.org/10.1007/JHEP08(2011)067}{\emph{JHEP}
  {\bfseries 08} (2011) 067} [\href{https://arxiv.org/abs/1106.3228}{{\ttfamily
  1106.3228}}].

\bibitem{GhostCondensate}
N.~Arkani-Hamed, H.-C.~Cheng, M.A.~Luty and S.~Mukohyama, \emph{{Ghost
  condensation and a consistent infrared modification of gravity}},
  \href{https://doi.org/10.1088/1126-6708/2004/05/074}{\emph{JHEP} {\bfseries
  05} (2004) 074} [\href{https://arxiv.org/abs/hep-th/0312099}{{\ttfamily
  hep-th/0312099}}].

\bibitem{GhostInflation}
N.~Arkani-Hamed, P.~Creminelli, S.~Mukohyama and M.~Zaldarriaga, \emph{{Ghost
  inflation}}, \href{https://doi.org/10.1088/1475-7516/2004/04/001}{\emph{JCAP}
  {\bfseries 04} (2004) 001}
  [\href{https://arxiv.org/abs/hep-th/0312100}{{\ttfamily hep-th/0312100}}].

\bibitem{Qin:2022fbv}
Z.~Qin and Z.-Z.~Xianyu, \emph{{Helical Inflation Correlators: Partial
  Mellin-Barnes and Bootstrap Equations}},
  \href{https://arxiv.org/abs/2208.13790}{{\ttfamily 2208.13790}}.

\bibitem{snowmass}
G.~Cabass, M.M.~Ivanov, M.~Lewandowski, M.~Mirbabayi and M.~Simonovi\'c,
  \emph{{Snowmass White Paper: Effective Field Theories in Cosmology}},  in
  \emph{{2022 Snowmass Summer Study}}, 3, 2022
  [\href{https://arxiv.org/abs/2203.08232}{{\ttfamily 2203.08232}}].

\bibitem{Maldacena:2002vr}
J.M.~Maldacena, \emph{{Non-Gaussian features of primordial fluctuations in
  single field inflationary models}},
  \href{https://doi.org/10.1088/1126-6708/2003/05/013}{\emph{JHEP} {\bfseries
  05} (2003) 013} [\href{https://arxiv.org/abs/astro-ph/0210603}{{\ttfamily
  astro-ph/0210603}}].

\bibitem{Cusin:2017mzw}
G.~Cusin, M.~Lewandowski and F.~Vernizzi, \emph{{Nonlinear Effective Theory of
  Dark Energy}},
  \href{https://doi.org/10.1088/1475-7516/2018/04/061}{\emph{JCAP} {\bfseries
  04} (2018) 061} [\href{https://arxiv.org/abs/1712.02782}{{\ttfamily
  1712.02782}}].

\bibitem{Senatore:2009gt}
L.~Senatore, K.M.~Smith and M.~Zaldarriaga, \emph{{Non-Gaussianities in Single
  Field Inflation and their Optimal Limits from the WMAP 5-year Data}},
  \href{https://doi.org/10.1088/1475-7516/2010/01/028}{\emph{JCAP} {\bfseries
  01} (2010) 028} [\href{https://arxiv.org/abs/0905.3746}{{\ttfamily
  0905.3746}}].

\bibitem{Alvarez:2014vva}
M.~Alvarez et~al., \emph{{Testing Inflation with Large Scale Structure:
  Connecting Hopes with Reality}},
  \href{https://arxiv.org/abs/1412.4671}{{\ttfamily 1412.4671}}.

\bibitem{Cabass:2022wjy}
G.~Cabass, M.M.~Ivanov, O.H.E.~Philcox, M.~Simonovi\'c and M.~Zaldarriaga,
  \emph{{Constraints on Single-Field Inflation from the BOSS Galaxy Survey}},
  \href{https://doi.org/10.1103/PhysRevLett.129.021301}{\emph{Phys. Rev. Lett.}
  {\bfseries 129} (2022) 021301}
  [\href{https://arxiv.org/abs/2201.07238}{{\ttfamily 2201.07238}}].

\bibitem{DAmico:2022gki}
G.~D'Amico, M.~Lewandowski, L.~Senatore and P.~Zhang, \emph{{Limits on
  primordial non-Gaussianities from BOSS galaxy-clustering data}},
  \href{https://arxiv.org/abs/2201.11518}{{\ttfamily 2201.11518}}.

\bibitem{Burgess:2014eoa}
C.P.~Burgess, R.~Holman, G.~Tasinato and M.~Williams, \emph{{EFT Beyond the
  Horizon: Stochastic Inflation and How Primordial Quantum Fluctuations Go
  Classical}}, \href{https://doi.org/10.1007/JHEP03(2015)090}{\emph{JHEP}
  {\bfseries 03} (2015) 090} [\href{https://arxiv.org/abs/1408.5002}{{\ttfamily
  1408.5002}}].

\bibitem{tetrachirality}
A.~Rassat and P.W.~Fowler, \emph{{Is There a ``Most Chiral Tetrahedron''?}},
  \href{https://doi.org/https://doi.org/10.1002/chem.200400869}{\emph{Chemistry
  Europe} {\bfseries 10} (2004) 6575}.

\bibitem{Maldacena:2011nz}
J.M.~Maldacena and G.L.~Pimentel, \emph{{On graviton non-Gaussianities during
  inflation}}, \href{https://doi.org/10.1007/JHEP09(2011)045}{\emph{JHEP}
  {\bfseries 09} (2011) 045} [\href{https://arxiv.org/abs/1104.2846}{{\ttfamily
  1104.2846}}].

\bibitem{Mata:2012bx}
I.~Mata, S.~Raju and S.~Trivedi, \emph{{CMB from CFT}},
  \href{https://doi.org/10.1007/JHEP07(2013)015}{\emph{JHEP} {\bfseries 07}
  (2013) 015} [\href{https://arxiv.org/abs/1211.5482}{{\ttfamily 1211.5482}}].

\bibitem{Pajer:2016ieg}
E.~Pajer, G.L.~Pimentel and J.V.S.~Van~Wijck, \emph{{The Conformal Limit of
  Inflation in the Era of CMB Polarimetry}},
  \href{https://doi.org/10.1088/1475-7516/2017/06/009}{\emph{JCAP} {\bfseries
  06} (2017) 009} [\href{https://arxiv.org/abs/1609.06993}{{\ttfamily
  1609.06993}}].

\bibitem{Bzowski:2013sza}
A.~Bzowski, P.~McFadden and K.~Skenderis, \emph{{Implications of conformal
  invariance in momentum space}},
  \href{https://doi.org/10.1007/JHEP03(2014)111}{\emph{JHEP} {\bfseries 03}
  (2014) 111} [\href{https://arxiv.org/abs/1304.7760}{{\ttfamily 1304.7760}}].

\bibitem{CosmoBootstrap3}
D.~Baumann, C.~Duaso~Pueyo, A.~Joyce, H.~Lee and G.L.~Pimentel, \emph{{The
  Cosmological Bootstrap: Spinning Correlators from Symmetries and
  Factorization}},  \href{https://arxiv.org/abs/2005.04234}{{\ttfamily
  2005.04234}}.

\bibitem{Goodhew:2021oqg}
H.~Goodhew, S.~Jazayeri, M.H.~Gordon~Lee and E.~Pajer, \emph{{Cutting
  cosmological correlators}},
  \href{https://doi.org/10.1088/1475-7516/2021/08/003}{\emph{JCAP} {\bfseries
  08} (2021) 003} [\href{https://arxiv.org/abs/2104.06587}{{\ttfamily
  2104.06587}}].

\bibitem{Cespedes:2020xqq}
S.~C\'espedes, A.-C.~Davis and S.~Melville, \emph{{On the time evolution of
  cosmological correlators}},
  \href{https://doi.org/10.1007/JHEP02(2021)012}{\emph{JHEP} {\bfseries 02}
  (2021) 012} [\href{https://arxiv.org/abs/2009.07874}{{\ttfamily
  2009.07874}}].

\bibitem{Baumann:2021fxj}
D.~Baumann, W.-M.~Chen, C.~Duaso~Pueyo, A.~Joyce, H.~Lee and G.L.~Pimentel,
  \emph{{Linking the Singularities of Cosmological Correlators}},
  \href{https://arxiv.org/abs/2106.05294}{{\ttfamily 2106.05294}}.

\bibitem{sCOTt}
S.~Melville and E.~Pajer, \emph{{Cosmological Cutting Rules}},
  \href{https://doi.org/10.1007/JHEP05(2021)249}{\emph{JHEP} {\bfseries 05}
  (2021) 249} [\href{https://arxiv.org/abs/2103.09832}{{\ttfamily
  2103.09832}}].

\bibitem{Bordin:2018pca}
L.~Bordin, P.~Creminelli, A.~Khmelnitsky and L.~Senatore, \emph{{Light
  Particles with Spin in Inflation}},
  \href{https://doi.org/10.1088/1475-7516/2018/10/013}{\emph{JCAP} {\bfseries
  10} (2018) 013} [\href{https://arxiv.org/abs/1806.10587}{{\ttfamily
  1806.10587}}].

\bibitem{Arkani-Hamed:2017fdk}
N.~Arkani-Hamed, P.~Benincasa and A.~Postnikov, \emph{{Cosmological Polytopes
  and the Wavefunction of the Universe}},
  \href{https://arxiv.org/abs/1709.02813}{{\ttfamily 1709.02813}}.

\bibitem{Raju:2012zr}
S.~Raju, \emph{{New Recursion Relations and a Flat Space Limit for AdS/CFT
  Correlators}}, \href{https://doi.org/10.1103/PhysRevD.85.126009}{\emph{Phys.
  Rev. D} {\bfseries 85} (2012) 126009}
  [\href{https://arxiv.org/abs/1201.6449}{{\ttfamily 1201.6449}}].

\bibitem{Chen:2017ryl}
X.~Chen, Y.~Wang and Z.-Z.~Xianyu, \emph{{Schwinger-Keldysh Diagrammatics for
  Primordial Perturbations}},
  \href{https://doi.org/10.1088/1475-7516/2017/12/006}{\emph{JCAP} {\bfseries
  12} (2017) 006} [\href{https://arxiv.org/abs/1703.10166}{{\ttfamily
  1703.10166}}].

\bibitem{Musso:2006pt}
M.~Musso, \emph{{A new diagrammatic representation for correlation functions in
  the in-in formalism}},
  \href{https://doi.org/10.1007/JHEP11(2013)184}{\emph{JHEP} {\bfseries 11}
  (2013) 184} [\href{https://arxiv.org/abs/hep-th/0611258}{{\ttfamily
  hep-th/0611258}}].

\bibitem{Melville:2021lst}
S.~Melville and E.~Pajer, \emph{{Cosmological Cutting Rules}},
  \href{https://doi.org/10.1007/JHEP05(2021)249}{\emph{JHEP} {\bfseries 05}
  (2021) 249} [\href{https://arxiv.org/abs/2103.09832}{{\ttfamily
  2103.09832}}].

\bibitem{Sleight:2021plv}
C.~Sleight and M.~Taronna, \emph{{From dS to AdS and back}},
  \href{https://arxiv.org/abs/2109.02725}{{\ttfamily 2109.02725}}.

\bibitem{DiPietro:2021sjt}
L.~Di~Pietro, V.~Gorbenko and S.~Komatsu, \emph{{Analyticity and unitarity for
  cosmological correlators}},
  \href{https://doi.org/10.1007/JHEP03(2022)023}{\emph{JHEP} {\bfseries 03}
  (2022) 023} [\href{https://arxiv.org/abs/2108.01695}{{\ttfamily
  2108.01695}}].

\bibitem{BBBB}
E.~Pajer, \emph{{Building a Boostless Bootstrap for the Bispectrum}},
  \href{https://doi.org/10.1088/1475-7516/2021/01/023}{\emph{JCAP} {\bfseries
  01} (2021) 023} [\href{https://arxiv.org/abs/2010.12818}{{\ttfamily
  2010.12818}}].

\bibitem{CabassBordin}
L.~Bordin and G.~Cabass, \emph{{Graviton non-Gaussianities and Parity Violation
  in the EFT of Inflation}},
  \href{https://doi.org/10.1088/1475-7516/2020/07/014}{\emph{JCAP} {\bfseries
  07} (2020) 014} [\href{https://arxiv.org/abs/2004.00619}{{\ttfamily
  2004.00619}}].

\bibitem{Shiraishi:2011st}
M.~Shiraishi, D.~Nitta and S.~Yokoyama, \emph{{Parity Violation of Gravitons in
  the CMB Bispectrum}}, \href{https://doi.org/10.1143/PTP.126.937}{\emph{Prog.
  Theor. Phys.} {\bfseries 126} (2011) 937}
  [\href{https://arxiv.org/abs/1108.0175}{{\ttfamily 1108.0175}}].

\bibitem{Ivanov:2014yla}
M.M.~Ivanov and S.~Sibiryakov, \emph{{UV-extending Ghost Inflation}},
  \href{https://doi.org/10.1088/1475-7516/2014/05/045}{\emph{JCAP} {\bfseries
  05} (2014) 045} [\href{https://arxiv.org/abs/1402.4964}{{\ttfamily
  1402.4964}}].

\bibitem{Ashoorioon:2018uey}
A.~Ashoorioon, R.~Casadio, M.~Cicoli, G.~Geshnizjani and H.J.~Kim,
  \emph{{Extended Effective Field Theory of Inflation}},
  \href{https://doi.org/10.1007/JHEP02(2018)172}{\emph{JHEP} {\bfseries 02}
  (2018) 172} [\href{https://arxiv.org/abs/1802.03040}{{\ttfamily
  1802.03040}}].

\bibitem{Dubovsky:2006vk}
S.L.~Dubovsky and S.M.~Sibiryakov, \emph{{Spontaneous breaking of Lorentz
  invariance, black holes and perpetuum mobile of the 2nd kind}},
  \href{https://doi.org/10.1016/j.physletb.2006.05.074}{\emph{Phys. Lett. B}
  {\bfseries 638} (2006) 509}
  [\href{https://arxiv.org/abs/hep-th/0603158}{{\ttfamily hep-th/0603158}}].

\bibitem{Arkani-Hamed:2007ryv}
N.~Arkani-Hamed, S.~Dubovsky, A.~Nicolis, E.~Trincherini and G.~Villadoro,
  \emph{{A Measure of de Sitter entropy and eternal inflation}},
  \href{https://doi.org/10.1088/1126-6708/2007/05/055}{\emph{JHEP} {\bfseries
  05} (2007) 055} [\href{https://arxiv.org/abs/0704.1814}{{\ttfamily
  0704.1814}}].

\bibitem{Mukohyama:2009rk}
S.~Mukohyama, \emph{{Ghost condensate and generalized second law}},
  \href{https://doi.org/10.1088/1126-6708/2009/09/070}{\emph{JHEP} {\bfseries
  09} (2009) 070} [\href{https://arxiv.org/abs/0901.3595}{{\ttfamily
  0901.3595}}].

\bibitem{Mukohyama:2009um}
S.~Mukohyama, \emph{{Can ghost condensate decrease entropy?}},
  \href{https://doi.org/10.2174/1874381101003020030}{\emph{Open Astron. J.}
  {\bfseries 3} (2010) 30} [\href{https://arxiv.org/abs/0908.4123}{{\ttfamily
  0908.4123}}].

\bibitem{Jazayeri:2016jav}
S.~Jazayeri, S.~Mukohyama, R.~Saitou and Y.~Watanabe, \emph{{Ghost inflation
  and de Sitter entropy}},
  \href{https://doi.org/10.1088/1475-7516/2016/08/002}{\emph{JCAP} {\bfseries
  08} (2016) 002} [\href{https://arxiv.org/abs/1602.06511}{{\ttfamily
  1602.06511}}].

\bibitem{Son:2002zn}
D.T.~Son, \emph{{Low-energy quantum effective action for relativistic
  superfluids}},  \href{https://arxiv.org/abs/hep-ph/0204199}{{\ttfamily
  hep-ph/0204199}}.

\bibitem{Pajer:2018egx}
E.~Pajer and D.~Stefanyszyn, \emph{{Symmetric Superfluids}},
  \href{https://doi.org/10.1007/JHEP06(2019)008}{\emph{JHEP} {\bfseries 06}
  (2019) 008} [\href{https://arxiv.org/abs/1812.05133}{{\ttfamily
  1812.05133}}].

\bibitem{Nicolis:2015sra}
A.~Nicolis, R.~Penco, F.~Piazza and R.~Rattazzi, \emph{{Zoology of condensed
  matter: Framids, ordinary stuff, extra-ordinary stuff}},
  \href{https://doi.org/10.1007/JHEP06(2015)155}{\emph{JHEP} {\bfseries 06}
  (2015) 155} [\href{https://arxiv.org/abs/1501.03845}{{\ttfamily
  1501.03845}}].

\bibitem{Delacretaz:2014oxa}
L.V.~Delacr\'etaz, S.~Endlich, A.~Monin, R.~Penco and F.~Riva,
  \emph{{(Re-)Inventing the Relativistic Wheel: Gravity, Cosets, and Spinning
  Objects}}, \href{https://doi.org/10.1007/JHEP11(2014)008}{\emph{JHEP}
  {\bfseries 11} (2014) 008} [\href{https://arxiv.org/abs/1405.7384}{{\ttfamily
  1405.7384}}].

\bibitem{Higuchi:1986py}
A.~Higuchi, \emph{{Forbidden Mass Range for Spin-2 Field Theory in De Sitter
  Space-time}}, \href{https://doi.org/10.1016/0550-3213(87)90691-2}{\emph{Nucl.
  Phys. B} {\bfseries 282} (1987) 397}.

\bibitem{WFCtoCorrelators2}
G.~Goon, K.~Hinterbichler, A.~Joyce and M.~Trodden, \emph{{Shapes of gravity:
  Tensor non-Gaussianity and massive spin-2 fields}},
  \href{https://doi.org/10.1007/JHEP10(2019)182}{\emph{JHEP} {\bfseries 10}
  (2019) 182} [\href{https://arxiv.org/abs/1812.07571}{{\ttfamily
  1812.07571}}].

\bibitem{Arkani-Hamed:2018bjr}
N.~Arkani-Hamed and P.~Benincasa, \emph{{On the Emergence of Lorentz Invariance
  and Unitarity from the Scattering Facet of Cosmological Polytopes}},
  \href{https://arxiv.org/abs/1811.01125}{{\ttfamily 1811.01125}}.

\bibitem{Jazayeri:2022kjy}
S.~Jazayeri and S.~Renaux-Petel, \emph{{Cosmological Bootstrap in Slow
  Motion}},  \href{https://arxiv.org/abs/2205.10340}{{\ttfamily 2205.10340}}.

\bibitem{Pimentel:2022fsc}
G.L.~Pimentel and D.-G.~Wang, \emph{{Boostless Cosmological Collider
  Bootstrap}},  \href{https://arxiv.org/abs/2205.00013}{{\ttfamily
  2205.00013}}.

\bibitem{Tong:2021wai}
X.~Tong, Y.~Wang and Y.~Zhu, \emph{{Cutting rule for cosmological collider
  signals: a bulk evolution perspective}},
  \href{https://doi.org/10.1007/JHEP03(2022)181}{\emph{JHEP} {\bfseries 03}
  (2022) 181} [\href{https://arxiv.org/abs/2112.03448}{{\ttfamily
  2112.03448}}].

\bibitem{Lee:2016vti}
H.~Lee, D.~Baumann and G.L.~Pimentel, \emph{{Non-Gaussianity as a Particle
  Detector}}, \href{https://doi.org/10.1007/JHEP12(2016)040}{\emph{JHEP}
  {\bfseries 12} (2016) 040}
  [\href{https://arxiv.org/abs/1607.03735}{{\ttfamily 1607.03735}}].

\end{thebibliography}\endgroup
\end{document}